\documentclass[11pt]{article}
\usepackage[a4paper,hmargin=1.0in,vmargin=1.0in]{geometry}
\usepackage{mathtools,amsthm,amssymb,setspace,sectsty,bbm}
\usepackage[round]{natbib}
\usepackage[usenames,dvipsnames]{xcolor}
\usepackage[linktocpage=true,pagebackref=true,colorlinks,allcolors=blue,bookmarks=true,bookmarksopen,bookmarksnumbered]{hyperref}

\newtheoremstyle{DStheorem}
  {\topsep}
  {\topsep}
  {\itshape}
  {0pt}
  {\scshape}
  {.}
  { }
  {\thmname{#1}\thmnumber{ #2}\thmnote{ (#3)}}
\theoremstyle{DStheorem}
\newtheorem{theorem}{Theorem}[section]
\newtheorem{lemma}[theorem]{Lemma}
\newtheorem{claim}[theorem]{Claim}
\newtheorem{observation}[theorem]{Observation}
\newtheorem{corollary}[theorem]{Corollary}

\let\oldproofname=\proofname
\renewcommand{\proofname}{\rm\sc{\oldproofname}}

\newcommand{\bstitle}[1]{\texorpdfstring{$\boldsymbol{#1}$}{}}
\newcommand{\MyAbove}[2]{\genfrac{}{}{0pt}{}{#1}{#2}}
\newcommand{\bs}[1]{\boldsymbol{#1}}
\newcommand{\bbR}{\mathbbm{R}}
\newcommand{\bbN}{\mathbbm{N}}
\newcommand{\eps}{\epsilon}

\newcommand{\opt}{\mathrm{OPT}}
\newcommand{\ex}[1]{\mathbbm{E}\left[#1\right]}
\newcommand{\expar}[1]{\mathbbm{E}[#1]}
\newcommand{\exsub}[2]{\mathbbm{E}_{#1}\left[#2\right]}
\newcommand{\exsubnop}[1]{\mathbbm{E}_{#1}}
\newcommand{\exsubpar}[2]{\mathbbm{E}_{#1}[#2]}
\newcommand{\pr}[1]{\mathrm{Pr}\left[#1\right]}

\newcommand{\prsub}[2]{\mathrm{Pr}_{#1}\left[#2\right]}
\newcommand{\prsubpar}[2]{\mathrm{Pr}_{#1}[#2]}

\newcommand{\mydense}{\mathrm{dense}}
\newcommand{\mycomb}{\mathrm{combined}}
\newcommand{\mycyc}{\mathrm{cycle}}
\newcommand{\mynext}{\mathrm{next}}
\newcommand{\mydp}{\mathrm{DP}}
\newcommand{\mylb}{\mathrm{LB}}
\newcommand{\myub}{\mathrm{UB}}
\newcommand{\mypref}{\mathrm{prefix}}
\newcommand{\mysuff}{\mathrm{suffix}}
\newcommand{\mymid}{\mathrm{mid}}

\sectionfont{\large} \subsectionfont{\normalsize}
\allowdisplaybreaks
\makeindex

\begin{document}

\begin{titlepage}

\title{New Approximation Guarantees for \\
The Economic Warehouse Lot Scheduling Problem}
\author{%
Danny Segev\thanks{School of Mathematical Sciences and Coller School of Management, Tel Aviv University, Tel Aviv 69978, Israel. Email: {\tt segevdanny@tauex.tau.ac.il}. Supported by Israel Science Foundation grant 1407/20.}}
\date{}
\maketitle

\setcounter{page}{200}
\thispagestyle{empty}

\begin{abstract}
In this paper, we present long-awaited algorithmic advances toward the efficient construction of near-optimal replenishment policies for a true inventory management classic, the economic warehouse lot scheduling problem. While this paradigm has accumulated a massive body of surrounding literature since its inception in the late '50s, we are still very much in the dark as far as basic computational questions are concerned, perhaps due to the evasive nature of dynamic policies in this context. The latter feature forced earlier attempts to either study highly-structured classes of policies or to forgo provably-good performance guarantees altogether; to this day, rigorously analyzable results have been few and far between. 

The current paper develops novel analytical foundations for directly competing against dynamic policies. Combined with further algorithmic progress and newly-gained insights, these ideas culminate to a polynomial-time approximation scheme for constantly-many commodities as well as to a proof-of-concept $(2-\frac{17}{5000} + \eps )$-approximation for general problem instances. In this regard, the efficient design of $\eps$-optimal dynamic policies appeared to have been out of reach, since beyond algorithmic challenges by themselves, even the polynomial-space representation of such policies has been a fundamental open question. On the other front, our sub-$2$-approximation constitutes  the first improvement over the performance guarantees achievable via ``stationary order sizes and stationary intervals'' (SOSI) policies, which have been state-of-the-art since the mid-'90s \citep*{Anily91, GallegoQS96}.
\end{abstract}


\bigskip \bigskip \bigskip \bigskip

\begin{center}
Dedicated to Prof.\ Shoshi Anily, on occasion of her retirement.
\end{center}

\end{titlepage}

\pagestyle{empty}
\tableofcontents

\newpage
\pagestyle{plain}
\setcounter{page}{1}
\section{Introduction}

The primary intent of this paper is to present long-awaited algorithmic advances toward the efficient construction of near-optimal replenishment policies for a true inventory management classic, the economic warehouse lot scheduling problem. Appearing since the late '50s under a wide range of nicknames, decades-long investigations of this fundamental paradigm have lead to a massive body of surrounding literature. However, we are still very much in the dark as far as basic computational questions are concerned, perhaps due to the evasive nature of dynamic replenishment policies in this context, forcing earlier attempts to either study highly-structured classes of policies or to forgo provably-good performance guarantees altogether. To this day, rigorously analyzable results have been few and far between, to say the least. The current paper develops novel analytical foundations for directly competing against dynamic policies; combined with further algorithmic progress and newly-gained insights, these ideas culminate to a polynomial-time approximation scheme for constantly-many commodities as well as to a proof-of-concept sub-$2$-approximation for general problem instances. As explained later on, the efficient design of $\eps$-optimal dynamic policies appeared to have been out of reach, since beyond  algorithmic challenges by themselves, even the polynomial-space representation of such policies has been a major open question. On the other front, our sub-$2$-approximation constitutes  the first improvement over the performance guarantees achievable via ``stationary order sizes and stationary intervals'' (SOSI) policies, which have been state-of-the-art since the mid-'90s 

Given the extensive body of work that has accumulated around the economic warehouse lot scheduling problem, including heuristic approaches, experimental studies, and concrete  applications, there is no way we can do justice and present an in-depth overview of this literature. For a broader perspective on these topics, readers are referred to excellent book chapters \citep{HadleyW1963, JohnsonM74, HaxC84, Zipkin00, SimchiLeviCB14, NahmiasL15}, as well as to the references therein, noting that it may be especially interesting to revisit pioneering papers such as those of \cite{Holt58}, \cite{Homer1966}, \cite{PageP76}, and \cite{Zoller77}. From a bird's-eye view, our problem of interest and most of its nearby relatives are all inherently concerned with  lot-sizing and scheduling decisions for multiple commodities over a given planning horizon, where we wish to minimize long-run average operating costs. However, on top of marginal ordering and inventory holding costs, a particularly problematic feature of this model is the interaction between different commodities, incurred due to sharing a common resource; the latter will be referred to as ``warehouse space'', just to align ourselves with classic terminology. In this regard, capacity-feasible replenishment policies should ensure that, at any point in time, the momentary inventory levels of all commodities can be jointly packed within our warehouse's capacity, with each commodity contributing toward this space requirement at an individual rate. Unfortunately, this seemingly-innocent constraint generally forces optimal policies to violate many fundamental properties that have been explored and exploited for single-commodity models. Consequently, we have very sparse theoretical grounds to rely on, leading to long-standing algorithmic questions about how provably-good capacity-feasible policies can be efficiently identified, as well as to a deep analytical void regarding their structural characterization.

Moving forward, to delve into the finer details of these questions, Section~\ref{subsec:model_definition} provides a complete mathematical description of the economic warehouse lot scheduling problem in its broadest  form. Subsequently, Section~\ref{subsec:related_work} sheds light on cornerstone theoretical advances, including the well-known SOSI-driven approach, whose approximation guarantees have been state-of-the-art to this day. Concurrently, we will elaborate on the primary open questions that motivated our work. Finally, Section~\ref{subsec:contributions} discusses the main contributions of this paper, leaving their structural characterization, algorithmic ideas, and analytical arguments
to be presented in subsequent sections.

\subsection{Model formulation} \label{subsec:model_definition}

\paragraph{The economic order quantity model.} To describe the inner-workings of our model in an accessible way, we start off by  introducing its basic building block, the economic order quantity (EOQ) model. At a high level, our supposed objective is to identify an optimal dynamic replenishment policy for a single commodity, aiming to minimize its long-run average cost over the continuous planning horizon $[0,\infty)$. Without loss of generality, this commodity is assumed to be characterized by a stationary demand rate of $1$, to be completely met upon occurrence, meaning that lost sales or back orders are not permitted. In this context, the notion of a ``dynamic'' policy ${\cal P}$ will be captured by two ingredients: 
\begin{itemize}
    \item An sequence of ordering points $0 = \tau_0 < \tau_1 < \tau_2 < \cdots$ covering the entire planning horizon $[0,\infty)$, in the sense that $\lim_{k \to \infty} \tau_k = \infty$.

    \item The respective real-valued ordering quantities $q_0, q_1, q_2, \ldots$ at these points.
\end{itemize}
For ease of notation, $I( {\cal P}, t )$ will stands for the inventory level we end up with at time $t$, when operating under the replenishment policy ${\cal P}$. One straightforward way to express this function is via the relation $I( {\cal P}, t ) = \sum_{k \geq 0: \tau_k \leq t} q_k - t$. To parse this expression, note that $\sum_{k \geq 0: \tau_k \leq t} q_k$ represents the total ordering quantity up to and including time $t$; this term consists of finitely-many summands, since $\lim_{k \to \infty} \tau_k = \infty$. In addition, the second term, $t$, is precisely the cumulative consumption up until this time, due to having a unit demand rate. As such, we are indeed meeting our demand at any point in time via previously-placed orders if and only if $I( {\cal P}, t ) \geq 0$ for all $t \in [0,\infty)$. Any policy satisfying this condition is said to be feasible, and we designate the collection of such policies by ${\cal F}$.

We proceed by explaining how the long-run average cost of any policy ${\cal P} \in {\cal F}$ is structured. For this purpose, each of the ordering points $\tau_0, \tau_1 ,\tau_2, \ldots$ incurs a fixed set-up cost of $K$, regardless of its quantity. In other words, letting $N( {\cal P}, [0,t)) = | \{ k \in \bbN_0 : \tau_k < t \} |$ be the number of orders placed along $[0,t)$, this interval is associated with a total ordering cost of ${\cal K}( {\cal P}, [0,t)) = K \cdot N( {\cal P}, [0,t))$. At least intuitively, these costs by themselves motivate us to place infrequent orders. On the other hand, we are concurrently facing a linear holding cost of $2H$, incurred per time unit for each inventory unit in stock. Technically speaking, ${\cal H}( {\cal P}, [0,t))$ will denote our total holding cost across the interval $[0,t)$, given by ${\cal H}( {\cal P}, [0,t)) = 2H \cdot \int_{[0,t)} I( {\cal P}, \tau ) \mathrm{d} \tau$. These costs by themselves are pulling us in the opposite direction, as we generally wish to avoid high inventory levels via frequent orders. Putting both ingredients into a single objective, the combined cost of a feasible policy ${\cal P}$ along $[0,t)$ will be designated by $C( {\cal P}, [0,t)) = {\cal K}( {\cal P}, [0,t)) + {\cal H}( {\cal P}, [0,t))$. In turn, the long-run average cost of this policy can be specified by
\begin{equation} \label{eqn:long_run_cost_single}
C( {\cal P} ) ~~=~~ \limsup_{t \to \infty} \frac{ C( {\cal P}, [0,t)) }{ t } \ .    
\end{equation}
As a side note, one can easily come up with feasible policies ${\cal P}$ for which $\lim_{t \to \infty} \frac{ C( {\cal P}, [0,t)) }{ t }$ does not exist, meaning that $\limsup$ above is indeed required.

Based on the preceding description, in the economic order quantity problem, our goal is to identify a feasible replenishment policy ${\cal P}$ whose long-run average cost is minimized, in the sense that $C( {\cal P} ) = \min_{ \hat{\cal P} \in {\cal F}} C( \hat{\cal P} )$. By consulting  relevant textbooks, such as those of \citet[Sec.~3]{Zipkin00}, \citet[Sec.~2]{MuckstadtS10}, or \citet[Sec.~7.1]{SimchiLeviCB14}, unfamiliar readers will quickly find out that the latter minimum is indeed attained, via the  extremely simple class of stationary order sizes and stationary intervals (SOSI) policies. These policies are characterized by a single parameter, $T$, standing for the uniform time interval between successive orders. Having fixed this parameter, orders will be placed at the time points $0, T, 2T, 3T, \ldots$, each consisting of exactly $T$ units, meaning that zero-inventory levels are reached at each of these points. As such, since the long-run average cost of a SOSI policy $T$ can be written as $C(T) = \frac{ K }{ T } + HT$, the economic order quantity problem admits an elegant closed-form solution. The next claim summarizes several well-known properties exhibited by this function, all following from elementary
calculus arguments.

\begin{claim} \label{clm:EOQ_properties}
The cost function $C(T) = \frac{ K }{ T } + HT$ satisfies the next few properties:
\begin{enumerate}
    \item $C$ is strictly convex.

    \item The unique minimizer of $C$ is $T^* = \sqrt{ K / H }$.

    \item For every $\alpha > 0$ and $T > 0$,
    \[ C( \alpha T ) ~~\leq~~ \max \left\{ \alpha, \frac{ 1 }{ \alpha } \right\} \cdot C(T) \qquad \text{and} \qquad C( \alpha T ) + C ( T / \alpha ) ~~=~~ \frac{ \alpha^2 + 1 }{ \alpha }  \cdot C(T) \ . \]
\end{enumerate}
\end{claim}

\paragraph{The economic warehouse lot scheduling problem.} Given  the optimality of SOSI policies for the economic order quantity model, one can retrospectively wonder whether placing dynamic policies at the heart of our exposition makes any sense. Circling back to this issue in the sequel, we proceed by explaining how these foundations allow for a smooth transition to the economic warehouse lot scheduling problem, whose essence can be encapsulated into the next high-level question:
\begin{quote}
{\em How should we coordinate multiple economic order quantity models, when different commodities are tied together by sharing a common resource?}
\end{quote}
Specifically, we wish to synchronize the lot sizing of $n$ distinct commodities, where each commodity $i \in [n]$ is coupled with its own EOQ model, parameterized by ordering
and holding costs $K_i$ and $2H_i$, respectively. As explained above, deciding to replenish this commodity via the dynamic policy ${\cal P}_i \in {\cal F}$ would lead to a marginal long-run average cost of $C( {\cal P}_i )$, prescribed by equation~\eqref{eqn:long_run_cost_single}. However, the complicating caveat is that we are concurrently facing a ``warehouse space'' constraint, stating that the momentary inventory levels of all commodities can be jointly packed within our warehouse's capacity, with each commodity contributing toward this space requirement at an individual rate. 

To formalize this notion, for every commodity $i \in [n]$, each inventory unit requires $\gamma_i$ amount of space to be stocked in a common warehouse, whose overall capacity will be denoted by ${\cal V}$. Consequently, for any dynamic replenishment policy ${\cal P} = ( {\cal P}_1, \ldots, {\cal P}_n ) \in {\cal F}^n$, by recalling that $\{ I( {\cal P}_i, t) \}_{i \in [n]}$ designate the underlying inventory levels at time $t \in [0, \infty)$, it follows that the  warehouse space occupied at that point in time is given by $V( {\cal P}, t) = \sum_{i \in [n]} \gamma_i \cdot I( {\cal P}_i, t )$. In turn, the maximal space ever occupied by this policy can be written as $V_{\max}( {\cal P} ) = \sup_{t \in [0, \infty)}  V( {\cal P}, t)$, and we say that ${\cal P}$ is capacity-feasible when $V_{\max}( {\cal P} ) \leq {\cal V}$. Clearly, this condition is much stronger than merely asking each of the marginal policies ${\cal P}_1, \ldots, {\cal P}_n$ to be feasible by itself.

Putting everything together, in the economic warehouse lot scheduling problem, our goal is
to determine a capacity-feasible replenishment policy ${\cal P} = ( {\cal P}_1, \ldots, {\cal P}_n )$ whose long-run average
cost $C( {\cal P} ) = \sum_{i \in [n]} C( {\cal P}_i )$ is minimized. Moving forward, it will be convenient to succinctly formulate this model as 
\begin{equation} \label{eqn:model_warehouse}
\tag{$\Pi$}
\begin{array}{ll}
{\displaystyle \min_{{\cal P} \in {\cal F}^n}} & C( {\cal P} ) \\
\text{s.t.} & V_{\max}( {\cal P} ) \leq {\cal V} 
\end{array}
\end{equation}
It is worth pointing out that the above-mentioned minimum is indeed attained by some capacity-feasible policy. The proof of this claim is rather straightforward, and we leave its details to be rediscovered by avid readers.

\subsection{Known results and open questions} \label{subsec:related_work}

In what follows, we survey cornerstone results regarding the economic warehouse lot scheduling problem, with a particular emphasis on those that are directly related to our basic research questions. For this purpose, we will be going in two complementary directions, one centered around rigorous algorithmic methods for efficiently identifying provably-good replenishment policies, and the other briefly discussing known intractability results; as previously mentioned, such findings have been extremely rare events. In regard to tangential research directions, given the extensive literature in this context, we are not seeing a reasonable way to present an all-inclusive overview. Further background on historical developments, heuristic approaches, and experimental studies can be attained by consulting numerous book chapters dedicated to these topics \citep{HadleyW1963, JohnsonM74, HaxC84, Zipkin00, SimchiLeviCB14, NahmiasL15}. 

\paragraph{Hardness results.} While our work is algorithmically driven for the most part, to thoroughly understand the boundaries of provably-good performance guarantees, it is worth briefly mentioning known intractability results. Along these lines, unlike many deterministic inventory models whose computational status is very much unclear, \cite*{GallegoSS92} were  successful in rigorously examining the plausibility of computing optimal replenishment policies in an efficient way. Their main finding in this context consists of a polynomial-time reduction from the infamous $3$-partition problem \citep{GareyJ75, GareyJ79}, showing that economic warehouse lot scheduling is strongly NP-hard in its decision problem setup. As an immediate byproduct, we know that optimal replenishment policies cannot be computed in polynomial time unless $\mathrm{P} = \mathrm{NP}$; moreover, this setting does not admit a fully polynomial-time approximation scheme under the same complexity assumption \citep[Sec.~8.3]{Vazirani01}.

\paragraph{Representation issues.} On top of complexity-related findings in terms of computation time, a seemingly unsurpassable obstacle on the road to efficiently identifying truly near-optimal policies can be captured by the next two  questions: 
\begin{quote}
\begin{itemize}
    \item {\em Is there any way we can represent optimal or $\eps$-optimal dynamic policies in polynomially-bounded memory space?}

    \item {\em Even more basically, can we address this challenge for $O(1)$ commodities?}
\end{itemize}
\end{quote}
In other words, suppose we are not worried about running times  to the slightest degree, and instead, we are only wondering about the feasibility of efficiently describing near-optimal dynamic policies. One could attempt to tackle this fundamental issue, for example, by proving that there exists a polynomially-representable pattern behind the sequence of ordering points $\tau_0^i, \tau_1^i, \tau_2^i, \ldots$ and quantities $q_0^i, q_1^i, q_2^i, \ldots$ associated with each commodity $i \in [n]$. Alternatively, one may wish to argue about the existence of a short-duration cyclic policy, where we will be seeing only a polynomial number of ordering points. To our knowledge, these questions have been wide open for decades, even for the most basic setting of only two commodities!

\paragraph{Constant-factor approximations via SOSI policies.} Prior to diving into state-of-the-art approximation guarantees, it is worth pointing out that, by superficially browsing through the abstracts of many early papers, we could mistakenly get the impression that economic warehouse lot scheduling is well-understood, at least in stylized settings with two or three commodities. However, the intrinsic caveat is that provably-good and efficient algorithmic results along these lines are focusing on very specific classes of policies rather than considering arbitrarily-structured dynamic replenishments, . 

The first breach in this wall should be attributed to the seminal work of \cite{Anily91}. While  her  objective was to optimize over the class of SOSI policies, she has been able to rigorously establish a bridge between economic warehouse lot scheduling and the SOSI-restricted economic order quantity model, where the underlying commodities are interacting via a warehouse space constraint on their  peak inventory levels. As a byproduct of these ideas, Anily  devised a polynomial-time construction of a SOSI policy whose long-run average cost is within factor $2$ of the minimum-possible within this class. Subsequently, the ingenious work of \cite*{GallegoQS96} fused Anily's approach with further insights, prescribing a lower bound on the long-run average cost of any capacity-feasible dynamic policy via an elegant convex relaxation. As a direct consequence, the authors have been successful at approximating the economic warehouse lot scheduling problem in its utmost generality, proposing a polynomial-time algorithm for identifying a SOSI policy whose cost is within factor $2$ of optimal; this time, ``optimal'' is referring to the best-possible dynamic policy! In Section~\ref{sec:anily_approx}, we will revisit these two papers in greater detail, due to their important role in addressing several easy-to-handle regimes within our overall approach.

\paragraph{Main open questions.} Quite surprisingly, while we have witnessed a continuous stream of literature on this topic across the last three decades, mostly revolving around stylized settings, heuristic approaches, and experimental studies, the algorithmic ideas of \cite{Anily91} and \cite{GallegoQS96} along with their lower-bounding mechanism still represent the best-known approximation guarantees for computing dynamic replenishment policies. Moreover, at present time, there is absolutely no distinction between a constant number and an arbitrary number of commodities in terms of deriving improved guarantees. Given this state of affairs, the fundamental
questions that lie at the heart of our work, as stated in many papers, books, and course materials, can be succinctly summarized as follows:
\begin{quote}
\begin{itemize}
    \item {\em Can we outperform these long-standing results, in any shape or form?}

    \item {\em Which algorithmic techniques and analytical ideas could be useful for this purpose?}

    \item {\em In spite of basic representation issues, can we leverage the flexibility of dynamic policies to derive a sub-$2$-approximation for general problem instances?}

    \item {\em Can we take advantage of settings with $O(1)$ commodities for computing truly near-optimal policies?}
\end{itemize}
\end{quote}

\subsection{Main contributions} \label{subsec:contributions}

The first and foremost contribution of this paper resides in developing long-awaited algorithmic advances --- some being completely novel and some repurposing seemingly-unrelated inventory management techniques --- for  resolving all open questions listed in Section~\ref{subsec:related_work}. For each of these questions, our work constitutes the first improvement over existing SOSI-driven approaches to economic warehouse lot scheduling, whose approximation guarantees have been unbeatable since the mid-90’s. Beyond this quantitative progress, we establish new-fashioned analytical foundations for directly competing against dynamic policies, which may be of even greater importance and applicability. In what follows, we provide a concise description of our main
findings, leaving their structural characterization, algorithmic techniques, and analytical arguments
to be presented in subsequent sections. 

\paragraph{Polynomial-time approximation scheme for $\bs{O(1)}$ commodities.} In Section~\ref{sec:exponential_approx}, we devise an exponential-in-$n/\eps$ dynamic programming approach for computing a near-optimal capacity-feasible replenishment policy. Specifically, as stated in Theorem~\ref{thm:PTAS_exponential} below, our $O( | {\cal I} |^{O(n)} \cdot 2^{ O( n^{6} / \eps^{5} ) } )$-time algorithm constructs an efficiently-representable cyclic policy whose long-run average cost is within factor $1 + \eps$ of optimal. While  this outcome  applies to all problem instances, it identifies with the notion of a polynomial-time approximation scheme (PTAS) for constantly-many commodities, i.e., when $n = O(1)$. 

\begin{theorem} \label{thm:PTAS_exponential}
For any $\eps > 0$, we can compute a capacity-feasible policy ${\cal P}$ whose long-run average cost is $C({\cal P}) \leq (1 + \eps) \cdot \opt\eqref{eqn:model_warehouse}$. The running time of our algorithm is $O( | {\cal I} |^{O(n)} \cdot 2^{ O( n^{6} / \eps^{5} ) } )$, where $| {\cal I} |$ stands for the input length in its binary specification.
\end{theorem}

Put differently, for $O(1)$ commodities, this result allows us to approach the best-possible cost achievable by dynamic policies within any degree of accuracy; at the same time, we are still able to  represent our near-optimal policy in polynomial space. On a different note, it is important to emphasize that we have not attempted to optimize any of the running time exponents, shooting for the most accessible presentation rather than concentrating on technical minutiae. By diving into the nuts-and-bolts of Section~\ref{sec:exponential_approx} and Appendix~\ref{app:proofs_sec_exponential_approx}, expert readers will discover that such enhancements are doable, yet come at the cost of requiring tedious analysis.

\paragraph{Sub-$\bs{2}$-approximation for general instances.} In Sections~\ref{sec:sub2-approx} and~\ref{sec:po2-sync}, we develop a novel ensemble of analytical ideas and algorithmic advances, culminating to a proof-of-concept improvement on the long-standing performance guarantees of \cite{Anily91} and \cite{GallegoQS96}. Specifically, as outlined in Theorem~\ref{thm:2_minus_delta} below, our main finding for general instances of the economic warehouse lot scheduling problem consists in devising a polynomial-time construction of a random capacity-feasible policy whose expected long-run average cost is within factor $2-\frac{17}{6250}$ of optimal. 

\begin{theorem} \label{thm:2_minus_delta}
For any $\eps \in (0,\frac{ 1 }{ 10 })$, we can compute in $O( | {\cal I} |^{\tilde{O}( 1/\eps^5)} \cdot 2^{ \tilde{O}( 1 / \eps^{35} ) } )$ time a random replenishment policy ${\cal P}$ satisfying the next two properties:
\begin{enumerate}
    \item {\em Feasibility}: ${\cal P}$ is almost surely capacity-feasible.

    \item {\em Expected cost}: $\expar{ C({\cal P}) }\leq (2-\frac{17}{5000} + \eps ) \cdot \opt\eqref{eqn:model_warehouse}$.
\end{enumerate}
\end{theorem}

Once again, it is important to point out that these developments are by no means genuine attempts to
arrive at either the best-possible approximation ratio or the most efficient implementation. Rather,  our presentation  aims to offer a balanced tradeoff between attaining  long-awaited improvements over classic approximation guarantees on the one hand, and introducing novel technical ideas in their simplest form on the other hand. 
\section{Background: Basics of Known \bstitle{2}-Approximations} \label{sec:anily_approx}

In this section, we revisit the work of \cite{Anily91} and \cite{GallegoQS96} on computing a capacity-feasible replenishment policy whose long-run average cost is within factor $2$ of optimal. Even though the upcoming analysis and its related notation are slightly different, the intrinsic ideas should be fully attributed to these authors. We are bringing the reader up to speed on their approach since, once we peel away the initial layers of complexity toward our sub-$2$-approximation, it will play an important role in addressing several easy-to-handle parametric regimes.

\subsection{The average-space bound and its induced relaxation} \label{subsec:average_bound_relaxation}

\paragraph{The average-space bound.} In what follows, we elaborate on the well-known average-space bound, tying between the overall capacity ${\cal V}$ and the average inventory levels of the underlying commodities with respect to any capacity-feasible cyclic policy ${\cal P}$. Such policies can be viewed as those repeating precisely the same actions across a bounded-length cycle, say $[0, \tau_\mycyc)$. To arrive at the desired bound,  suppose we sample a uniformly distributed random variable $\Theta \sim U[0,\tau_\mycyc)$. Then, over the randomness in $\Theta$, the expected space occupied by ${\cal P}$  can be written as
\begin{eqnarray}
\exsub{ \Theta }{V( {\cal P}, \Theta)} & = & \frac{ 1 }{ \tau_\mycyc } \cdot \int_{[0,\tau_\mycyc)} V( {\cal P}, t ) \mathrm{d} t \nonumber \\
& = & \frac{ 1 }{ \tau_\mycyc } \cdot \sum_{i \in [n]} \gamma_i \cdot \int_{[0,\tau_\mycyc)} I( {\cal P}_i, t ) \mathrm{d} t \nonumber \\
& = & \sum_{i \in [n]} \gamma_i \cdot \bar{I}( {\cal P}_i ) \ , \label{eqn:expect_uniform_sample}
\end{eqnarray}
with the convention that $\bar{I}( {\cal P}_i ) = \frac{ 1 }{ \tau_\mycyc } \cdot \int_{[0,\tau_\mycyc)} I( {\cal P}_i, t ) \mathrm{d} t$ stands for the expected inventory level of commodity $i$ across a single cycle. Now, since ${\cal P}$ is capacity-feasible, $V( {\cal P}, \Theta ) \leq {\cal V}$ almost surely, implying in conjunction with representation~\eqref{eqn:expect_uniform_sample} that 
\begin{equation} \label{eqn:average-space-bound}
\sum_{i \in [n]} \gamma_i \cdot \bar{I}( {\cal P}_i ) ~~\leq~~ {\cal V} \ .  
\end{equation}

\paragraph{The resulting relaxation.} Now, suppose we wish to compute a cyclic replenishment policy ${\cal P}$ of minimum long-run average cost, when the capacity constraint $V_{\max}( {\cal P} ) \leq {\cal V}$ is replaced by inequality~\eqref{eqn:average-space-bound}. In other words, we consider the next formulation:
\begin{equation} \label{eqn:relax_warehouse}
\tag{$\tilde{\Pi}_{\mycyc}$}
\begin{array}{lll}
{\displaystyle \inf_{{\cal P} \text{ cyclic}}} & C( {\cal P} ) \\
\text{s.t.} & {\displaystyle \sum_{i \in [n]} \gamma_i \cdot \bar{I}( {\cal P}_i ) \leq {\cal V}} 
\end{array}
\end{equation}
Based on the preceding discussion, it follows that any capacity-feasible  cyclic policy for our original problem is a feasible solution to~\eqref{eqn:relax_warehouse}. Moreover, it is not difficult to verify that the non-necessarily-cyclic optimum, $\opt\eqref{eqn:model_warehouse}$, can be approximated by a cyclic policy within any degree of accuracy. In other words, for any $\eps > 0$, there exists a capacity-feasible cyclic policy ${\cal P}^{\eps}$ with a long-run average cost of $C( {\cal P}^{\eps} ) \leq (1 + \eps) \cdot \opt\eqref{eqn:model_warehouse}$. The latter claim can be ascertained by jumping ahead to Section~\ref{sec:exponential_approx}, where Theorem~\ref{thm:good_cyclic} actually proves a much stronger result, with specific limits on the cycle length and number of orders per commodity. Consequently, it follows that~\eqref{eqn:relax_warehouse} forms a relaxation of~\eqref{eqn:model_warehouse}.

\begin{corollary} \label{cor:rel_vs_opt_ware}
$\opt\eqref{eqn:relax_warehouse} \leq \opt\eqref{eqn:model_warehouse}$.   
\end{corollary}

\paragraph{Simplifying~\eqref{eqn:relax_warehouse}.} Next, let us temporarily restrict relaxation~\eqref{eqn:relax_warehouse} to stationary order sizes and stationary intervals (SOSI) policies. As explained in Section~\ref{subsec:model_definition}, any such policy is fully specified by a given set of ordering intervals, $T_1, \ldots, T_n$. In turn, for each commodity $i \in [n]$, its long-run average inventory level is precisely $\bar{I}( {\cal P}_i ) = \frac{ T_i }{ 2 }$; similarly, its long-run average cost is $C_i( T_i ) = \frac{ K_i }{ T_i } + H_i T_i$. Consequently, when specializing formulation~\eqref{eqn:relax_warehouse} to SOSI policies, our resulting problem can be written as:
\begin{equation} \label{eqn:modify_relax_warehouse}
\tag{$\tilde{\Pi}_{\text{SOSI}}$}
\begin{array}{lll}
{\displaystyle \min_T} & {\displaystyle \sum_{i \in [n] } C_i( T_i )} \\
\text{s.t.} & {\displaystyle \sum_{i \in [n]} \gamma_i T_i \leq 2{\cal V}} 
\end{array}
\end{equation}
Here, one can easily show that the optimum value is indeed attained via basic convexity arguments. Interestingly, as shown in Lemma~\ref{lem:sosi_optimal} below, by limiting attention to this class of policies, we are not compromising on optimality in any way. The proof of this result appears in Section~\ref{subsec:proof_lem_sosi_optimal}; as previously mentioned, these details will be important for arguing about several parametric regimes in Section~\ref{sec:sub2-approx}.

\begin{lemma} \label{lem:sosi_optimal}
Problem~\eqref{eqn:relax_warehouse} admits a SOSI optimal replenishment policy, obtained by solving~\eqref{eqn:modify_relax_warehouse}.
\end{lemma}

\paragraph{Optimal solution to~\eqref{eqn:modify_relax_warehouse}?} One approach to solving this relaxation in polynomial time is straightforward, since it is a convex optimization problem. Alternatively, it is not difficult to verify that we can compute a  $(1+\eps)$-approximate policy for~\eqref{eqn:modify_relax_warehouse} in $O( \frac{ n^3 }{ \eps } )$ time by means of dynamic programming. In essence, by observing that commodities are only linked together by the additive  constraint $\sum_{i \in [n]} \gamma_i T_i \leq 2{\cal V}$, we can discretize the basic units of this budget to integer multiples of $\frac{ \eps }{ n } \cdot {\cal V}$. As such, a knapsack-like dynamic program will process one commodity after the other, solving in each step a single-variable subproblem of the form ``minimize $C_i(T_i)$ subject to an upper bound on $T_i$'', which admits a closed-form solution (see, for example, Section~\ref{subsec:useful_partition_dense}).

\subsection{The approximate policy} \label{subsec:final_2app_anily}

Let $T^* = (T_1^*, \ldots, T_n^*)$ be an optimal solution to~\eqref{eqn:modify_relax_warehouse}, noting that this policy is generally not capacity-feasible, since its maximal space requirement is $V_{\max}( T^* ) = \sum_{i \in [n]} \gamma_i T_i^*$, which could be greater than ${\cal V}$. To correct this issue, consider the scaled-down SOSI policy $\hat{T} = (\hat{T}_1, \ldots, \hat{T}_n)$, defined by setting $\hat{T}_i = \frac{ T_i^* }{ 2 }$ for every commodity $i \in [n]$. As such, $\hat{T}$ is capacity-feasible, since 
\begin{equation} \label{eqn:fix_cap_scale}
V_{\max}( \hat{T} ) ~~=~~ \sum_{i \in [n]} \gamma_i \hat{T}_i ~~=~~ \frac{ 1 }{ 2 } \cdot \sum_{i \in [n]} \gamma_i T_i^* ~~\leq~~ {\cal V} \ ,
\end{equation}
where the last inequality holds since $T^*$ is in particular a feasible solution to~\eqref{eqn:modify_relax_warehouse}, implying that $\sum_{i \in [n]} \gamma_i T_i^* \leq 2{\cal V}$. In terms of cost,
\begin{eqnarray}
C( \hat{T} ) & = & \sum_{i \in [n] } \left( \frac{ K_i }{ \hat{T}_i } + H_i \hat{T}_i \right) \nonumber \\
& = &  \sum_{i \in [n] } \left( \frac{ 2K_i }{ T_i^* } + \frac{ H_i T_i^* }{ 2 } \right) \nonumber \\
& \leq & 2 \cdot \sum_{i \in [n] } C_i( T_i^* ) \nonumber \\
& = & 2 \cdot \opt\eqref{eqn:modify_relax_warehouse} \label{eqn:final_2app_eq1} \\
& = & 2 \cdot \opt\eqref{eqn:relax_warehouse} \label{eqn:final_2app_eq2} \\
& \leq & 2 \cdot \opt\eqref{eqn:model_warehouse} \label{eqn:final_2app_eq3} \ .
\end{eqnarray}
Here, inequality~\eqref{eqn:final_2app_eq1} holds since $T^*$ is an optimal solution to~\eqref{eqn:modify_relax_warehouse}. Inequality~\eqref{eqn:final_2app_eq2} is obtained by recalling that formulations~\eqref{eqn:relax_warehouse} and~\eqref{eqn:modify_relax_warehouse} are equivalent, as explained in Section~\ref{subsec:average_bound_relaxation}. Finally, inequality~\eqref{eqn:final_2app_eq3} follows by noting that~\eqref{eqn:relax_warehouse} is a relaxation of~\eqref{eqn:model_warehouse}, as shown in Corollary~\ref{cor:rel_vs_opt_ware}.

\subsection{Proof of Lemma~\ref{lem:sosi_optimal}} \label{subsec:proof_lem_sosi_optimal}

To establish the desired claim, let ${\cal P}^{\mycyc}$ be a feasible  policy with respect to formulation~\eqref{eqn:relax_warehouse}, noting once again that this problem is restricted to considering cyclic policies. Our proof will show that ${\cal P}^{\mycyc}$ can be iteratively converted to a SOSI policy, feasible with respect to~\eqref{eqn:modify_relax_warehouse}, without increasing its long-run average cost. To this end, we examine how ${\cal P}^{\mycyc}$ operates along a single cycle, $[0, \tau_\mycyc)$. Clearly, one can easily verify that, for any commodity $i \in [n]$, we may assume without loss of generality that ${\cal P}^{\mycyc}_i$ is a zero-inventory ordering (ZIO) policy, ending with zero inventory at time $\tau_\mycyc$. As such, let us separately consider each commodity $i \in [n]$.

Suppose that ${\cal P}^{\mycyc}_i$ places $m$ orders along $[0,\tau_\mycyc)$, with between-order times of $\Delta_1, \ldots, \Delta_m$. Then, we will convert ${\cal P}^{\mycyc}_i$ into a SOSI policy $\hat{\cal P}_i$ for commodity $i$ such that, by defining
\[ \hat{\cal P} ~~=~~ ( \underbrace{{\cal P}^{\mycyc}_1, \ldots, {\cal P}^{\mycyc}_{i-1}}_{ \text{unchanged} }, \underbrace{ \hat{\cal P}_i }_{ \text{new} }, \underbrace{ {\cal P}^{\mycyc}_{i+1}, \ldots, {\cal P}^{\mycyc}_n }_{ \text{unchanged} }) \ , \]
we obtain a feasible solution to~\eqref{eqn:relax_warehouse}, with $C(\hat{\cal P}) \leq C({\cal P}^{\mycyc})$. For this purpose, $\hat{\cal P}_i$ will simply be the policy where we place our $m$ orders at  $0, \frac{ \tau_\mycyc }{ m }, \frac{ 2\tau_\mycyc }{ m }, \ldots$; namely, their between-order times are identical, with $\hat{\Delta} = \cdots = \hat{\Delta}_m = \frac{ \tau_\mycyc }{ m }$.

\paragraph{Feasibility.} We first show that $\sum_{j \in [n]} \gamma_j \cdot \bar{I}( \hat{\cal P}_j ) \leq {\cal V}$. To this end, it suffices to argue that $\bar{I}( \hat{\cal P}_i ) \leq \bar{I}( {\cal P}^{\mycyc}_i )$, which is indeed the case since
\begin{eqnarray}
\bar{I}( {\cal P}^{\mycyc}_i ) & = & \frac{ 1 }{ 2\tau_\mycyc } \cdot \sum_{ \mu \in [m]} \Delta_{ \mu }^2 \nonumber  \\
& \geq & \frac{ 1 }{ 2\tau_\mycyc } \cdot \min \left\{ \sum_{ \mu \in [m]} \bar{\Delta}_{ \mu }^2 : \sum_{ \mu \in [m]} \bar{\Delta}_{ \mu } = \tau_\mycyc, \bar{\Delta} \in \bbR^m_+ \right\} \nonumber \\
& = & \frac{ 1 }{ 2\tau_\mycyc } \cdot m \cdot \left( \frac{ \tau_\mycyc }{m} \right)^2 \nonumber  \\
& = & \frac{ 1 }{ 2\tau_\mycyc } \cdot \sum_{ \mu \in [m]} \hat{\Delta}_{ \mu }^2 \nonumber \\
& = & \bar{I}( \hat{\cal P}_i ) \ . \label{eqn:proof_sosi_optimal_1}
\end{eqnarray}

\paragraph{Objective value.} In terms of ordering costs, since $\hat{\cal P}_i$ and ${\cal P}^{\mycyc}_i$ place the same number of orders across $[0,\tau_\mycyc)$, this ingredient remains unchanged, i.e., ${\cal K}( \hat{\cal P}_i, [0,\tau_\mycyc)) = {\cal K}( {\cal P}^{\mycyc}_i, [0,\tau_\mycyc))$. In terms of holding costs, 
\[ {\cal H}( \hat{\cal P}_i, [0,\tau_\mycyc)) ~~=~~ 2H_i \tau_\mycyc \cdot  \bar{I}( \hat{\cal P}_i ) ~~\leq~~ 2H_i  \tau_\mycyc \cdot  \bar{I}( {\cal P}^{\mycyc}_i ) ~~=~~ {\cal H}( {\cal P}^{\mycyc}_i, [0,\tau_\mycyc)) \ , \]
where the inequality above holds since $\bar{I}( \hat{\cal P}_i ) \leq \bar{I}( {\cal P}^{\mycyc}_i )$, as shown in~\eqref{eqn:proof_sosi_optimal_1}. 
\section{PTAS for \bstitle{O(1)} Commodities} \label{sec:exponential_approx}

In this section, we devise a dynamic programming approach for computing a near-optimal capacity-feasible policy. Specifically, as stated in Theorem~\ref{thm:PTAS_exponential}, given any $\eps > 0$, our $O( | {\cal I} |^{O(n)} \cdot 2^{ O( n^{6} / \eps^{5} ) } )$-time algorithm constructs an efficiently-representable cyclic policy whose long-run average cost is within factor $1 + \eps$ of optimal. Moving forward, Sections~\ref{subsec:structured_cyclic} and~\ref{subsec:structure_thm} will be dedicated to proving the existence of highly-structured near-optimal cyclic policies and  to examining some of their properties in regard to a hierarchical timeline partition, culminating to the Alignment Theorem. Subsequently, Sections~\ref{subsec:recusive_main_result}-\ref{subsec:DP_analysis}  will present the algorithmic implications of this theorem, by unveiling its optimal substructure and exploiting it within an approximate dynamic programming framework to compute an efficiently-representable near-optimal policy.

\subsection{Existence of low-cost short-cycle policies} \label{subsec:structured_cyclic}

For the purpose of limiting our solution space to cyclic policies, we begin by investigating the next  question: How good could  cyclic policies be, when we place certain bounds on their cycle length and number of orders per commodity? As formally stated in Theorem~\ref{thm:good_cyclic} below, we prove that such policies are capable of nearly matching the long-run average cost of optimal fully-dynamic polices, with useful bounds on their cycle length and number of orders per commodity. It is worth mentioning that the existence of a near-optimal cyclic policy by itself is a rather trivial question. The crux of our approach will be to concurrently attain specific bounds on its cycle length and ordering pattern, which will be useful for algorithmic purposes later on. Due to its very technical nature, we provide the proof of this result in Appendix~\ref{app:proof_lem_good_cyclic}. In addition, to avoid  cumbersome expressions in items~2 and~3 below, we make use of the shorthand notation ${\cal M} = \sum_{i \in [n]} ( \frac{ K_i }{ T^{\cal V}_i } + H_i T^{\cal V}_i )$, where $T^{\cal V}_i = \min \{ \sqrt{K_i/H_i}, {\cal V}/ \gamma_i \}$.

\begin{theorem} \label{thm:good_cyclic}
There exists a capacity-feasible cyclic policy $\tilde{\cal P}$ satisfying the following properties:
\begin{enumerate}
    \item {\em Near-optimality}: $C( \tilde{\cal P} ) \leq (1 + 3\eps) \cdot \opt\eqref{eqn:model_warehouse}$.

    \item {\em Cycle length}: $\tilde{\cal P}$ has a cycle length of $\tau_{\mycyc} \in [\frac{ K_{\max} }{ 2\eps n {\cal M} }, \frac{ 2 n {\cal M} }{ \eps^2 H_{\min} }]$.    

    \item {\em Orders per cycle}:     $N( \tilde{\cal P}_i, [0,\tau_{\mycyc})) \in [\frac{ 1 }{ \eps }, \frac{ 4 n^2 {\cal M}^2 }{ \eps^2 K_{\min} H_{\min} }]$, for every commodity $i \in [n]$. Additionally, $N( \tilde{\cal P}_i, [0,\tau_{\mycyc})) = \frac{ 1 }{ \eps }$ for at least one commodity $i \in [n]$.
\end{enumerate}
\end{theorem}

\subsection{Frequency classes and the Alignment Theorem} \label{subsec:structure_thm}

\paragraph{Frequency classes.} It is important to emphasize that Theorem~\ref{thm:good_cyclic}  provides a proof of existence and nothing more; the near-optimal cyclic policy $\tilde{\cal P}$ is obviously unknown from an algorithmic perspective. Toward arriving at a constructive result, let us focus on a single cycle $[0, \tau_\mycyc)$ of this policy. By Theorem~\ref{thm:good_cyclic}(3), this segment has at most ${\cal U} = \frac{ 4 n^2 {\cal M}^2 }{ \eps^2 K_{\min} H_{\min} }$ orders of each commodity, and we can therefore partition the set of commodities into frequency classes $\tilde{\cal F}_1, \ldots, \tilde{\cal F}_Q$ as follows:
\begin{itemize}
    \item The class $\tilde{\cal F}_1$ consists of commodities with $N( \tilde{\cal P}_i, [0,\tau_{\mycyc})) \in [1,(\frac{n}{\eps})^{3}]$ orders.

    \item The class $\tilde{\cal F}_2$ is comprised of those with $N( \tilde{\cal P}_i, [0,\tau_{\mycyc})) \in ((\frac{n}{\eps})^{3}, (\frac{n}{\eps})^{6}]$ orders.

    \item So on and so forth, up to $\tilde{\cal F}_Q$, consisting of commodities with $N( \tilde{\cal P}_i, [0,\tau_{\mycyc})) \in ((\frac{n}{\eps})^{3(Q-1)}, (\frac{n}{\eps})^{3Q}]$. Here, we set $Q = \lceil \log_{(n/\eps)^{3}} {\cal U}  \rceil$, thereby ensuring that the union of $\tilde{\cal F}_1, \ldots, \tilde{\cal F}_Q$  contains all commodities.
\end{itemize}

\paragraph{Subdivisions and the Alignment Theorem.} For every $q \in [Q]$, suppose we subdivide the segment $[0,\tau_{\mycyc})$ into
$(\frac{n}{\eps})^{3q + 1}$ equal-length subintervals; the collection of breakpoints here will be denoted by $B_q^+$. Concurrently, we define an additional subdivision of $[0,\tau_{\mycyc})$, partitioning this segment into $(\frac{n}{\eps})^{3q-4}$ equal-length subintervals. Here, $B_q^-$ will designate the set of resulting breakpoints, noting that $B_q^- \subseteq B_q^+$. With respect to these subdivisions, we will be limiting our attention to a structured family of policies, referred to as being $B$-aligned. To this end, we say that a zero-inventory ordering policy ${\cal P}$ for the segment $[0,\tau_{\mycyc})$ is $B$-aligned when, for every frequency class $q \in [Q]$  and commodity $i \in \tilde{\cal F}_q$, the next two properties are simultaneously satisfied:
\begin{enumerate}
    \item \label{item:B_aligned_1} {\em Zero inventory at $B_q^-$-points:} $I( {\cal P}_i, b^-) = \lim_{t \uparrow b} I( {\cal P}_i, t) = 0$, for every $b \in B_q^-$. 

    \item \label{item:B_aligned_2} {\em Orders only at $B_q^+$-points:} ${\cal P}_i$ places orders only at points in $B_q^+$.  
\end{enumerate}
Our main structural result in this context, called the Alignment Theorem, proves the existences of $B$-aligned policies that are near-feasible and near-optimal at the same time. The specifics of this finding are summarized in Theorem~\ref{thm:exist_aligned}, whose proof is deferred to Appendix~\ref{app:proof_lem_exist_aligned} due to being rather lengthy. While the nuts-and-bolts of our proof are subtle, in the grand scheme of things, property~\ref{item:B_aligned_1} will be attained by placing suitable orders at $B_q^-$-points, whereas property~\ref{item:B_aligned_2} will require stretching and possibly discarding certain orders. Combined with Theorem~\ref{thm:good_cyclic}, we will argue that these alterations have minor consequences in terms of space and cost.

\begin{theorem}[Alignment]\label{thm:exist_aligned}
There exists a $B$-aligned policy $\hat{\cal P}$ satisfying the next two properties:
\begin{enumerate}
    \item {\em Space}: $V_{\max}( \hat{\cal P} ) \leq (1+\eps) \cdot {\cal V}$.
    
    \item {\em Cost}: $C( \hat{\cal P}, [0,\tau_{\mycyc}) ) \leq (1 + \eps) \cdot C( \tilde{\cal P}, [0,\tau_{\mycyc}))$. 
\end{enumerate}
\end{theorem}

\subsection{Main result, optimal substructure, and approximate statistics} \label{subsec:recusive_main_result}

\paragraph{Intent.} In what follows, we exploit the Alignment Theorem for the purpose of computing a capacity-feasible replenishment policy whose long-run average cost is within factor $1 + O(\eps)$ of $\opt\eqref{eqn:model_warehouse}$. Toward this objective, letting $\tilde{\cal P}$ and $\hat{\cal P}$ be the policies whose existence has been respectively established in Theorems~\ref{thm:good_cyclic} and~\ref{thm:exist_aligned}, we know that  
\[ C( \hat{\cal P} ) ~~\leq~~ (1 + \eps) \cdot C( \tilde{\cal P}) ~~\leq~~ (1 + 7\eps) \cdot \opt\eqref{eqn:model_warehouse} \ . \]
Therefore, in terms of space requirement and cost, it suffices to obtain a $(1 + O(\eps))$-feasible policy ${\cal P}$ for the interval $[0, \tau_\mycyc)$ with $C( {\cal P}, [0,\tau_{\mycyc}) ) = (1+O(\eps)) \cdot C( \hat{\cal P}, [0,\tau_{\mycyc}) )$. Scaling such a policy by a factor of $1 + O(\eps)$, we end up with a capacity-feasible policy whose long-run average cost is within factor $1 + O(\eps)$ of optimal. In terms of efficiency, we will show that our algorithm can be implemented in $O( | {\cal I} |^{O(n)} \cdot 2^{ O( n^{6} / \eps^{5} ) } )$ time. 

\paragraph{Notation.} For the remainder of this section, let  ${\cal P}^{B*}$ be an optimal $B$-aligned $(1+\eps)$-feasible policy for the segment $[0,\tau_{\mycyc})$, noting that $C( {\cal P}^{B*}, [0,\tau_{\mycyc}) ) \leq (1 + \eps) \cdot C( \hat{\cal P}, [0,\tau_{\mycyc}))$ by Theorem~\ref{thm:exist_aligned}. We remind the reader that, for any $q \in [Q]$, the segment $[0, \tau_\mycyc)$ was subdivided into $(\frac{n}{\eps})^{3q-4}$ equal-length subintervals, thereby creating the set of breakpoints $B_q^-$. By property~\ref{item:B_aligned_1} of $B$-aligned policies, we know that for every commodity $i \in \tilde{\cal F}_q$, the policy ${\cal P}^{B*}$ reaches zero inventory at all $B_q^-$-points. Moreover, since $B_1^- \subseteq \cdots \subseteq B_Q^-$, this policy actually  reaches zero inventory at $B_q^-$-points for every commodity belonging to one of the classes $\tilde{\cal F}_q, \ldots, \tilde{\cal F}_Q$. We use $\tilde{\cal F}_{\geq q}$ to designate the union of these classes, with $\tilde{\cal F}_{\leq q-1}$ being the union of $\tilde{\cal F}_1, \ldots, \tilde{\cal F}_{q-1}$.

\paragraph{Optimal substructure.} To unveil the optimal substructure that stands behind our dynamic program, consider the following question: 
\begin{quote}
{\em For any $q \in [Q]$, what is the dependency between the actions taken by ${\cal P}^{B*}$ for $\tilde{\cal F}_{\geq q}$-commodities and its actions for $\tilde{\cal F}_{\leq q-1}$-commodities?} 
\end{quote}
To address this question, let us examine an interval $[b_\mathrm{entry}, b_\mathrm{exit}]$ that stretches between two successive points in $B_q^-$. Since ${\cal P}^{B*}$ is $B$-aligned, it follows that $I({\cal P}^{B*}_i, b_\mathrm{entry}^-) = I({\cal P}^{B*}_i, b_\mathrm{exit}^-) = 0$ for every commodity $i \in \tilde{\cal F}_{\geq q}$. Thus, we can assume without loss of generality that the actions taken by ${\cal P}^{B*}$ for $\tilde{\cal F}_{\geq q}$-commodities across $[b_\mathrm{entry}, b_\mathrm{exit}]$ only depend on two types of statistics, summarizing its actions for $\tilde{\cal F}_{\leq q-1}$-commodities: 
\begin{itemize}
    \item {\em $\tilde{\cal F}_{q-1}$-inventory levels}: By property~\ref{item:B_aligned_2} of $B$-aligned policies, it follows that ${\cal P}^{B*}$ places orders for every commodity $i \in \tilde{\cal F}_{q-1}$ only at $B_{q-1}^+$-points. Since $|B_q^-| = (\frac{n}{\eps})^{3q-4}$ and $|B_{q-1}^+| = (\frac{n}{\eps})^{3(q-1)+1}$, we know that $[b_\mathrm{entry}, b_\mathrm{exit}]$ has exactly $(\frac{n}{\eps})^2$ $B_{q-1}^+$-points where $i$-orders can be placed. Once the inventory levels of commodity $i$ at these points (including $b_\mathrm{entry}$ and $b_\mathrm{exit}$) are known, they uniquely determine its inventory level across the entire interval $[b_\mathrm{entry}, b_\mathrm{exit}]$.
    
    \item {\em $\tilde{\cal F}_{\leq q-2}$-entry-inventory levels}: By the same argument, for every commodity $i \in \tilde{\cal F}_{\leq q-2}$, any $i$-order of ${\cal P}^{B*}$ resides in $B_{q-2}^+$. Since $B_{q-2}^+ \subseteq B_q^-$, it follows that the interval $[b_\mathrm{entry}, b_\mathrm{exit}]$ does not have any $i$-order in its interior. As a result, for such commodities, the optimal policy ${\cal P}^{B*}$ only has to account for the inventory level $I({\cal P}^{B*}_i, b_\mathrm{entry}^+) = \lim_{t \downarrow b_\mathrm{entry}} I({\cal P}^{B*}_i,t)$  immediately after entering this interval.
\end{itemize}
In fact, once we know the above-mentioned statistics, the interval $[b_\mathrm{entry}, b_\mathrm{exit}]$ becomes superfluous, in the sense that ${\cal P}^{B*}$ would act in an identical way between any pair of successive $B_q^-$-points  with these statistics.

\paragraph{Exact representation for $\bs{\tilde{\cal F}_{q-1}}$-inventory levels.} In terms of their representation within our dynamic program, $\tilde{\cal F}_{q-1}$-inventory levels will be relatively easy to keep track of. As mentioned above, for any commodity $i \in \tilde{\cal F}_{q-1}$, there are $(\frac{n}{\eps})^2$ $B_{q-1}^+$-points where $i$-orders can be placed along $[b_\mathrm{entry}, b_\mathrm{exit}]$. In addition, it is not difficult to verify that the latest $i$-order to the left of $b_\mathrm{entry}$ is restricted to being one of 
$(\frac{n}{\eps})^{5}$ $B_{q-1}^+$-points. Indeed, by property~\ref{item:B_aligned_1} of $B$-aligned policies, we know that this commodity reaches zero inventory at $B_{q-1}^-$-points. Therefore, there must be at least one $i$-order between the latest $B_{q-1}^-$-point to the left of $b_\mathrm{entry}$ and $b_\mathrm{entry}$ itself. This  interval has at most $\frac{ | B_{q-1}^+ | }{ | B_{q-1}^- | } = (\frac{n}{\eps})^{5}$ $B_{q-1}^+$-points, which are the only ones where $i$-orders can be placed, by property~\ref{item:B_aligned_2}. Precisely the same claim applies to the earliest $i$-order to the right of $b_\mathrm{exit}$. Once we determine whether there is an order at each of these points, for which there are $( 2^{ (n/\eps)^2 } \cdot (\frac{n}{\eps})^{10} )^n = 2^{ O( n^3 / \eps^2 ) }$ options over all $\tilde{\cal F}_{q-1}$-commodities, they uniquely determine $\tilde{\cal F}_{q-1}$-inventory levels.

\paragraph{Proxy representation for $\bs{\tilde{\cal F}_{\leq q-2}}$-entry-inventory levels.} In contrast, regarding the number of possible configurations that can be taken by $\tilde{\cal F}_{\leq q-2}$-entry-inventory levels $\{ I({\cal P}^{B*}_i, b_\mathrm{entry}^+) \}_{i \in \tilde{\cal F}_{\leq q-2}}$,  we are not seeing how this quantity can be upper-bounded to meet the intended running time of Theorem~\ref{thm:PTAS_exponential}. For this reason, we propose a simple one-dimensional statistic that will serve as a sufficiently-good alternative. To understand where this statistic is coming from, note that since the interval $[b_\mathrm{entry}, b_\mathrm{exit}]$ does not have  $\tilde{\cal F}_{\leq q-2}$-orders in its interior, the inventory level of each commodity $i \in \tilde{\cal F}_{\leq q-2}$ linearly drops by exactly $b_\mathrm{exit} - b_\mathrm{entry}$ units. As such, the total space requirement of the $\tilde{\cal F}_{\leq q-2}$-commodities just before exiting $[b_\mathrm{entry}, b_\mathrm{exit}]$ is related to the analogous quantity immediately after entering this interval through  
\begin{eqnarray}
\sum_{i \in \tilde{\cal F}_{\leq q-2}} \gamma_i \cdot I({\cal P}^{B*}_i, b_\mathrm{exit}^-) & = & \sum_{i \in \tilde{\cal F}_{\leq q-2}} \gamma_i \cdot \left( I({\cal P}^{B*}_i, b_\mathrm{entry}^+) - (b_\mathrm{exit} - b_\mathrm{entry}) \right)  \nonumber \\
& = & \sum_{i \in \tilde{\cal F}_{\leq q-2}} \gamma_i \cdot  I({\cal P}^{B*}_i, b_\mathrm{entry}^+) - \frac{ \tau_\mycyc }{ (n/\eps)^{3q-4} } \cdot \sum_{i \in \tilde{\cal F}_{\leq q-2}} \gamma_i \ . \label{eqn:recursive_before_after}
\end{eqnarray}
Given this observation, rather than accurately keeping track of $\{ I({\cal P}^{B*}_i, b_\mathrm{entry}^+) \}_{i \in \tilde{\cal F}_{\leq q-2}}$, our proxy statistic will be a lower bound on the total space requirement of these commodities at any point in $[b_\mathrm{entry}, b_\mathrm{exit})$. This statistic, termed ``$\tilde{\cal F}_{\leq q-2}$-space bound'', is given by $\sum_{i \in \tilde{\cal F}_{\leq q-2}} \gamma_i \cdot  I({\cal P}^{B*}_i, b_\mathrm{exit}^-)$, which is exactly the left-hand-side of inequality~\eqref{eqn:recursive_before_after}. Interestingly, our dynamic program will falsely assume that the space requirement of these commodities remains unchanged at this statistic across $[b_\mathrm{entry}, b_\mathrm{exit})$. While this false assumption may lead to a capacity violation, we will prove that the maximum possible violation is negligible.

\subsection{The dynamic program: Guessing and state space} \label{subsec:dp_guess_State}

\paragraph{Guessing.} Having laid the theoretical foundations of our algorithmic approach, we move on to describe its concrete implementation. As a preliminary step, we begin by guessing two ingredients related to the policies that were discussed up until now:
\begin{itemize}
    \item {\em Cycle length $\tau_\mycyc$.} We first obtain an over-estimate $\tilde{\tau}_\mycyc$ for the cycle length $\tau_\mycyc$ within a factor of $1 + \eps$, meaning that $\tilde{\tau}_\mycyc \in [\tau_\mycyc, (1+\eps) \cdot \tau_\mycyc]$. Given Theorem~\ref{thm:good_cyclic}(2), we know that $\tau_{\mycyc} \in [\frac{ K_{\max} }{ 2\eps n {\cal M} }, \frac{ 2 n {\cal M} }{ \eps^2 H_{\min} }]$, meaning that the number of values to be tested as possible estimates is $O( \log_{ 1 + \eps } ( \frac{ 4n^2 {\cal M}^2 }{ \eps K_{\max}  H_{\min} } ) ) = O( (\frac{ |{\cal I}| }{ \eps } )^{ O(1) })$. For ease of presentation, we make direct use of $\tau_\mycyc$; the negligible implications of this simplification will be discussed as part of our analysis.

    \item {\em Frequency classes $\tilde{\cal F}_1, \ldots, \tilde{\cal F}_Q$.} Second, for each commodity $i \in [n]$, we guess the frequency class to which it belongs, leading to $Q^n$ values to be enumerated over. As mentioned in Section~\ref{subsec:structure_thm}, we have $Q = \lceil \log_{(n/\eps)^{3}} {\cal U}  \rceil$ and ${\cal U} = \frac{ 4 n^2 {\cal M}^2 }{ \eps^2 K_{\min} H_{\min} }$, implying that $Q^n = O( (\frac{ |{\cal I}| }{ \eps } )^n)$. 
\end{itemize}

\paragraph{State space.} Each state $(q, {\cal I}_{q-1}, \mylb_{ \leq q-2 })$ of our dynamic program consists of the next three parameters:
\begin{enumerate}
    \item {\em Class index $q$:} This index corresponds to the frequency class $\tilde{\cal F}_q$ for which we are currently placing orders, across an interval stretching between two successive $B_q^-$-points. We emphasize that this interval is not explicitly specified within our state description. Rather, in conjunction with ${\cal I}_{q-1}$ and $\mylb_{ \leq q-2 }$, we will apply the exact same policy for any interval between two successive $B_q^-$-points; for clarity, we symbolically denote such an interval by $[b_\mathrm{entry}, b_\mathrm{exit}]$. It is important to point out that only non-empty classes are ever considered by our dynamic program, meaning that $q$ will be taking $O( n )$ possible values out of $1, \ldots, Q$, even though $Q$ itself could be much larger.

    \item {\em $\tilde{\cal F}_{q-1}$-inventory levels ${\cal I}_{q-1}$}: Let $x_1, \ldots, x_R$ be the sequence of $(\frac{n}{\eps})^2$ $B_{q-1}^+$-points where $B$-aligned policies are allowed to place $\tilde{\cal F}_{q-1}$-orders along $[b_\mathrm{entry}, b_\mathrm{exit}]$. That is, 
    \[ x_1 ~~=~~ b_\mathrm{entry}, \quad x_2 ~~=~~ b_\mathrm{entry} + \frac{ b_\mathrm{exit} - b_\mathrm{entry} }{ ( n/\eps )^2 }, \qquad \ldots, \quad x_R ~~=~~ b_\mathrm{exit} \ . \]    
    Then, the $(|\tilde{\cal F}_{q-1}| \cdot R)$-dimensional  vector ${\cal I}_{q-1}$ will represent the already-decided inventory level of each commodity $i \in \tilde{\cal F}_{q-1}$ at each of the point $x_1, \ldots, x_R$. As explained in Section~\ref{subsec:recusive_main_result}, this vector can take $2^{ O( n^3 / \eps^2 ) }$ different values.    

    \item {\em $\tilde{\cal F}_{\leq q-2}$-space bound $\mylb_{ \leq q-2 }$}: This parameter serves as a lower bound on the already-decided total space requirement of $\tilde{\cal F}_{\leq q-2}$-commodities at any point in  $[b_\mathrm{entry}, b_\mathrm{exit})$. As explained in Section~\ref{subsec:recursive_details}, we will restrict $\mylb_{ \leq q-2 }$ to integer multiples of $\frac{ \eps }{ n } \cdot {\cal V}$ within $[0, (1 + \eps) \cdot {\cal V}]$, meaning that this parameter only takes $O( \frac{ n }{ \eps } )$ values.
\end{enumerate}
In light of this description, the overall number of states  is $O(n \cdot 2^{ O( n^3 / \eps^2 ) } \cdot \frac{ n }{ \eps } ) = O( 2^{ O( n^3 / \eps^2 ) } )$.

\subsection{The dynamic program: Recursive equations} \label{subsec:recursive_details}

In what follows, we describe how our dynamic program operates at each state $(q, {\cal I}_{q-1}, \mylb_{ \leq q-2 })$. Broadly speaking, this procedure will compute a $B$-aligned policy for all $\tilde{\cal F}_{ \geq q }$-commodities over an interval $[b_\mathrm{entry}, b_\mathrm{exit})$ that stretches between two successive $B_q^-$-points, assuming that our actions regarding $\tilde{\cal F}_{\leq q-1}$-commodities resulted in $\tilde{\cal F}_{q-1}$-inventory levels ${\cal I}_{q-1}$ and $\tilde{\cal F}_{\leq q-2}$-space bound $\mylb_{ \leq q-2 }$ for this interval. We make use of $C(q, {\cal I}_{q-1}, \mylb_{ \leq q-2 })$ to denote the total cost of this policy across $[b_\mathrm{entry}, b_\mathrm{exit})$.

\paragraph{Step 1: Action space for $\bs{\tilde{\cal F}_q}$.} Let us begin by observing that, given property~\ref{item:B_aligned_2} of $B$-aligned policies, for every commodity $i \in \tilde{\cal F}_q$, such policies are allowed to place $i$-orders only at $B_q^+$-points. Since successive $B_q^+$-points are positioned within distance $\frac{ \tau_\mycyc }{ | B_q^+| }$ of each other, and since $[b_\mathrm{entry}, b_\mathrm{exit}]$ is of length $\frac{ \tau_\mycyc }{ | B_q^-| }$, it follows that the number of $B_q^+$-points in this interval is $\frac{ | B_q^+ | }{ |B_q^-| } = ( \frac{ n }{ \eps } )^{ 5 }$. For future reference, we use $y_1, \ldots, y_S$ to denote these points in left-to-right order. Consequently, jointly over all commodities in $\tilde{\cal F}_q$, there are only $2^{ |\tilde{\cal F}_q| \cdot S } = 2^{ O( n^{6} / \eps^{5} ) }$ possible $B$-aligned policies to be considered. For any such policy ${\cal P}^q$, we go through steps~2 and~3 below, where ${\cal P}^q$ will either be extended to a $B$-aligned policy ${\cal P}^{\geq q}$ for all $\tilde{\cal F}_{ \geq q }$-commodities or tagged as being ``unextendable''. In the event where all policies tested are unextendable, our algorithm will report this fact by returning $C(q, {\cal I}_{q-1}, \mylb_{ \leq q-2 }) = \bot$. In the opposite scenario, we return the extendable policy ${\cal P}^{\geq q}$ whose cost $C( {\cal P}^{\geq q}, [b_\mathrm{entry}, b_\mathrm{exit}))$ is minimal, setting $C(q, {\cal I}_{q-1}, \mylb_{ \leq q-2 })$ to this value.

\paragraph{Step 2: Verifying acceptability.} Due to limiting our attention to $B$-aligned policies, once the vector ${\cal I}_{q-1}$ of $\tilde{\cal F}_{q-1}$-inventory levels is fixed, it determines a unique replenishment policy ${\cal P}^{q-1}$ for the $\tilde{\cal F}_{q-1}$-commodities across $[b_\mathrm{entry}, b_\mathrm{exit})$. In turn, we say that the policy ${\cal P}^q$ is acceptable when, for every point $t \in [b_\mathrm{entry}, b_\mathrm{exit})$, we have
\begin{equation} \label{eqn:def_acceptable}   
\underbrace{ \sum_{i \in \tilde{\cal F}_q} \gamma_i \cdot I( {\cal P}^q_i, t ) }_{ \text{exact space due to $\tilde{\cal F}_q$} } + \underbrace{ \sum_{i \in \tilde{\cal F}_{q-1}} \gamma_i \cdot I( {\cal P}^{ q-1}_i, t ) }_{ \text{exact space due to $\tilde{\cal F}_{q-1}$} } + \underbrace{ \mylb_{ \leq q-2 } }_{ \MyAbove{ \text{lower bound on} }{ \text{space due to $\tilde{\cal F}_{\leq q-2}$} } } ~~\leq~~ (1 +\eps) \cdot {\cal V} \ . 
\end{equation}
It is important to point out that this definition is weaker than $(1 +\eps)$-feasibility, since $\mylb_{ \leq q-2 }$ is only a lower bound on the space requirement of $\tilde{\cal F}_{\leq q-2}$-commodities. Another important observation is that, even though condition~\eqref{eqn:def_acceptable} is written for all points $t \in [b_\mathrm{entry}, b_\mathrm{exit})$, we can verify whether ${\cal P}^q$ is acceptable in polynomial time. Specifically, it suffices to examine whether this condition holds for the $B_q^+$-points in this interval, $y_1, \ldots, y_S$, since the inventory level of any commodity in $\tilde{\cal F}_q \cup \tilde{\cal F}_{q-1}$ is decreasing between any pair of successive points, $y_s$ and $y_{s+1}$.  We proceed to step~3 only when ${\cal P}^q$ is acceptable; otherwise, this policy is eliminated from being further considered, and we tag it as unextendable. 

\paragraph{Step 3A: Recursive policy for $\bs{\tilde{\cal F}_{\geq q+1}}$, when $\bs{\tilde{\cal F}_{q+1} \neq \emptyset}$.} Recalling that $|B_{q+1}^-| = (\frac{n}{\eps})^{3(q+1) - 4} $ and $| B_q^-| = (\frac{n}{\eps})^{3q-4}$, it follows that $[b_\mathrm{entry}, b_\mathrm{exit})$ is subdivided by $B_{q+1}^-$-points into $( \frac{n}{\eps})^3$ equal-length intervals, say $I_1, \ldots, I_M$ in left-to-right order. In this step, we consider the case where $\tilde{\cal F}_{q+1} \neq \emptyset$, while step~3B will discuss the one where $\tilde{\cal F}_{q+1} = \emptyset$.

Let us focus on a single  interval $I_m = [b_\mathrm{entry}^m, b_\mathrm{exit}^m)$, obviously belonging to the next level of our recursion, $q+1$, which indeed exists since $\tilde{\cal F}_{q+1} \neq \emptyset$ by the hypothesis of step~3A. The important observation is that, since we have already fixed the policy ${\cal P}^q$, it uniquely determines the vector ${\cal I}_q^m$ of $\tilde{\cal F}_q$-inventory levels  associated with this interval. In addition, since the policy ${\cal P}^{q-1}$ has been fixed as well (see step~2), a straightforward lower bound $\mylb_{ \leq q-1 }^m$ on the total space requirement of $\tilde{\cal F}_{\leq q-1}$-commodities at any point in  $I_m = [b_\mathrm{entry}^m, b_\mathrm{exit}^m)$ is given by
\begin{equation} \label{eqn:recur_lb_case1}
\mylb_{ \leq q-1 }^m ~~=~~ \mylb_{ \leq q-2 } + \left\lfloor \sum_{i \in \tilde{\cal F}_{q-1} } \gamma_i \cdot I( {\cal P}^{q-1}_i, b_\mathrm{exit}^{m-}) \right\rfloor_{ \eps {\cal V} / n} \ .         
\end{equation}    
Here, $\lfloor \cdot \rfloor_{ \eps {\cal V} / n}$ is an operator that rounds its argument down to the nearest integer multiple of $\frac{ \eps }{ n } \cdot {\cal V}$. As such, for every interval $I_m$, we examine the actions taken by our dynamic program at state $(q+1, {\cal I}_q^m, \mylb_{ \leq q-1 }^m)$. When $C(q+1, {\cal I}_q^m, \mylb_{ \leq q-1 }^m) = \bot$ for some $m \in [M]$, we report that ${\cal P}^q$ is unextendable. In the opposite case, for every $m \in [M]$, we obtain a $B$-aligned policy ${\cal P}^{\geq q+1,m}$ for all $\tilde{F}_{ \geq q+1 }$-commodities over $I_m$, with cost $C( {\cal P}^{\geq q+1, m}, I_m ) = C(q+1, {\cal I}_q^m, \mylb_{ \leq q-1 }^m)$. Finally, we create a $B$-aligned policy ${\cal P}^{ \geq q}$ for all $\tilde{F}_{ \geq q }$-commodities by gluing ${\cal P}^q$ together with the concatenation of ${\cal P}^{\geq q+1,1}, \ldots, {\cal P}^{\geq q+1,M}$ in this order.

\paragraph{Step 3B: Recursive policy for $\bs{\tilde{\cal F}_{\geq q+1}}$, when $\bs{\tilde{\cal F}_{q+1} = \emptyset}$.} Again, let us focus on a single interval $I_m = [b_\mathrm{entry}^m, b_\mathrm{exit}^m)$. As opposed to step~3A, we cannot employ recursive calls to level $q+1$, since $(q+1, \cdot, \cdot)$ are not states of our dynamic program when $\tilde{\cal F}_{q+1} = \emptyset$, which is exactly the hypothesis of step~3B. Toward circumventing this issue, let $q_\mynext$ be the minimum index of a non-empty frequency class out of $\tilde{\cal F}_{q+2}, \ldots, \tilde{\cal F}_{Q}$. Rather than descending to the non-existing level $q+1$, we will jump right to level $q_\mynext$. Technically speaking, we fill $I_m$ by gluing identical copies of the policy ${\cal P}^{ \geq q_\mynext, m }$, obtained by employing the actions our dynamic program takes at the single state $(q_\mynext, {\cal I}_{q_\mynext - 1}^m, \mylb_{ \leq q_\mynext-2 }^m)$, where ${\cal I}_{q_\mynext - 1}^m$ and $\mylb_{ \leq q_\mynext-2 }^m$ are determined as follows:
\begin{itemize}
    \item Since $\tilde{\cal F}_{q+1} = \emptyset$ by the case hypothesis, we know that $\tilde{\cal F}_{q_\mynext-1} = \emptyset$ as well. Consequently, our vector  ${\cal I}_{q_\mynext - 1}^m$ of $\tilde{\cal F}_{q_\mynext-1}$-inventory levels does not even exist (equivalently, it is the zero vector), which will be indicated by ${\cal I}_{q_\mynext - 1}^m = \vec{0}$.

    \item Along the lines of step~3A, our lower bound $\mylb_{ \leq q_\mynext-2 }^m$ on the total space requirement of $\tilde{\cal F}_{\leq q_\mynext-2}$-commodities at any point will be
    \begin{equation} \label{eqn:recur_lb_case2}
    \mylb_{ \leq q_\mynext-2 }^m ~~=~~ \mylb_{ \leq q-2 } + \left\lfloor \sum_{i \in \tilde{\cal F}_q } \gamma_i \cdot I( {\cal P}^{q}_i, b_\mathrm{exit}^{m-}) + \sum_{i \in \tilde{\cal F}_{q-1} } \gamma_i \cdot I( {\cal P}^{q-1}_i, b_\mathrm{exit}^{m-}) \right\rfloor_{ \eps {\cal V} / n} \ . 
    \end{equation}
\end{itemize}
Of course, when $C(q_\mynext, {\cal I}_{q_\mynext - 1}^m, \mylb_{ \leq q_\mynext-2 }^m) = \bot$, we report that ${\cal P}^q$ is unextendable. It is important to notice that, in order to fill $I_m$, the number of ${\cal P}^{ \geq q_\mynext, m }$-copies needed is $\frac{ | B_{q_\mynext}^-| }{ | B_{q-1}^-| } = ( \frac{ n }{ \eps } )^{ 3(q_\mynext - q+1) }$. This quantity can be as large as $( \frac{ n }{ \eps } )^{\Omega(Q)}$, which is precisely the type of running time dependency we wish to avoid. However, since we  make use of identical copies, only one needs to be computed and represented. Finally, as in step~3A, we create a $B$-aligned policy ${\cal P}^{ \geq q}$ for all $\tilde{F}_{ \geq q }$-commodities by gluing ${\cal P}^q$ together with the concatenation of the above-mentioned policies for $I_1, \ldots, I_M$. 

\paragraph{Top level.} To describe the overall policy we return for the entire cycle $[0, \tau_\mycyc)$, it is convenient to assume that $\tilde{\cal F}_1 \neq \emptyset$. This assumption is without loss of generality, since we can add a dummy commodity, indexed $0$, with $H_0 = K_0 = \gamma_0 = 0$, which is ordered by the policy $\tilde{\cal P}$ so that $0 \in \tilde{\cal F}_1$. As such, our overall policy is the one computed at state $(1, \vec{0}, 0)$.

\subsection{Analysis} \label{subsec:DP_analysis}

In what follows, we begin by addressing the most basic question: Regardless of capacity-feasibility and cost, will our dynamic program even compute a replenishment policy for the entire cycle? As explained below, this naive question is a-priori unclear. Subsequently, letting ${\cal P}^{ \mydp }$ be the resulting policy, we will show that it indeed meets the performance guarantees stated in the opening paragraph of Section~\ref{subsec:recusive_main_result}. In other words, we will prove that ${\cal P}^{ \mydp }$ is $(1 + O(\eps))$-feasible and that its cost is $C( {\cal P}^{ \mydp }, [0,\tau_{\mycyc}) ) = (1+O(\eps)) \cdot C( \hat{\cal P}, [0,\tau_{\mycyc}) )$. We will also discuss the running time of our particular implementation.

\paragraph{Extendability and cost.} Clearly, 
our dynamic program computes a replenishment policy for the entire cycle if and only if  $C(1, \vec{0}, 0) \neq \bot$. The crucial observation is that, by going through each and every algorithmic step of Section~\ref{subsec:recursive_details}, it is easy to verify that the optimal $B$-aligned $(1+\eps)$-feasible policy ${\cal P}^{B*}$ for the segment $[0,\tau_{\mycyc})$ is a feasible solution to our recursion. In other words, at level $q=1$, one of the policies examined in step~1 as part of the action space for $\tilde{\cal F}_1$ is that of duplicating the orders made by $\{ {\cal P}^{B*}_i \}_{ i \in \tilde{\cal F}_1 }$, which obviously passes our acceptability check in step~2, since this condition is weaker than $(1+\eps)$-feasibility.  Once we descend to the second level (say $q=2$) of the recursion  with this policy, one of the actions examined is that of duplicating $\{ {\cal P}^{B*}_i \}_{ i \in \tilde{\cal F}_2 }$, so on and so forth. Consequently, we are guaranteed to have $C(1, \vec{0}, 0) \neq \bot$, implying that our dynamic program indeed computes a replenishment policy ${\cal P}^{ \mydp }$ for the entire cycle. Moreover, in terms of cost, these observations supposedly imply that $C( {\cal P}^{ \mydp }, [0,\tau_{\mycyc}) ) \leq C( {\cal P}^{B*}, [0,\tau_{\mycyc}) )$. However, due to making a direct use of $\tau_{\mycyc}$ in place of its estimate $\tilde{\tau}_\mycyc \in [\tau_\mycyc, (1+\eps) \cdot \tau_\mycyc]$, we actually have 
\[ C( {\cal P}^{ \mydp }, [0,\tau_{\mycyc}) ) ~~\leq~~ (1 + \eps) \cdot C( {\cal P}^{B*}, [0,\tau_{\mycyc}) ) ~~\leq~~ (1 + \eps) \cdot
 C( \hat{\cal P}, [0,\tau_{\mycyc}) ) \ , \]
where the second inequality holds since, by Theorem~\ref{thm:exist_aligned}, we know that $\hat{\cal P}$ is a $B$-aligned $(1+\eps)$-feasible policy for the segment $[0,\tau_{\mycyc})$, and since ${\cal P}^{B*}$ is a minimum-cost such policy.

\paragraph{Space requirement.} The next question we address is regarding the maximal space requirement of ${\cal P}^{ \mydp }$, noting that our dynamic program does not explicitly enforce capacity-feasibility. The next claim shows that ${\cal P}^{ \mydp }$ is $(1 + 2\eps)$-feasible.

\begin{lemma}
$V_{\max}( {\cal P}^{ \mydp } ) \leq (1 + 2\eps) \cdot {\cal V}$.
\end{lemma}
\begin{proof}
To bound the space requirement of ${\cal P}^{ \mydp }$ at any point in $[0,\tau_{\mycyc})$, suppose that throughout our dynamic program, we would have augmented each state $(q, {\cal I}_{q-1}, \mylb_{ \leq q-2 })$ with a fourth parameter, $\myub_{ \leq q-2 }$. Operating in the opposite direction of $\mylb_{ \leq q-2 }$, this parameter would have served as an upper bound on the already-decided total space requirement of $\tilde{\cal F}_{\leq q-2}$-commodities at any point in  $[b_\mathrm{entry}, b_\mathrm{exit})$. To recursively update this parameter, one can mimic equations~\eqref{eqn:recur_lb_case1} and~\eqref{eqn:recur_lb_case2}, such that:
\[ \myub_{ \leq q-1 }^m ~~=~~ \begin{cases}
    \myub_{ \leq q-2 } +  \sum_{i \in \tilde{\cal F}_{q-1} } \gamma_i \cdot I( {\cal P}^{q-1}_i, b_\mathrm{entry}^{m+}), & \text{if } \tilde{\cal F}_{q+1} \neq \emptyset \\
    \myub_{ \leq q-2 } +  \sum_{i \in \tilde{\cal F}_q } \gamma_i \cdot I( {\cal P}^{q}_i, b_\mathrm{entry}^{m+}) + \sum_{i \in \tilde{\cal F}_{q-1} } \gamma_i \cdot I( {\cal P}^{q-1}_i, b_\mathrm{entry}^{m+}),  & \text{if } \tilde{\cal F}_{q+1} = \emptyset
\end{cases}\]
Now, by combining this definition with equation~\eqref{eqn:recursive_before_after}, we conclude that $\myub_{ \leq q-2 }^m - \mylb_{ \leq q-2 }^m \leq \frac{ \tau_\mycyc }{ (n/\eps)^{3q-4} } \cdot \sum_{i \in \tilde{\cal F}_{\leq q-2}} \gamma_i + \frac{ \eps {\cal V} }{ n }$, for every class index $q \geq 3$ with $\tilde{\cal F}_q \neq \emptyset$. As such, the maximum-possible space violation of ${\cal P}^{ \mydp }$ across the cycle $[0,\tau_{\mycyc})$ is at most 
\begin{eqnarray}
\sum_{q \geq 3: \tilde{\cal F}_q \neq \emptyset } \left( \myub_{ \leq q-2 }^m - \mylb_{ \leq q-2 }^m  \right) & \leq & \sum_{q \geq 3: \tilde{\cal F}_q \neq \emptyset } \frac{ \tau_\mycyc }{ (n/\eps)^{3q-4} } \cdot \sum_{i \in \tilde{\cal F}_{\leq q-2}} \gamma_i + \eps {\cal V} \nonumber \\
& \leq & \sum_{q \geq 3: \tilde{\cal F}_q \neq \emptyset } | \tilde{\cal F}_{\leq q-2} | \cdot \frac{ {\cal V} }{ (n/\eps)^2 } \cdot + \eps {\cal V} \label{eqn:rec_space_bound} \\
& \leq & \frac{ \eps^2 }{ n } \cdot {\cal V} + \eps {\cal V} \nonumber \\
& \leq & 2\eps {\cal V} \ . \nonumber
\end{eqnarray}
Here, inequality~\eqref{eqn:rec_space_bound} is obtained by arguing that $\gamma_i \cdot \frac{ \tau_\mycyc }{ (n/\eps)^{3(q-2)} } \leq {\cal V}$ for every commodity $i \in {\cal F}_{\leq q-2}$. To this end, by definition of ${\cal F}_1, \ldots, {\cal F}_{q-2}$, we know that with respect to the policy $\tilde{\cal P}_i$, this commodity has $N( \tilde{\cal P}_i, [0,\tau_{\mycyc})) \leq (\frac{n}{\eps})^{3(q-2)}$ orders across the entire cycle. Therefore, at least one these orders consists of at least $\frac{ \tau_\mycyc }{ (n/\eps)^{3(q-2)} }$ units, which require $\gamma_i \cdot \frac{ \tau_\mycyc }{ (n/\eps)^{3(q-2)} }$ amount of space by themselves. As shown in Theorem~\ref{thm:good_cyclic}, the policy $\tilde{\cal P}$ is capacity-feasible, implying in particular that $\gamma_i \cdot \frac{ \tau_\mycyc }{ (n/\eps)^{3(q-2)} } \leq {\cal V}$.
\end{proof}

\paragraph{Running time.} As far as efficiency is concerned, as explained in Section~\ref{subsec:dp_guess_State}, our preliminary guessing step involves enumerating over $O( (\frac{ |{\cal I}| }{ \eps } )^n)$ possible configuration of the cycle length estimate $\tilde{\tau}_{ \mycyc }$ and the frequency classes $\tilde{\cal F}_1, \ldots, \tilde{\cal F}_Q$. Given these guesses, our dynamic program consists of $O( 2^{ O( n^3 / \eps^2 ) } )$ states. Following the discussion in Section~\ref{subsec:recursive_details}, at each of these states, there are only $2^{ O( n^{6} / \eps^{5} ) }$ possible $B$-aligned policies to be examined. Any such policy can be tested for acceptability in polynomial time, and subsequently extended via $( \frac{n}{\eps})^3$ recursive calls. Putting these pieces together, a straightforward implementation requires $O( | {\cal I} |^{O(n)} \cdot 2^{ O( n^{6} / \eps^{5} ) } )$ time overall.

\section{Polynomial Time \bstitle{(2-\frac{17}{5000} + \eps)}-Approximation} \label{sec:sub2-approx}

In what follows, we provide an in-depth description of our approach for devising an $O( | {\cal I} |^{\tilde{O}( 1/\eps^5)} \cdot 2^{ \tilde{O}( 1 / \eps^{35} ) } )$-time construction of a random capacity-feasible policy whose expected long-run average cost is within factor $2-\frac{17}{5000}+ \eps$ of optimal, as formally stated in Theorem~\ref{thm:2_minus_delta}. For this purpose, Section~\ref{subsec:alg_outline_2minus} is intended to present a high-level outline of these developments. On the one hand, as explained in Section~\ref{subsec:high_level_easy}, we will isolate relatively easy parametric regimes that can be handled by fusing the algorithmic ideas of Sections~\ref{sec:anily_approx} and~\ref{sec:exponential_approx}. On the other hand, we will expose the truly difficult regime, whose finer details will be discussed in Sections~\ref{subsec:high_level_difficult}-\ref{susbec:dense_class_analysis}. The latter analysis will heavily rely on an essential structural finding, referred to as the Po2-Synchronization Theorem, whose proof will be outsourced to Section~\ref{sec:po2-sync}.

\subsection{Algorithmic outline} \label{subsec:alg_outline_2minus}

Moving forward, rather than directly comparing ourselves to an optimal replenishment policy, which could be convolutedly  structured, it will be convenient to have a cyclic policy as an intermediate benchmark. To this end, as explained in Sections~\ref{subsec:average_bound_relaxation} and~\ref{subsec:structured_cyclic}, we know that for any $\eps > 0$, there exists a capacity-feasible cyclic policy ${\cal P}$ with a long-run average cost of $C( {\cal P} ) \leq (1 + \eps) \cdot \opt\eqref{eqn:model_warehouse}$. From this point on, ${\cal P}^{\eps}$ will stands for one such policy.

\paragraph{Volume classes.} By writing the average space bound~\eqref{eqn:average-space-bound} in terms of ${\cal P}^{\eps}$, we infer that its average occupied space must be upper-bounded by our overall capacity, i.e., $\sum_{i \in [n]} \gamma_i \cdot \bar{I}( {\cal P}_i^{\eps} ) \leq {\cal V}$. Here, the average inventory levels $\{ \bar{I}( {\cal P}_i^{\eps} ) \}_{i \in [n]}$ across a single cycle are well-defined, which would not have been the case had we been focusing on non-cyclic policies. Motivated by this bound, for purposes of analysis, let us  partition the underlying set of commodities into $O( \frac{ 1 }{ \eps } \log \frac{ n }{ \eps } )$ volume classes ${\cal V}_1^{\eps}, \ldots, {\cal V}_L^{\eps}, {\cal V}_{\infty}^{\eps}$ based on their average occupied space $\{ \gamma_i \cdot \bar{I}( {\cal P}^{\eps}_i ) \}_{i \in [n]}$. This partition is defined as follows:
\begin{itemize}
    \item The class ${\cal V}_1^{\eps}$ consists of commodities with $\gamma_i \cdot \bar{I}( {\cal P}^{\eps}_i ) \in ( \frac{1}{1+\eps} \cdot {\cal V}, {\cal V}]$.

    \item The class ${\cal V}_2^{\eps}$ consists of those with $\gamma_i \cdot \bar{I}( {\cal P}^{\eps}_i ) \in ( \frac{1}{(1+\eps)^2} \cdot {\cal V}, \frac{1}{1+\eps} \cdot{\cal V}]$.

    \item So on and so forth, up to ${\cal V}_L^{\eps}$, containing commodities with $\gamma_i \cdot \bar{I}( {\cal P}^{\eps}_i ) \in ( \frac{1}{(1+\eps)^L} \cdot {\cal V}, \frac{1}{(1+\eps)^{L-1}} \cdot{\cal V}]$. Here, we set $L = \lceil \log_{1+\eps} (\frac{ n }{ \eps }) \rceil$, meaning that $\frac{ 1 }{ (1+\eps)^L } \leq \frac{ \eps }{ n }$.

    \item In addition, the remaining class ${\cal V}_{\infty}^{\eps}$ is comprised of commodities with $\gamma_i \cdot \bar{I}( {\cal P}^{\eps}_i ) \leq \frac{1}{(1+\eps)^L} \cdot {\cal V}$.
\end{itemize}

\paragraph{Sparsity and density.}
For every $\ell \in [L]_{\infty}$, we say that class ${\cal V}_{\ell}^{\eps}$ is sparse when $| {\cal V}_{\ell}^{\eps} | \leq \frac{ 100 \ln (1/\eps) }{ \eps^4 }$; otherwise, this class is called dense. We make use of ${\cal S}$ and ${\cal D}$ to denote the index sets of sparse and dense classes. Now, suppose that ${\cal S} = \{ \ell_1, \ldots, \ell_M \}$, with the convention that indices are listed by increasing order. We proceed by decomposing this set into
\[ \underbrace{ \ell_1 \quad , \quad \ldots \quad , \quad \ell_{\mymid} }_{ {\cal S}_{\mypref} } \quad , \quad \underbrace{ \ell_{\mymid+1} \quad \ldots \quad , \quad \ell_M }_{ {\cal S}_{\mysuff} } \ .  \]
Here, $\ell_{\mymid}$ is the unique index up to which $\Delta = \lceil \log_{1 + \eps} ( \frac{ 125 \ln(1/\eps) }{ \eps^6 } ) \rceil$  of the sparse classes ${\cal V}^{\eps}_{\ell_1}, \ldots, {\cal V}^{\eps}_{\ell_{\mymid}}$ are not empty; when there are fewer than $\Delta$ such classes, $\ell_{\mymid} = \ell_M$.

\paragraph{Guessing.} As our first algorithmic step, we guess the next few ingredients, related to the unknown $\eps$-optimal policy ${\cal P}^{\eps}$:
\begin{itemize}
    \item {\em Types of  classes}: For every $\ell \in [L]_{\infty}$, we guess whether class ${\cal V}_{\ell}^{\eps}$ is prefix-sparse, suffix-sparse, or dense by enumerating over all  $3^{L+1} = n^{ \tilde{O}(1 / \eps) }$ options. From this point on, the sets ${\cal S}_{\mypref}$, ${\cal S}_{\mysuff}$, and ${\cal D}$ will be assumed to be known.

    \item {\em Commodities within prefix-sparse classes}: For every $\ell \in {\cal S}_{\mypref}$, we guess  the precise identity of ${\cal V}_{\ell}^{\eps}$. Noting that ${\cal S}_{\mypref}$ consists of at most $\Delta = O( \frac{ 1 }{ \eps } \log \frac{ 1 }{ \eps } )$ non-empty classes, and that any such class contains at most $\frac{ 100 \ln (1/\eps) }{ \eps^4 }$ commodities, the total number of guesses across all prefix-sparse classes is $n^{ \tilde{O}( 1/\eps^5 ) } $.

    \item {\em Size of suffix-sparse classes}: For every $\ell \in {\cal S}_{\mysuff}$, we guess the cardinality $| {\cal V}_{\ell}^{\eps} |$ of this  class. Here, there are at most $(\frac{ 100 \ln (1/\eps) }{ \eps^4 }+1)^{|{\cal S}_{\mysuff}|} = O( n^{ \tilde{O}( 1/ \eps ) } )$ options to be considered.
\end{itemize}

\paragraph{The easy scenario: High volume of sparse classes.} For simplicity of notation, let $\bar{V}^{\eps}_{\cal S} = \sum_{\ell \in {\cal S}} \sum_{i \in {\cal V}_{\ell}^{\eps}} \gamma_i \cdot \bar{I}( {\cal P}_i^{\eps} )$ and $\bar{V}^{\eps}_{\cal D} = \sum_{\ell \in {\cal D}} \sum_{i \in {\cal V}_{\ell}^{\eps}} \gamma_i \cdot \bar{I}( {\cal P}_i^{\eps} )$ be the total average space occupied by the commodities in sparse and dense classes, both with respect to the $\eps$-optimal policy ${\cal P}^{\eps}$. In Section~\ref{subsec:high_level_easy}, we consider the retrospectively easy scenario, where the average space $\bar{V}^{\eps}_{\cal S}$ of sparse commodities forms a sufficiently large fraction of the capacity bound ${\cal V}$, specifically meaning that $\bar{V}^{\eps}_{\cal S} \geq (\frac{1}{2} + \delta) \cdot {\cal V}$; here, $\delta \in (0, \frac{ 1 }{ 2 })$ is an absolute constant whose value will be determined later on. As formally stated below, by judiciously combining the algorithmic ideas of Sections~\ref{sec:anily_approx} and~\ref{sec:exponential_approx}, this scenario will allow us to approach the optimal long-run average cost within a factor of essentially $2(1-\delta)$. 

\begin{lemma} \label{lem:large_V_sparse}
Suppose that $\bar{V}^{\eps}_{\cal S} \geq (\frac{1}{2} + \delta) \cdot {\cal V}$. Then, we can compute in $O( | {\cal I} |^{\tilde{O}( 1/\eps^6)} \cdot 2^{ \tilde{O}( 1 / \eps^{41} ) } )$ time a deterministic capacity-feasible policy ${\cal P}$ with 
$C( {\cal P} ) \leq (2 - 2\delta + 8\eps) \cdot C( {\cal P}^{\eps} )$. 
\end{lemma} 

\paragraph{The difficult scenario: Low volume of sparse classes.} In Section~\ref{subsec:high_level_difficult}, we will operate in the complementary scenario, where $\bar{V}^{\eps}_{\cal S} < (\frac{1}{2} + \delta) \cdot {\cal V}$. Along the way, we will discover that this regime captures the essence of why sub-$2$-approximations do not seem plausible via existing methods. Since our approach in this context involves quite a few moving parts, its specifics will be discussed in Sections~\ref{subsec:useful_partition_dense}-\ref{susbec:dense_class_analysis} as well as in Section~\ref{sec:po2-sync}. Technically speaking, up to $\eps$-dependent terms, these ideas will culminate to a polynomial-time approximation for the optimal long-run average cost within factor 
$\max \{ 2(1-\delta), 1.9932 + 3\delta \}$. 

\begin{lemma} \label{lem:large_V_dense}
Suppose that $\bar{V}^{\eps}_{\cal S} < (\frac{1}{2} + \delta) \cdot {\cal V}$. Then, we can compute in $O( n^{ \tilde{O}( 1 / \eps^2 ) } )$ time a random capacity-feasible policy ${\cal P}$ with an expected cost of
\[ \ex{ C({\cal P}) } ~~\leq~~ \max \{ 2 - 2\delta + 4\eps, 1.9932 + 2\delta + 3\eps \} \cdot C( {\cal P}^{\eps} ) \ . \]
\end{lemma}

\paragraph{Choosing $\bs{\delta}$.} Needless to say, we have no control over whether any given instance falls within the easy scenario or the difficult one. Thus, for the purpose of minimizing the worst-case performance guarantee over these two scenarios, we pick our yet-unspecified threshold parameter as $\delta = \frac{17}{10000}$. As such, the approximation ratios stated in Lemmas~\ref{lem:large_V_sparse} and~\ref{lem:large_V_dense} respectively become $2 - \frac{17}{5000} + 8\eps$ and $2 - \frac{17}{5000} + 4\eps$, thereby concluding the proof of  Theorem~\ref{thm:2_minus_delta}.

\subsection{The easy scenario: \bstitle{\bar{V}^{\eps}_{\cal S} \geq (\frac{1}{2} + \delta) \cdot {\cal V}}} \label{subsec:high_level_easy}

\paragraph{Policy for prefix-sparse classes.} Following the guessing procedure of Section~\ref{subsec:alg_outline_2minus}, we already know the index set ${\cal S}_{\mypref}$ of prefix-sparse classes, as well as the set of commodities that belong to each of these classes. We begin by recalling that, while ${\cal S}_{\mypref}$ may be comprised of $\Omega( \frac{ 1 }{ \eps } \log \frac{ n }{ \eps } )$ classes, at most $\Delta = O( \frac{ 1 }{ \eps} \log \frac{ 1 }{ \eps })$ of these classes are  non-empty. Since each such class contains up to $\frac{ 100 \ln (1/\eps) }{ \eps^4 }$ commodities, we have $\sum_{\ell \in {\cal S}_{\mypref}} | {\cal V}^{\eps}_{ \ell } | = O( \frac{ 1 }{ \eps^5 } \log^2 (\frac{ 1 }{ \eps }) ) $. Therefore, according to Theorem~\ref{thm:PTAS_exponential}, by employing the approximation scheme proposed in Section~\ref{sec:exponential_approx}, we can compute a $(1+\eps)$-approximate capacity-feasible policy ${\cal P}_{\mypref}$ for these commodities in $O( | {\cal I} |^{\tilde{O}( 1/\eps^5)} \cdot 2^{ \tilde{O}( 1 / \eps^{35} ) } )$ time. In terms of cost, since restricting the policy ${\cal P}^{\eps}$ to these commodities forms a feasible solution in this context, we have 
\begin{equation} \label{eqn:easy_sparse_both}
C( {\cal P}_{\mypref} ) ~~\leq~~ (1+\eps) \cdot \sum_{\ell \in {\cal S}_{\mypref}} \sum_{i \in {\cal V}_{\ell}^{\eps}} C_i( {\cal P}_i^{\eps} ) \ . 
\end{equation}

\paragraph{Policy for suffix+dense classes.} Moving forward, it is important to emphasize that our guessing procedure does not inform us about how the set of commodities outside of prefix-sparse classes are partitioned between suffix-sparse classes and dense classes. To go around this obstacle, we first argue that the total average space occupied by the commodities in ${\cal S}_{\mysuff}$ is rather negligible. The proof of this result is given in Appendix~\ref{app:proof_lem_avg_space_prefix}.

\begin{lemma} \label{lem:avg_space_prefix}
$\sum_{\ell \in {\cal S}_{\mysuff}} \sum_{i \in {\cal V}_{\ell}^{\eps}} \gamma_i \cdot \bar{I}( {\cal P}_i ) \leq \eps {\cal V}$.
\end{lemma}

Motivated by this observation, for the collection of commodities within the union of suffix-sparse classes and dense classes, we design a joint policy by adapting the algorithmic approach of  Section~\ref{sec:anily_approx} as follows:
\begin{itemize}
    \item Recalling that $\bar{V}^{\eps}_{\cal D} = \sum_{\ell \in {\cal D}} \sum_{i \in {\cal V}_{\ell}^{\eps}} \gamma_i \cdot \bar{I}( {\cal P}_i^{\eps} )$, we begin by guessing an overestimate $\tilde{V}_{\cal D}$ for this quantity, such that $\tilde{V}_{\cal D} \in [\bar{V}^{\eps}_{\cal D}, (1+\eps) \cdot \bar{V}^{\eps}_{\cal D} + \eps {\cal V}]$. For this purpose, it suffices to enumerate over $\eps {\cal V}, (1 + \eps) \cdot \eps {\cal V}, (1 + \eps)^2 \cdot \eps {\cal V}, \ldots$, meaning that there are only $O( \frac{ 1 }{ \eps } \log \frac{ 1 }{ \eps } )$ values to be tested.

    \item In light of Lemma~\ref{lem:avg_space_prefix}, it follows that by restricting ${\cal P}^{\eps}$ to commodities in suffix-sparse classes and dense classes, we obtain a feasible solution to the next formulation:
    \begin{equation} \label{eqn:relax_warehouse_dense}
    \tag{$\tilde{\Pi}^{\mysuff+\mydense}$}
    \begin{array}{lll}
    {\displaystyle \min_{\cal P}} & {\displaystyle \sum_{\ell \in {\cal S}_{\mysuff} \cup {\cal D}} \sum_{i \in {\cal V}_{\ell}^{\eps}} C( {\cal P}_i ) } \\
    \text{s.t.} & {\displaystyle \sum_{\ell \in {\cal S}_{\mysuff} \cup {\cal D}} \sum_{i \in {\cal V}_{\ell}^{\eps}} \gamma_i \cdot \bar{I}( {\cal P}_i ) \leq \tilde{V}_{\cal D}} + \eps {\cal V}
    \end{array}
    \end{equation}
    Consequently, $\opt\eqref{eqn:relax_warehouse_dense} \leq \sum_{\ell \in {\cal S}_{\mysuff} \cup {\cal D}} \sum_{i \in {\cal V}_{\ell}^{\eps}} C( {\cal P}^{\eps}_i )$.

    \item By duplicating the proof of Lemma~\ref{lem:sosi_optimal} to the letter, one can verify that problem~\eqref{eqn:relax_warehouse_dense} admits a SOSI optimal replenishment policy, meaning that $\opt\eqref{eqn:relax_warehouse_dense} = \opt\eqref{eqn:modify_relax_warehouse_dense}$, where the latter formulation is given by
    \begin{equation} \label{eqn:modify_relax_warehouse_dense} \tag{$\tilde{\Pi}^{\mysuff+\mydense}_{\text{SOSI}}$}
    \begin{array}{lll}
    {\displaystyle \min_T} & {\displaystyle \sum_{\ell \in {\cal S}_{\mysuff} \cup {\cal D}} \sum_{i \in {\cal V}_{\ell}^{\eps}} C_i(T_i) } \\
    \text{s.t.} & {\displaystyle \sum_{\ell \in {\cal S}_{\mysuff} \cup {\cal D}} \sum_{i \in {\cal V}_{\ell}^{\eps}} \gamma_i T_i \leq 2 \cdot (\tilde{V}_{\cal D}} + \eps {\cal V}) 
    \end{array}
    \end{equation}

    \item Finally, by computing an optimal solution to this relaxation (see Section~\ref{subsec:average_bound_relaxation}), we obtain a policy ${\cal P}_{\mysuff+\mydense}$ whose maximal space requirement is
    \begin{equation} \label{eqn:easy_sparse_space}
    V_{\max}( {\cal P}_{\mysuff+\mydense} ) ~~\leq~~ 2 \cdot (\tilde{V}_{\cal D} + \eps {\cal V}) ~~\leq~~ 2(1+\eps) \cdot \bar{V}^{\eps}_{\cal D} + 4\eps {\cal V} \ .
    \end{equation}
    At the same time, the long-run average cost of this policy is
    \begin{eqnarray} 
    C( {\cal P}_{\mysuff+\mydense} ) & = &  \opt\eqref{eqn:modify_relax_warehouse_dense} \nonumber \\
    & = &  \opt\eqref{eqn:relax_warehouse_dense} \nonumber \\
    & \leq & \sum_{\ell \in {\cal S}_{\mysuff} \cup {\cal D}} \sum_{i \in {\cal V}_{\ell}^{\eps}} C( {\cal P}^{\eps}_i ) \ . \label{eqn:easy_sparse_cost}
    \end{eqnarray}
\end{itemize}

\paragraph{The combined policy.} To obtain a single policy ${\cal P}_{\mycomb}$ for all commodities, we first glue ${\cal P}_{\mypref}$ and ${\cal P}_{\mysuff+\mydense}$ together. Since ${\cal P}_{\mypref}$ is capacity-feasible by itself, in conjunction with inequality~\eqref{eqn:easy_sparse_space}, we infer that the maximal space requirement of ${\cal P}_{\mycomb}$ is
\[V_{\max}( {\cal P}_{\mycomb} ) ~~\leq~~ {\cal V} + 2(1+\eps) \cdot \bar{V}^{\eps}_{\cal D} + 4\eps {\cal V} ~~\leq~~ (2 - 2\delta + 5\eps) \cdot {\cal V} \ , \]
where the last inequality holds since 
$\bar{V}^{\eps}_{\cal D} \leq {\cal V} - \bar{V}^{\eps}_{\cal S} \leq (\frac{1}{2} - \delta) \cdot {\cal V}$, due to operating in the scenario where $\bar{V}^{\eps}_{\cal S} \geq (\frac{1}{2} + \delta) \cdot {\cal V}$. In terms of cost, by inequalities~\eqref{eqn:easy_sparse_both} and~\eqref{eqn:easy_sparse_cost}, we have $C( {\cal P}_{\mycomb} ) \leq (1+\eps) \cdot C( {\cal P}^{\eps} )$. Therefore, scaling down ${\cal P}_{\mycomb}$ by a factor of $2 - 2\delta + 5\eps$, we obtain a capacity-feasible policy $\hat{\cal P}_{\mycomb}$ with
\begin{eqnarray*}
C( \hat{\cal P}_{\mycomb} ) & \leq & (2 - 2\delta + 5\eps) \cdot C( {\cal P}_{\mycomb} ) \\
&\leq & (2 - 2\delta + 5\eps) \cdot (1+\eps) \cdot C( {\cal P}^{\eps} )\\
&\leq & (2 - 2\delta + 8\eps) \cdot C( {\cal P}^{\eps} ) \ .
\end{eqnarray*}

\subsection{The difficult scenario: \bstitle{\bar{V}^{\eps}_{\cal S} < (\frac{1}{2} + \delta) \cdot {\cal V}}} \label{subsec:high_level_difficult}

\paragraph{Lower bound on $\bs{\bar{V}^{\eps}_{\cal D}}$.} Let us first put aside an easily-addressable case, where  $\bar{V}^{\eps}_{\cal D} < (\frac{1}{2} - 2\delta) \cdot {\cal V}$, implying that  $\bar{V}^{\eps}_{\cal S} + \bar{V}^{\eps}_{\cal D} \leq (\frac{1}{2} + \delta) \cdot {\cal V} + (\frac{1}{2} - 2\delta) \cdot {\cal V} = (1 - \delta) \cdot {\cal V}$. In this case, similarly to how  the policy ${\cal P}_{\mysuff+\mydense}$ was constructed in Section~\ref{subsec:high_level_easy}, we can apply precisely the same arguments for the entire collection of commodities, ending up with a policy ${\cal P}_{\mycomb}$ whose maximal space requirement is 
\[ V_{\max}( {\cal P}_{\mycomb} ) ~~\leq~~ 2(1+\eps) \cdot (\bar{V}^{\eps}_{\cal S} + \bar{V}^{\eps}_{\cal D}) + 2\eps {\cal V} ~~\leq~~ (2 - 2\delta + 4\eps) \cdot {\cal V} \ , \]
and whose long-run average cost is $C( {\cal P}_{\mycomb} ) \leq C( {\cal P}^{\eps} )$. Subsequently, scaling down ${\cal P}_{\mycomb}$ by a factor of $2 - 2\delta + 4\eps$, we obtain a capacity-feasible policy $\hat{\cal P}_{\mycomb}$ with $C( \hat{\cal P}_{\mycomb} ) \leq (2 - 2\delta + 4\eps) \cdot C( {\cal P}^{\eps} )$. The latter expression is precisely the first term mentioned in Lemma~\ref{lem:large_V_dense}. In the remainder of this section, we consider the regime where $\bar{V}^{\eps}_{\cal D} \geq (\frac{1}{2} - 2\delta) \cdot {\cal V}$.

\paragraph{Policy for prefix-sparse classes.} The treatment of these commodities will be identical to our actions in the suffix+dense case of Section~\ref{subsec:high_level_easy}. By running through this construction, with ``suffix+dense'' and ${\cal S}_{\mysuff} \cup {\cal D}$ replaced by ``prefix-sparse'' and ${\cal S}_{\mypref}$, we can compute in polynomial time a policy ${\cal P}_{\mypref}$ whose maximal space requirement and long-run average cost are  
\begin{equation} \label{eqn:diff_sparse_space_cost}
V_{\max}( {\cal P}_{\mypref} ) ~~\leq~~ 2(1+\eps) \cdot \bar{V}^{\eps}_{\cal S} + 2\eps {\cal V} \qquad \text{and} \qquad C( {\cal P}_{\mypref} ) ~~\leq~~ \sum_{\ell \in {\cal S}_{\mypref}} \sum_{i \in {\cal V}_{\ell}^{\eps}} C( {\cal P}^{\eps}_i ) \ .
\end{equation}

\paragraph{Policy for suffix+dense classes.}  We have just landed at the most challenging part of our algorithmic approach. Since handling these classes is particularly involved, we provide the finer details of our construction in Sections~\ref{subsec:useful_partition_dense}-\ref{susbec:dense_class_analysis} as well as in Section~\ref{sec:po2-sync}, where the next result is established. 

\begin{lemma} \label{lem:diff_case_dense_main}
We can compute in $O( n^{ \tilde{O}( 1 / \eps^2 ) } )$  time a random policy ${\cal P}_{\mysuff+\mydense}$ satisfying the next two properties: 
\begin{enumerate}
    \item {\em Space requirement:} $V_{\max}( {\cal P}_{\mysuff+\mydense} ) \leq (1 + 8\eps) \cdot \frac{ 7/4 }{ \sqrt{2} \ln 2} \cdot \bar{V}^{\eps}_{\cal D} + 10\eps  {\cal V}$ almost surely. 

    \item {\em Expected cost}: $\expar{ C( {\cal P}_{\mysuff+\mydense} ) } \leq  ( 1 + \frac{2\eps}{5} ) \cdot \frac{ 32/31 }{ \sqrt{2} \ln 2} \cdot \sum_{\ell \in {\cal S}_{\mysuff} \cup {\cal D}} \sum_{i \in {\cal V}_{\ell}^{\eps}} C_i( {\cal P}_i^{\eps} )$. 
\end{enumerate}
\end{lemma}

\paragraph{The combined policy.} To obtain a single policy ${\cal P}_{\mycomb}$ for all commodities, we first glue   ${\cal P}_{\mypref}$ and ${\cal P}_{\mysuff+\mydense}$ together. By inequality~\eqref{eqn:diff_sparse_space_cost} and Lemma~\ref{lem:diff_case_dense_main}(2), we infer that the expected  long-run average cost of this policy is
\begin{eqnarray*}
\ex{ C( {\cal P}_{\mycomb} ) } & = & C( {\cal P}_{\mypref} ) + \ex{ C( {\cal P}_{\mysuff+\mydense} ) } \\
& \leq & \sum_{\ell \in {\cal S}_{\mypref}} \sum_{i \in {\cal V}_{\ell}^{\eps}} C( {\cal P}^{\eps}_i ) + \left( 1 + \frac{2\eps}{5} \right) \cdot \frac{ 32/31 }{ \sqrt{2} \ln 2} \cdot \sum_{\ell \in {\cal S}_{\mysuff} \cup {\cal D}} \sum_{i \in {\cal V}_{\ell}^{\eps}} C_i( {\cal P}_i^{\eps} ) \\
& \leq & \left( 1 + \frac{2\eps}{5} \right) \cdot 1.0531 \cdot C( {\cal P}^{\eps} ) \ . 
\end{eqnarray*}
At the same time, the maximal space requirement of ${\cal P}_{\mycomb}$ is almost surely 
\begin{eqnarray}
V_{\max}( {\cal P}_{\mycomb} ) & \leq & V_{\max}( {\cal P}_{\mypref} ) + V_{\max}( {\cal P}_{\mysuff+\mydense} ) \nonumber \\
& \leq & 2(1+\eps) \cdot \bar{V}^{\eps}_{\cal S} + 2\eps {\cal V} + (1+8\eps) \cdot \frac{ 7/4 }{ \sqrt{2} \ln 2} \cdot  \bar{V}^{\eps}_{\cal D} + 10\eps  {\cal V}\label{eqn:diff_combined_vmax_1} \\
& = &(1+8\eps) \cdot \left( 2 \cdot \left( \bar{V}^{\eps}_{\cal S} + \bar{V}^{\eps}_{\cal D} \right) - \left( 2 - \frac{ 7/4 }{ \sqrt{2} \ln 2}  \right) \cdot \bar{V}^{\eps}_{\cal D} \right) + 12\eps {\cal V} \nonumber \\
& \leq & (1+8\eps) \cdot \left( 2  {\cal V} - \left( 2 - \frac{ 7/4 }{ \sqrt{2} \ln 2} \cdot  \right) \cdot \bar{V}^{\eps}_{\cal D} \right) + 12\eps {\cal V}\label{eqn:diff_combined_vmax_2} \\
& \leq & (1+8\eps) \cdot \left( 1 +  \frac{ 7/8 }{ \sqrt{2} \ln 2} + \frac{ \delta }{ 2 } + 12\eps \right) \cdot {\cal V} \ . \label{eqn:diff_combined_vmax_3}
\end{eqnarray}
Here, inequality~\eqref{eqn:diff_combined_vmax_1} follows from inequality~\eqref{eqn:diff_sparse_space_cost} and Lemma~\ref{lem:diff_case_dense_main}(1). Inequality~\eqref{eqn:diff_combined_vmax_2} is obtained by recalling that the $\eps$-optimal policy ${\cal P}^{\eps}$ is in particular capacity-feasible, implying that $\bar{V}^{\eps}_{\cal S} + \bar{V}^{\eps}_{\cal D} \leq {\cal V}$ due to the average-space bound (see Section~\ref{subsec:average_bound_relaxation}). Finally, inequality~\eqref{eqn:diff_combined_vmax_3} holds since $\bar{V}^{\eps}_{\cal D} \geq (\frac{1}{2} - 2\delta) \cdot {\cal V}$.

Based on these observations, when scaling down ${\cal P}_{\mycomb}$ by a factor of $(1+8\eps) \cdot ( 1 +  \frac{ 7/8 }{ \sqrt{2} \ln 2} + \frac{ \delta }{ 2 } + 12\eps )$, we obtain a capacity-feasible policy $\hat{\cal P}_{\mycomb}$ with an expected long-run average cost of
\begin{eqnarray*}
\ex{ C( \hat{\cal P}_{\mycomb} ) } & \leq & (1+8\eps) \cdot \left( 1 +  \frac{ 7/8 }{ \sqrt{2} \ln 2} + \frac{ \delta }{ 2 } + 12\eps \right) \cdot \ex{ C( {\cal P}_{\mycomb} ) } \\
&\leq & (1+9\eps) \cdot \left( 1 +  \frac{ 7/8 }{ \sqrt{2} \ln 2} + \frac{ \delta }{ 2 } + 12\eps \right) \cdot \left( 1 + \frac{2\eps}{5} \right) \cdot 1.0531 \cdot C( {\cal P}^{\eps} )\\
&\leq & (1.9932 + 2\delta + 3\eps) \cdot C( {\cal P}^{\eps} ) \ .
\end{eqnarray*}

\subsection{The suffix+dense classes algorithm: Computing a mimicking partition} \label{subsec:useful_partition_dense}

As previously mentioned, the next few sections are dedicated to establishing  Lemma~\ref{lem:diff_case_dense_main}. To this end, let us recall that, subsequently to our guessing procedure (see Section~\ref{subsec:alg_outline_2minus}), the index sets ${\cal S}_{\mysuff}$ and ${\cal D}$  as well as the combined set  of commodities $\bigcup_{\ell \in {\cal S}_{\mysuff} \cup {\cal D}} {\cal V}_{\ell}^{\eps}$ that reside within these classes are already known. However, we have no idea about how these commodities are partitioned between the underlying classes.

\paragraph{Additional guessing step.} To simplify several  future arguments, for every class index $\ell \in {\cal D} \setminus \{ \infty \}$, let $\bar{V}^{\eps}_{\ell} = \sum_{i \in {\cal V}_{\ell}^{\eps}} \gamma_i \cdot \bar{I}( {\cal P}_i^{\eps} )$ be the total average space occupied by ${\cal V}_{\ell}^{\eps}$-commodities with respect to the policy ${\cal P}^{\eps}$. As our final guessing step, we obtain an over-estimate $\tilde{V}^{\eps}_{\ell}$ for this quantity, satisfying $\bar{V}^{\eps}_{\ell} \leq \tilde{V}^{\eps}_{\ell} \leq \bar{V}^{\eps}_{\ell} + \frac{ \eps }{ |{\cal D}| } \cdot {\cal V}$. By observing that $\sum_{\ell \in {\cal D}} \bar{V}^{\eps}_{\ell} = \bar{V}^{\eps}_{\cal D} \leq {\cal V}$, elementary balls-and-bins counting arguments show that there are only $2^{ O( |{\cal D}| / \eps ) } = n^{ \tilde{O}( 1 / \eps^2 ) }$ options to be jointly considered for $\{ \tilde{V}^{\eps}_{\ell} \}_{\ell \in {\cal D}}$. Furthermore, letting $\tilde{\cal N}_{\ell} = \frac{ (1 + \eps)^{\ell} }{ {\cal V} } \cdot \tilde{V}^{\eps}_{\ell}$, we argue that this term provides an upper bound on the cardinality of ${\cal V}_{ \ell }^{\eps}$. Indeed, 
\[ \tilde{V}^{\eps}_{\ell} ~~\geq~~  \bar{V}^{\eps}_{\ell} ~~=~~ \sum_{i \in {\cal V}_{\ell}^{\eps}} \gamma_i \cdot \bar{I}( {\cal P}_i^{\eps} )  ~~\geq~~ | {\cal V}_{\ell}^{\eps} | \cdot \frac{\cal V}{(1+\eps)^{\ell}} \ , \]
where the last inequality holds since $\gamma_i \cdot \bar{I}( {\cal P}^{\eps}_i ) \in ( \frac{1}{(1+\eps)^{\ell}} \cdot {\cal V}, \frac{1}{(1+\eps)^{\ell-1}} \cdot{\cal V}]$ for every commodity $i \in {\cal V}_{\ell}^{\eps}$. Multiplying both sides of the above inequality by $\frac{ (1 + \eps)^{\ell} }{ {\cal V} }$, we immediately get $\tilde{\cal N}_{\ell} \geq | {\cal V}_{ \ell }^{\eps} |$.

\paragraph{The mimicking partition.} In what follows, we devise a matching-based method to efficiently partition the commodities in $U = \bigcup_{\ell \in {\cal S}_{\mysuff} \cup {\cal D}} {\cal V}_{\ell}^{\eps}$ into subsets $\{ \tilde{\cal V}_{ \ell } \}_{\ell \in {\cal S}_{\mysuff} \cup {\cal D}}$ and  to identify SOSI replenishment policies $\{ \hat{T}_i \}_{i \in U}$ for these commodities, such that the next three properties are satisfied: 
\begin{enumerate}
    \item \label{prop:tildeV_size} {\em Class size:} $|\tilde{\cal V}_{ \ell }| = |{\cal V}_{\ell}^{\eps} |$, for every $\ell \in {\cal S}_{\mysuff}$, and $|\tilde{\cal V}_{ \ell }| \in [\frac{ 100 \ln (1/\eps) }{ \eps^4 }, \tilde{\cal N}_{\ell}]$ for every $\ell \in {\cal D}$.

    \item \label{prop:hatT_cost} {\em Total cost:} $\sum_{i \in U}  C_i( \hat{T}_i ) \leq  \sum_{i \in U} C_i( {\cal P}^{\eps}_i )$. 

    \item \label{prop:hatT_space} {\em Average space within class:} For every class $\ell \in {\cal S}_{\mysuff} \cup {\cal D}$ and commodity $i \in \tilde{\cal V}_{\ell}$,
    \begin{equation} \label{eqn:occupied_space_constraint}
    \gamma_i \cdot \bar{I}( \hat{T}_i ) ~~\leq~~ \begin{cases}
    \frac{1}{(1+\eps)^{\ell-1}} \cdot{\cal V}, \qquad & \text{if } \ell \in [L] \\
    \frac{ \eps }{ n } \cdot {\cal V}, & \text{if } \ell = \infty
    \end{cases}    
    \end{equation}
\end{enumerate}

\paragraph{Matching-like formulation.} Toward this objective, we define an instance ${\cal I}$ of the minimum-weight $b$-matching problem as follows:
\begin{itemize}
    \item {\em Graph:} The underlying graph $G$ is bipartite and complete, with the set of commodities in $U$ on one side, and with the index set ${\cal S}_{\mysuff} \cup {\cal D}$ on the other.

    \item {\em Edge weights:} The weight $w_{i \ell}$ of each edge $(i,\ell)$ is set as the minimum $C_i$-cost of a SOSI policy $\hat{T}_{i \ell}$ satisfying constraint~\eqref{eqn:occupied_space_constraint}. One can easily notice that $w_{i \ell}$ admits a closed-form expression. Indeed, since $\bar{I}( T_i ) = \frac{ T_i }{ 2 }$, we are minimizing $C_i( T_i ) = \frac{ K_i }{ T_i } + H_i T_i$ subject to an upper bound on $T_i$. As such, by consulting Claim~\ref{clm:EOQ_properties}, it is not difficult to verify that
    \begin{equation} \label{eqn:closed_form_Ti}
    \hat{T}_{i \ell} ~~=~~ \begin{cases}
    \min \{ \sqrt{K_i / H_i}, \frac{ 2{\cal V} }{(1+\eps)^{\ell-1} \cdot \gamma_i} \}, \qquad & \text{if } \ell \in [L] \\
    \min \{ \sqrt{K_i / H_i}, \frac{ 2\eps {\cal V} }{n \gamma_i} \}, & \text{if } \ell = \infty
    \end{cases}
    \end{equation}

    \item {\em Degree constraints:} Each commodity-vertex $i \in U$ should have a degree of exactly $1$. On the opposing side, the degree of each class-vertex $\ell \in {\cal D}$ should reside  within $[\frac{ 100 \ln (1/\eps) }{ \eps^4 }, \tilde{\cal N}_{\ell}]$. Finally, each class-vertex  $\ell \in {\cal S}_{\mysuff}$ should have a degree of exactly $|{\cal V}_{ \ell }^{\eps}|$, noting that the latter quantity is known, given the guessing procedure in Section~\ref{subsec:alg_outline_2minus}.
\end{itemize}

With respect to this  formulation, we compute a minimum-weight set of edges ${\cal E}^* \subseteq E(G)$ satisfying the above-mentioned degree constraints. Any polynomial-time bipartite $b$-matching algorithm (see, e.g., \citep[Chap.~21]{Schrijver03} or \citep[Chap.~12]{KorteV18}) works for our purposes, noting that this particular instance is clearly feasible. Indeed, one can easily notice that algorithmically-unknown set of edges $\{ (i, \ell) : i \in {\cal V}_{ \ell }^{\eps} \}$ forms a possible solution. Next, to define the subsets of commodities  $\{ \tilde{\cal V}_{ \ell } \}_{\ell \in {\cal S}_{\mysuff} \cup {\cal D}}$, each such set $\tilde{\cal V}_{ \ell }$ will be given by the ${\cal E}^*$-neighbors of the class-vertex $\ell$, i.e., $\tilde{\cal V}_{ \ell } = \{ i \in U: (i,\ell) \in {\cal E}^* \}$. In addition, for each commodity $i \in \tilde{\cal V}_{ \ell }$, its corresponding SOSI policy $\hat{T}_i = \hat{T}_{i \ell}$ is the one computed via the closed-form expression~\eqref{eqn:closed_form_Ti}.

\paragraph{Verifying properties~\ref{prop:tildeV_size}-\ref{prop:hatT_space}.} We begin by observing that $|\tilde{\cal V}_{ \ell }|$ is precisely the degree of each class-vertex $\ell \in {\cal S}_{\mysuff} \cup {\cal D}$ with respect to the edge set ${\cal E}^*$, meaning that property~\ref{prop:tildeV_size} is trivially guaranteed by our degree constraints. Property~\ref{prop:hatT_space} is satisfied as well, simply by definition of $\hat{T}_i$. The non-trivial argument is regarding the total cost of the policies $\{ \hat{T}_i \}_{i \in U}$, corresponding to property~\ref{prop:hatT_cost}. The next claim, whose proof is provided in Appendix~\ref{app:proof_lem_cost_U_dense}, establishes this property.

\begin{lemma} \label{lem:cost_U_dense}
$\sum_{i \in U}  C_i( \hat{T}_i ) \leq  \sum_{i \in U} C_i( {\cal P}^{\eps}_i )$.
\end{lemma}

\subsection{The suffix+dense classes algorithm: High-level outline} \label{subsec:V_ell_rounding}

\paragraph{Policy for suffix-sparse classes.} Let $\{ \tilde{\cal V}_{ \ell } \}_{\ell \in {\cal S}_{\mysuff} \cup {\cal D}}$ be the partition resulting from Section~\ref{subsec:useful_partition_dense}, and let $\{ \hat{T}_i \}_{i \in U}$ be the SOSI policies associated with its underlying set of commodities $U = \bigcup_{\ell \in {\cal S}_{\mysuff} \cup {\cal D}} {\cal V}_{\ell}^{\eps} = \bigcup_{\ell \in {\cal S}_{\mysuff} \cup {\cal D}} \tilde{\cal V}_{ \ell }$, jointly satisfying properties~\ref{prop:tildeV_size}-\ref{prop:hatT_space}.  Given these ingredients, our replenishment policy for the commodities in $\{ \tilde{\cal V}_{ \ell } \}_{\ell \in {\cal S}_{\mysuff}}$ is very simple: For each such commodity $i$, we directly make use of its corresponding SOSI policy, $\hat{T}_i$. The next claim, whose proof is provided in Appendix~\ref{app:proof_lem_bound_space_suffix}, shows that the maximal space requirement of this policy is negligible. 

\begin{lemma} \label{lem:bound_space_suffix}
Let ${\cal P}_{\mysuff}$ the deterministic  policy where ${\cal P}_{\mysuff,i} = \hat{T}_i$ for every $i \in \bigcup_{\ell \in {\cal S}_{\mysuff}} \tilde{\cal V}_{ \ell }$. Then, $V_{\max}( {\cal P}_{\mysuff} ) \leq 4 \eps {\cal V}$ and $C( {\cal P}_{\mysuff} ) = \sum_{\ell \in {\cal S}_{\mysuff}} \sum_{i \in \tilde{\cal V}_{ \ell } } C_i( \hat{T}_i )$.
\end{lemma}

\paragraph{Policy for dense classes.} In contrast, our policy for the commodities in $\{ \tilde{\cal V}_{ \ell } \}_{\ell \in {\cal D}}$ is significantly more involved. In what follows, for every $\ell \in {\cal D}$, we explain how to efficiently construct a possibly-random replenishment policy $\tilde{\cal P}^{\ell}$ for $\tilde{\cal V}_{ \ell }$-commodities that ``approximates'' the behavior of the $\eps$-optimal policy ${\cal P}^{\eps}$ with respect to ${\cal V}_{\ell}^{\eps}$ in a very specific sense. To make this objective more concrete, we say that the policy $\tilde{\cal P}^{\ell}$ guarantees an $(\alpha,\beta)$-ratio when 
\[ V_{\max}( \tilde{\cal P}^{\ell} ) ~~\leq~~ \alpha \cdot  | \tilde{\cal V}_{\ell} | \cdot \frac{1}{(1+\eps)^{\ell-1}} \cdot{\cal V} \qquad \text{and} \qquad \expar{ C( \tilde{\cal P}^{\ell} ) } ~~\leq~~ \beta \cdot \sum_{i \in \tilde{\cal V}_{ \ell } } C_i( \hat{T}_i ) \ , \]
with the former condition holding almost surely. We proceed by arguing that, up to $\eps$-dependent terms, a ratio of $(\frac{ 7/4 }{ \sqrt{2} \ln 2} , \frac{ 32/31 }{ \sqrt{2} \ln 2})$ is attainable for the  case where $\ell \in {\cal D} \setminus \{ \infty \}$. Subsequently, the special case of $\ell = \infty$ will be handled via a separate argument.

\paragraph{Light and heavy commodities.} Starting with the scenario where $\ell \in {\cal D} \setminus \{ \infty \}$, let us recall that  $\tilde{\cal V}_{ \ell }$ consists of at least $\frac{ 100 \ln (1/\eps) }{ \eps^4 }$ commodities, by property~\ref{prop:tildeV_size}. Another important point is that, by property~\ref{prop:hatT_space}, the SOSI policy $\hat{T}_i$ corresponding to each commodity $i \in \tilde{\cal V}_{ \ell }$  satisfies constraint~\eqref{eqn:occupied_space_constraint}. Namely, $\gamma_i \cdot \bar{I}( \hat{T}_i ) \leq \frac{1}{(1+\eps)^{\ell-1}} \cdot{\cal V}$, and as a result, we can decompose $\tilde{\cal V}_{ \ell }$ into commodities of two possible types: Commodity $i \in \tilde{\cal V}_{ \ell }$ is called light when $\gamma_i \cdot \bar{I}( \hat{T}_i ) \leq \frac{ 3 }{ 4 } \cdot \frac{1}{(1+\eps)^{\ell-1}} \cdot{\cal V}$; otherwise, $\gamma_i \cdot \bar{I}( \hat{T}_i ) \in (\frac{ 3 }{ 4 } \cdot \frac{1}{(1+\eps)^{\ell-1}} \cdot{\cal V}, \frac{1}{(1+\eps)^{\ell-1}} \cdot{\cal V}]$, and this commodity is called heavy. The collections of light and heavy commodities within $\tilde{\cal V}_{ \ell }$ will be denoted by ${\cal L}_{ \ell }$ and ${\cal H}_{ \ell }$, respectively. We proceed to construct a replenishment policy $\tilde{\cal P}^{\ell}$ for $\tilde{\cal V}_{ \ell }$-commodities, independently of any other class, by considering two cases, depending on the cardinality of ${\cal L}_{ \ell }$ and ${\cal H}_{ \ell }$.

\paragraph{Easy case: $\bs{ | {\cal L}_{ \ell } | \geq \frac{ | \tilde{\cal V}_{ \ell } | }{ 2 }}$.} We first consider the straightforward scenario, where at least half of the commodities in $\tilde{\cal V}_{ \ell }$ are light. The reason for referring to this case as ``straightforward'' is that, due to having a majority of light commodities, the SOSI policies $\{ \hat{T}_i \}_{i \in \tilde{\cal V}_{ \ell }}$ themselves lead to a $(\frac{ 7 }{ 4 }, 1)$-ratio, which is even better than the $(\frac{ 7/4 }{ \sqrt{2} \ln 2} , \frac{ 32/31 }{ \sqrt{2} \ln 2})$-ratio we are shooting for. The next claim, whose proof is provided in Appendix~\ref{app:proof_lem_main_result_Vell_light}, formalizes this statement. 

\begin{lemma} \label{lem:main_result_Vell_light}
Let $\tilde{\cal P}^{\ell} = \{ \tilde{\cal P}^{\ell}_i \}_{i \in \tilde{\cal V}_{ \ell }}$ be the deterministic policy where $\tilde{\cal P}^{\ell}_i = \hat{T}_i$ for every $i \in \tilde{\cal V}_{ \ell }$. When $| {\cal L}_{ \ell } | \geq \frac{ | \tilde{\cal V}_{ \ell } | }{ 2 }$, this policy guarantees a $(\frac{ 7 }{ 4 }, 1)$-ratio.
\end{lemma}

\paragraph{Difficult case: $\bs{ | {\cal H}_{ \ell } | > \frac{ | \tilde{\cal V}_{ \ell } | }{ 2 }}$.} As it turns out, handling the opposite scenario where ${ | {\cal H}_{ \ell } | > \frac{ | \tilde{\cal V}_{ \ell } | }{ 2 }}$ is significantly more challenging, forcing us to develop a new set of algorithmic tools and analytical ideas. For ease of exposition, these contents are discussed in Section~\ref{sec:po2-sync}, where we prove the following result, referred to as the Po2-Synchronization Theorem.

\begin{theorem}[Po2-Synchronization] \label{thm:main_result_Vell_heavy}
When ${ | {\cal H}_{ \ell } | > \frac{ | \tilde{\cal V}_{ \ell } | }{ 2 }}$, we can construct in polynomial time a random  policy $\tilde{\cal P}^{\ell} = \{ \tilde{\cal P}^{\ell}_i \}_{i \in \tilde{\cal V}_{ \ell }}$ that guarantees a $((1 + 6\eps) \cdot \frac{ 7/4 }{ \sqrt{2} \ln 2},(1 + \frac{2\eps}{5}) \cdot \frac{ 32/31 }{ \sqrt{2} \ln 2})$-ratio.
\end{theorem}

\paragraph{Constructing $\bs{\tilde{\cal P}^{\ell}}$ when $\bs{\ell = \infty}$.} In this case, it is unclear how to efficiently attain a $(\frac{ 7/4 }{ \sqrt{2} \ln 2} , \frac{ 32/31 }{ \sqrt{2} \ln 2})$-ratio, or any other useful ratio for that matter. However, by property~\ref{prop:hatT_space}, we know  that $\gamma_i \cdot \bar{I}( \hat{T}_i ) \leq \frac{ \eps }{ n } \cdot {\cal V}$ for every commodity $i \in \tilde{\cal V}_{\infty}$, implying that the SOSI policies $\{ \hat{T}_i \}_{i \in \tilde{\cal V}_{ \ell }}$ themselves carry a small additive error in terms of their space requirement, since $\sum_{i \in \tilde{\cal V}_{ \infty }} \gamma_i \cdot \hat{T}_i \leq | \tilde{\cal V}_{ \infty } | \cdot \frac{ 2\eps }{ n } \cdot {\cal V}  \leq 2\eps  {\cal V}$. As an immediate consequence, we obtain the next claim.

\begin{observation} \label{obs:main_result_Vinfty}
Let $\tilde{\cal P}^{\infty} = \{ \tilde{\cal P}^{\infty}_i \}_{i \in \tilde{\cal V}_{ \infty }}$ be the deterministic  policy where $\tilde{\cal P}^{\infty}_i = \hat{T}_i$ for every $i \in \tilde{\cal V}_{ \infty }$. Then, $V_{\max}( \tilde{\cal P}^{\infty} ) \leq 2\eps  {\cal V}$ and $C( \tilde{\cal P}^{\infty} ) = \sum_{i \in \tilde{\cal V}_{ \infty } } C_i( \hat{T}_i )$.
\end{observation}

\paragraph{Defining a combined policy.} Having just dealt with all possible cases, it remains to propose a single policy ${\cal P}_{\mysuff+\mydense}$ for the entire collection of commodities in $\{ \tilde{\cal V}_{ \ell } \}_{\ell \in {\cal S}_{\mysuff} \cup {\cal D}}$. To this end, we simply glue together ${\cal P}_{\mysuff}$ and ${\cal P}_{\mydense}$, where the latter is comprised of the policies $\{ \tilde{\cal P}_{ \ell } \}_{\ell \in {\cal D}}$ mentioned above. It is worth noting that ${\cal P}_{\mydense}$ is generally a random policy, since some of $\{ \tilde{\cal P}_{ \ell } \}_{\ell \in {\cal D}}$ are deterministic and some are random, depending on whether $\ell = \infty$ or not, and on whether ${ | {\cal L}_{ \ell } | \geq \frac{ | \tilde{\cal V}_{ \ell } | }{ 2 }}$ or ${ | {\cal H}_{ \ell } | > \frac{ | \tilde{\cal V}_{ \ell } | }{ 2 }}$.

\subsection{The suffix+dense classes algorithm: Analysis} \label{susbec:dense_class_analysis}

We conclude the proof of Lemma~\ref{lem:diff_case_dense_main} by deriving the next two claims, with the desired bounds on the maximal space requirement and the expected long-run average cost of our combined policy ${\cal P}_{\mysuff+\mydense}$.

\begin{lemma} 
$V_{\max}( {\cal P}_{\mysuff+\mydense} ) \leq (1 + 8\eps) \cdot \frac{ 7/4 }{ \sqrt{2} \ln 2} \cdot \bar{V}^{\eps}_{\cal D} + 10\eps  {\cal V}$, almost surely. 
\end{lemma}
\begin{proof}
Based on the preceding discussion, we almost surely have 
\begin{eqnarray}
V_{\max}( {\cal P}_{\mysuff+\mydense} ) & \leq & \sum_{ \MyAbove{ \ell \in {\cal D} \setminus \{ \infty \} : }{ | {\cal L}_{ \ell } | \geq \frac{ | \tilde{\cal V}_{ \ell } | }{ 2 } } } V_{\max}( \tilde{\cal P}^{\ell} ) + \sum_{ \MyAbove{ \ell \in {\cal D} \setminus \{ \infty \} : }{ | {\cal H}_{ \ell } | > \frac{ | \tilde{\cal V}_{ \ell } | }{ 2 } } } V_{\max}( \tilde{\cal P}^{\ell} ) + V_{\max}( \tilde{\cal P}^{\infty} ) + V_{\max}( {\cal P}_{\mysuff} ) \nonumber \\
& \leq & \frac{ 7 }{ 4 } \cdot \sum_{ \MyAbove{ \ell \in {\cal D} \setminus \{ \infty \} : }{ | {\cal L}_{ \ell } | \geq \frac{ | \tilde{\cal V}_{ \ell } | }{ 2 } } } | \tilde{\cal V}_{\ell} | \cdot \frac{1}{(1+\eps)^{\ell-1}} \cdot{\cal V} \nonumber \\
&& \mbox{} + (1 + 6\eps) \cdot \frac{ 7/4 }{ \sqrt{2} \ln 2} \cdot \sum_{ \MyAbove{ \ell \in {\cal D} \setminus \{ \infty \} : }{ | {\cal H}_{ \ell } | > \frac{ | \tilde{\cal V}_{ \ell } | }{ 2 } } }  | \tilde{\cal V}_{\ell} | \cdot \frac{1}{(1+\eps)^{\ell-1}} \cdot{\cal V} + 6\eps  {\cal V} \label{eqn:UB_Pdense_space_1} \\
& \leq & (1 + 6\eps) \cdot \frac{ 7/4 }{ \sqrt{2} \ln 2} \cdot \sum_{\ell \in {\cal D} \setminus \{ \infty \}} | \tilde{\cal V}_{\ell} | \cdot \frac{1}{(1+\eps)^{\ell-1}} \cdot{\cal V} + 6\eps  {\cal V} \nonumber \\
& \leq & (1 + 6\eps) \cdot \frac{ 7/4 }{ \sqrt{2} \ln 2} \cdot (1 + \eps) \cdot \sum_{\ell \in {\cal D} \setminus \{ \infty \}} \tilde{V}^{\eps}_{\ell} + 6\eps  {\cal V} \label{eqn:UB_Pdense_space_2}  \\
& \leq & (1 + 8\eps) \cdot \frac{ 7/4 }{ \sqrt{2} \ln 2} \cdot \sum_{\ell \in {\cal D} \setminus \{ \infty \}}  \left( \bar{V}^{\eps}_{\ell} + \frac{ \eps }{ |{\cal D}| } \cdot {\cal V}\right) + 6\eps  {\cal V} \label{eqn:UB_Pdense_space_3} \\
& \leq & (1 + 8\eps) \cdot \frac{ 7/4 }{ \sqrt{2} \ln 2} \cdot \bar{V}^{\eps}_{\cal D} + 10\eps  {\cal V} \ . \label{eqn:UB_Pdense_space_4}
\end{eqnarray}
Here, inequality~\eqref{eqn:UB_Pdense_space_1} is obtained by combining Lemmas~\ref{lem:bound_space_suffix} and~\ref{lem:main_result_Vell_light}, Theorem~\ref{thm:main_result_Vell_heavy}, and Observation~\ref{obs:main_result_Vinfty}. Inequality~\eqref{eqn:UB_Pdense_space_2} follows from property~\ref{prop:tildeV_size}, stating in particular that $|\tilde{\cal V}_{ \ell }| \leq  \tilde{\cal N}_{\ell} = \frac{ (1 + \eps)^{\ell} }{ {\cal V} } \cdot \tilde{V}^{\eps}_{\ell}$ for every $\ell \in {\cal D}$. Inequality~\eqref{eqn:UB_Pdense_space_3} holds since $\tilde{V}^{\eps}_{\ell} \leq \bar{V}^{\eps}_{\ell} + \frac{ \eps }{ |{\cal D}| } \cdot {\cal V}$, as explained in Section~\ref{subsec:useful_partition_dense}. Finally, we arrive at inequality~\eqref{eqn:UB_Pdense_space_4} by recalling that  $\bar{V}^{\eps}_{\cal D} = \sum_{\ell \in {\cal D}} \bar{V}^{\eps}_{\ell}$.
\end{proof}

\begin{lemma}
$\expar{ C( {\cal P}_{\mysuff+\mydense} ) } \leq ( 1 + \frac{2\eps}{5} ) \cdot \frac{ 32/31 }{ \sqrt{2} \ln 2} \cdot \sum_{\ell \in {\cal S}_{\mysuff} \cup {\cal D}} \sum_{i \in {\cal V}_{\ell}^{\eps}} C_i( {\cal P}_i^{\eps} )$.
\end{lemma}
\begin{proof}
Similarly to the proof of the previous lemma, we have
\begin{eqnarray}
&&\ex{ C( {\cal P}_{\mysuff+\mydense} ) } \nonumber\\
& & \qquad =~~ \sum_{ \MyAbove{ \ell \in {\cal D} \setminus \{ \infty \} : }{ | {\cal L}_{ \ell } | \geq \frac{ | \tilde{\cal V}_{ \ell } | }{ 2 } } }  C( \tilde{\cal P}^{\ell} ) + \sum_{ \MyAbove{ \ell \in {\cal D} \setminus \{ \infty \} : }{ | {\cal H}_{ \ell } | > \frac{ | \tilde{\cal V}_{ \ell } | }{ 2 } } } \ex{ C( \tilde{\cal P}^{\ell} ) } + C( \tilde{\cal P}^{\infty} ) +C( {\cal P}_{\mysuff} ) 
\nonumber \\
&& \qquad \leq~~ \sum_{ \MyAbove{ \ell \in {\cal D} \setminus \{ \infty \} : }{ | {\cal L}_{ \ell } | \geq \frac{ | \tilde{\cal V}_{ \ell } | }{ 2 } } } \sum_{i \in \tilde{\cal V}_{ \ell } } C_i( \hat{T}_i ) + \left( 1 + \frac{2\eps}{5} \right) \cdot \frac{ 32/31 }{ \sqrt{2} \ln 2} \cdot
\sum_{ \MyAbove{ \ell \in {\cal D} \setminus \{ \infty \} : }{ | {\cal H}_{ \ell } | > \frac{ | \tilde{\cal V}_{ \ell } | }{ 2 } } }  \sum_{i \in \tilde{\cal V}_{ \ell } } C_i( \hat{T}_i ) \nonumber \\
&& \qquad \qquad  \mbox{} + \sum_{i \in \tilde{\cal V}_{ \infty } } C_i( \hat{T}_i ) + \sum_{ \ell \in {\cal S}_{\mysuff} } \sum_{i \in \tilde{\cal V}_{ \ell } } C_i( \hat{T}_i ) \label{eqn:UB_Pdense_cost_1} \\
&& \qquad \leq~~ \left( 1 + \frac{2\eps}{5} \right) \cdot \frac{ 32/31 }{ \sqrt{2} \ln 2} \cdot \sum_{\ell \in {\cal S}_{\mysuff} \cup {\cal D}} \sum_{i \in {\cal V}_{\ell}^{\eps}} C_i( \hat{T}_i ) \nonumber \\
&& \qquad \leq~~ \left( 1 + \frac{2\eps}{5} \right) \cdot \frac{ 32/31 }{ \sqrt{2} \ln 2} \cdot \sum_{\ell \in {\cal S}_{\mysuff} \cup {\cal D}} \sum_{i \in {\cal V}_{\ell}^{\eps}} C_i( {\cal P}_i^{\eps} ) \ . \label{eqn:UB_Pdense_cost_2}
\end{eqnarray}
Here, inequality~\eqref{eqn:UB_Pdense_cost_1} is obtained by combining Lemmas~\ref{lem:bound_space_suffix} and~\ref{lem:main_result_Vell_light}, Theorem~\ref{thm:main_result_Vell_heavy}, and Observation~\ref{obs:main_result_Vinfty}. Inequality~\eqref{eqn:UB_Pdense_cost_2} holds since $\sum_{i \in U}  C_i( \hat{T}_i ) \leq  \sum_{i \in U} C_i( {\cal P}^{\eps}_i )$, by property~\ref{prop:hatT_cost}. 
\end{proof}

\section{Proof of the Po2-Synchronization Theorem} \label{sec:po2-sync}

The main objective of this section is to establish Theorem~\ref{thm:main_result_Vell_heavy}, stating that when ${ | {\cal H}_{ \ell } | > \frac{ | \tilde{\cal V}_{ \ell } | }{ 2 }}$, we can efficiently construct  a random policy $\tilde{\cal P}^{\ell} = \{ \tilde{\cal P}^{\ell}_i \}_{i \in \tilde{\cal V}_{ \ell }}$ satisfying 
\begin{eqnarray*}
&& V_{\max}( \tilde{\cal P}^{\ell} ) ~~\leq~~ (1 + 6\eps) \cdot \frac{ 7/4 }{ \sqrt{2} \ln 2} \cdot  | \tilde{\cal V}_{\ell} | \cdot \frac{1}{(1+\eps)^{\ell-1}} \cdot{\cal V} \\
&& \qquad \qquad \qquad \qquad \qquad \text{and} \qquad \expar{ C( \tilde{\cal P}^{\ell} ) } ~~\leq~~ \left( 1 + \frac{2\eps}{5} \right) \cdot \frac{ 32/31 }{ \sqrt{2} \ln 2} \cdot \sum_{i \in \tilde{\cal V}_{ \ell } } C_i( \hat{T}_i ) \ , 
\end{eqnarray*}
with the former property holding almost surely. To this end, Section~\ref{subsec:sync_hack} will flesh out a basic obstacle on the way to improving the approximation guarantees of \cite{Anily91} and \cite{GallegoQS96}, identifying a seemingly stylized setting where the latter barrier can be breached. Sections~\ref{subsec:pow-rounding} and~\ref{subsec:near_far_pairs} will describe a probabilistic reduction by which we the make this stylized setting applicable in proving Theorem~\ref{thm:main_result_Vell_heavy}. Finally, Sections~\ref{subsec:PO2_sync_policy} and~\ref{subsec:analysis_Aell_space_cost} will explain how our randomized policy is created, with a detailed analysis of its performance guarantees.

\subsection{The synchronization hack} \label{subsec:sync_hack}

\paragraph{Stuck with a $\bs{2}$-approximation?} Circling back to Section~\ref{sec:anily_approx}, let us develop some basic understanding of why improving on its approximation guarantee seems implausible without additional advancements. In essence, letting $T^* = (T_1^*, \ldots, T_n^*)$ be the optimal vector of SOSI policies with respect to formulation~\eqref{eqn:modify_relax_warehouse}, we noticed that these policies are generally not capacity-feasible, since their  maximal space requirement $V_{\max}( T^* ) = \sum_{i \in [n]} \gamma_i T_i^*$ could be as large as $2{\cal V}$. To correct this issue, we scaled-down each SOSI policy by a factor of $2$. As inequalities~\eqref{eqn:fix_cap_scale} and~\eqref{eqn:final_2app_eq3} show, the resulting policy becomes capacity-feasible, with the downside of potentially increasing its long-run average cost to $2 \cdot \opt\eqref{eqn:model_warehouse}$.

Along these lines, one may wish to be more sophisticated, in the sense of scaling each policy $T_i^*$ by a factor of $\alpha_i > 0$ and including a horizontal shift of $\tau_i \in [0, \alpha_i T_i^*)$. Such an alteration scales the peak space utilization of this commodity by $\alpha_i$; however, its cost could blow-up by $\max \{ \alpha_i, \frac{ 1 }{ \alpha_i } \}$. As a result, since $\alpha_i \cdot \max \{ \alpha_i, \frac{ 1 }{ \alpha_i } \} \geq 1$, each commodity by itself appears to be in worse shape, and it is  unclear whether there is a way to pick $( \alpha_i, \tau_i )_{i \in [n]}$ such that we are creating a capacity-feasible policy whose  cost is at most $(2 - \delta) \cdot \opt\eqref{eqn:model_warehouse}$, for some absolute constant $\delta > 0$. In light of these observations, formulation~\eqref{eqn:modify_relax_warehouse} appears to be a dead end in terms of obtaining a sub-$2$-approximation.

\paragraph{Additional structure and dynamic policies?} That said, the crux of our approach towards deriving the Po2-Synchronization Theorem resides in identifying a stylized setting where one could beat the space-cost tradeoff discussed above on a pairwise basis. Specifically, suppose we are given a pair of commodities, $A$ and $B$, along with corresponding SOSI policies $T_A$ and $T_B$, that miraculously satisfy the next two properties:
\begin{enumerate}
    \item \label{prop:sub1_couple_space} The peak space utilization of $T_A$ and $T_B$ are nearly equal, i.e., $\frac{ \gamma_A T_A }{ \gamma_B T_B } \in 1 \pm \eps$.

    \item \label{prop:sub1_couple_ratio} The ratio between $T_A$ and $T_B$ is an integer power of $2$.
\end{enumerate}
In this case, we say that $(A,T_A)$ and $(B, T_B)$ jointly form a sub-$1$ couple. Our main finding for such couples is summarized in the next claim, whose proof is given in Appendix~\ref{app:proof_lem_construction_pair}. 

\begin{lemma} \label{lem:construction_pair}
Let $(A,T_A)$ and $(B, T_B)$ be a given sub-$1$ couple. Then, we can construct in polynomial time dynamic replenishment policies ${\cal P}_A$ and ${\cal P}_B$ such that:
\begin{enumerate}
    \item {\em Joint occupied space:} $V_{\max}( {\cal P}_A, {\cal P}_B ) \leq (1 + \eps) \cdot \frac{ 7 }{ 8 } \cdot ( \gamma_A T_A  + \gamma_B T_B )$.    

    \item {\em Cost blow-up:} $\max \{ \frac{ C_A( {\cal P}_A ) }{ C_A( T_A )}, \frac{ C_B( {\cal P}_B ) }{ C_B( T_B)} \} \leq \frac{ 32 }{ 31}$.    
\end{enumerate}
\end{lemma}

To interpret this result, let us first point out the obvious: We take explicit advantage of our freedom to design dynamic policies, unlike the approach of Section~\ref{sec:anily_approx}, which solely focuses on SOSI policies. The not-so-obvious feature of this construction is that by item~1, the maximal space requirement of gluing ${\cal P}_A$ and ${\cal P}_B$ together is at most $ (1 + \eps) \cdot\frac{ 7 }{ 8 }$ times that of $(T_A, T_B)$. At the same time, by item~2, the combined cost $C_A({\cal P}_A) + C_B( {\cal P}_B )$ is within factor $\frac{ 32 }{ 31 }$ of $C_A( T_A ) + C_B( T_B )$. Consequently, while we have no guarantees on the space-cost tradeoff of commodities $A$ and $B$ by themselves, their pairwise tradeoff is upper-bounded by $ (1 + \eps) \cdot \frac{ 28 }{ 31 }$, motivating the ``sub-$1$ couple'' terminology.

\paragraph{Why is this setting useful?} At the moment, readers can only wonder about basic questions such as: With respect to our current SOSI policies, $\{ \hat{T}_i \}_{i \in \tilde{\cal V}_{ \ell }}$, why are we guaranteed to have sufficiently-many sub-$1$ couples? What is their combined contribution toward the overall space requirement and long-run average cost? How are we going to handle commodities outside of sub-$1$ couples? In what follows, we address these issues by probabilistically messaging the policies $\{ \hat{T}_i \}_{i \in \tilde{\cal V}_{ \ell }}$ into having many sub-$1$ couples, without overly perturbing their space requirement and cost. Subsequently, combining Lemma~\ref{lem:construction_pair} with additional ideas will allow us to derive the Po2-Synchronization Theorem.

\subsection{Power-of-2 rounding} \label{subsec:pow-rounding}

To explain our method for probabilistically altering $\{ \hat{T}_i \}_{i \in \tilde{\cal V}_{ \ell }}$, we remind the reader that $\tilde{\cal V}_{ \ell }$ was partitioned in Section~\ref{subsec:V_ell_rounding} into heavy and light commodities, corresponding to ${\cal H}_{\ell}$ and ${\cal L}_{\ell}$. Let us arbitrarily partition ${\cal H}_{ \ell }$ into $Q = \frac{ 20 \ln (1/\eps) }{ \eps^2 }$ subsets ${\cal H}_{ \ell , 1}, \ldots, {\cal H}_{\ell,Q}$, each of size either $\lfloor \frac{ |{\cal H}_{ \ell }| }{ Q } \rfloor$ or $\lceil  \frac{ |{\cal H}_{ \ell }| }{ Q } \rceil$, noting that
\begin{equation} \label{eqn:ratio_HQ_eps}
\left\lfloor \frac{ |{\cal H}_{ \ell }| }{ Q } \right\rfloor ~~>~~ \left\lfloor \frac{ | \tilde{\cal V}_{ \ell } |/2 }{ Q } \right\rfloor  ~~\geq~~ \left\lfloor \frac{ 50\ln (1/\eps) }{ Q \eps^4 } \right\rfloor ~~\geq~~ \frac{ 2 }{ \eps^2 } \ . 
\end{equation}
Here, the first inequality holds since  ${ | {\cal H}_{ \ell } | > \frac{ | \tilde{\cal V}_{ \ell } | }{ 2 }}$, by the hypothesis of Theorem~\ref{thm:main_result_Vell_heavy}, and the second inequality follows by recalling that $|\tilde{\cal V}_{ \ell }| \geq \frac{ 100 \ln (1/\eps) }{ \eps^4 }$ for every $\ell \in {\cal D}$, according to property~\ref{prop:tildeV_size}. 

Now, for each subset ${\cal H}_{ \ell ,q}$, independently of any other subset, we proceed by performing power-of-$2$ rounding on its SOSI policies $\{ \hat{T}_i \}_{i \in {\cal H}_{ \ell ,q}}$. While there are several different methods in this context (see, e.g., \cite{Roundy85, Roundy86} and \cite{JacksonMM85}), for our particular purposes, we employ  the dependent rounding procedure of \cite{TeoB01} to create the collection of random policies $\{ T_i^{ \Theta_{\ell, q} } \} _{i \in {\cal H}_{ \ell ,q}}$. This procedure operates as follows:
\begin{itemize}
    \item For each commodity $i \in {\cal H}_{ \ell ,q}$, we first express its corresponding SOSI policy as $\hat{T}_i = 2^{\alpha_i + \beta_i} \cdot \hat{T}_{\min,\ell ,q}$, for some integer $\alpha_i \geq 0$ and real $\beta_i \in [0,1)$, where $\hat{T}_{\min,\ell ,q} = \min_{i \in {\cal H}_{ \ell ,q}} \hat{T}_i$.

    \item Next, we generate a  uniform random variable $\Theta_{\ell, q} \sim U[-\frac{1}{2}, \frac{1}{2}]$. 

    \item Finally, for every commodity $i \in {\cal H}_{ \ell ,q}$, we set
    \[ T_i^{ \Theta_{\ell, q} } ~~=~~ \begin{cases}
    2^{\alpha_i + \Theta_{\ell, q}} \cdot \hat{T}_{\min,\ell ,q}, & \text{when } \Theta_{\ell, q} \geq \beta_i - \frac{ 1 }{ 2 } \\
    2^{\alpha_i + \Theta_{\ell, q}+1} \cdot \hat{T}_{\min,\ell ,q}, \qquad & \text{when } \Theta_{\ell, q} < \beta_i - \frac{ 1 }{ 2 }
    \end{cases} \]    
\end{itemize}
We mention in passing that readers who are thoroughly familiar with the work of \citet[Sec.~2.2]{TeoB01} will notice subtle differences between their procedure and the one described above. Our point in  presenting this slightly modified implementation is that it may be the simplest possible perspective on power-of-$2$ rounding. As such, it is easy to verify a number of structural properties exhibited by $\{ T_i^{ \Theta_{\ell, q} } \}_{i \in {\cal H}_{ \ell ,q}}$, formally summarized in Claim~\ref{clm:properties_PO2} below. The credit should be fully given to \citeauthor{TeoB01}; we are merely rephrasing their approach. 

\begin{claim} \label{clm:properties_PO2}
The random policies $\{ T_i^{ \Theta_{\ell, q} } \} _{i \in {\cal H}_{ \ell ,q}}$ satisfy the next three properties:
\begin{enumerate}
    \item $\exsubpar{ \Theta_{\ell, q} } { T_i^{ \Theta_{\ell, q} } } = \frac{ 1 }{ \sqrt{2} \ln 2 } \cdot \hat{T}_i$ and $\exsubpar{ \Theta_{\ell, q} }{ \frac{ 1 }{ T_i^{ \Theta_{\ell, q} } } } = \frac{ 1 }{ \sqrt{2} \ln 2 } \cdot \frac{ 1 }{ \hat{T}_i }$.

    \item $T_i^{ \Theta_{\ell, q} } \in [ \frac{ \hat{T}_i }{ \sqrt{2} } , \sqrt{2} \hat{T}_i ]$ almost surely.

    \item For every pair of commodities $i_1, i_2 \in {\cal H}_{ \ell ,q}$, the ratio between $T_{i_1}^{ \Theta_{\ell, q} }$ and $T_{i_2}^{ \Theta_{\ell, q} }$ is an integer power of $2$.
\end{enumerate}
\end{claim}

It is worth pointing out that, in terms of the random policies $\{ T_i^{ \Theta_{\ell, q} } \} _{i \in {\cal H}_{ \ell ,q}}$, every pair $(i_1, T_{i_1}^{ \Theta_{\ell, q} })$ and $(i_2, T_{i_2}^{ \Theta_{\ell, q} })$ satisfies characterization~\ref{prop:sub1_couple_ratio} of sub-$1$ couples, by property~3 above.

\subsection{Near and far pairs} \label{subsec:near_far_pairs}

To concurrently instill characterization~\ref{prop:sub1_couple_space} for a sufficiently large collection of couples, let us consider a single subset ${\cal H}_{ \ell ,q}$, whose commodities will be denoted by $1, \ldots, m$. Recalling that the corresponding power-of-$2$ policies $\{ T_i^{ \Theta_{\ell, q} } \} _{i \in {\cal H}_{ \ell ,q}}$ are random, let $\Pi_{ \ell, q } : [m] \to [m]$ be the random permutation of the commodities $1, \ldots, m$ for which 
\begin{equation} \label{eqn:sequence_gamma_T}
\gamma_{ \Pi_{ \ell, q }(1) } \cdot T_{ \Pi_{ \ell, q }(1)}^{ \Theta_{\ell, q} } ~~\geq~~ \cdots ~~\geq~~ \gamma_{ \Pi_{ \ell, q }(m) } \cdot T_{ \Pi_{ \ell, q }(m)}^{ \Theta_{\ell, q} } \ , 
\end{equation}
where ties are broken by some fixed decision rule, say by minimum index, just to have a concrete definition of $\Pi_{ \ell, q }$.

\paragraph{Far and near pairs.} Now, suppose we break ${\cal H}_{ \ell ,q}$ into successive pairs $\{ \Pi_{ \ell, q }(1),\Pi_{ \ell, q }(2) \}$, $\{ \Pi_{ \ell, q }(3), \Pi_{ \ell, q }(4) \}$, so on and so forth, leaving $\Pi_{ \ell, q }(m)$ out when $m$ is odd. We say that such a pair $\{ \Pi_{ \ell, q }(2i-1), \Pi_{ \ell, q }(2i) \}$ is far when $\gamma_{ \Pi_{ \ell, q }(2i-1) } \cdot T_{ \Pi_{ \ell, q }(2i-1)}^{ \Theta_{\ell, q} } \geq (1+\eps) \cdot \gamma_{ \Pi_{ \ell, q }(2i) } \cdot T_{ \Pi_{ \ell, q }(2i)}^{ \Theta_{\ell, q} }$. Namely, there is a multiplicative gap of at least $1+\eps$ between the $(2i-1)$-th and $2i$-th terms of the sequence~\eqref{eqn:sequence_gamma_T}. In the opposite case, where $\gamma_{ \Pi_{ \ell, q }(2i-1) } \cdot T_{ \Pi_{ \ell, q }(2i-1)}^{ \Theta_{\ell, q} } < (1+\eps) \cdot \gamma_{ \Pi_{ \ell, q }(2i) } \cdot T_{ \Pi_{ \ell, q }(2i)}^{ \Theta_{\ell, q} }$, we say that $\{ \Pi_{ \ell, q }(2i-1), \Pi_{ \ell, q }(2i) \}$ is near. We use 
${\cal F}_{ \ell ,q}^{ \Pi_{ \ell, q } }$ and ${\cal N}_{ \ell ,q}^{ \Pi_{ \ell, q } }$ to denote the collection of far and near pairs in ${\cal H}_{ \ell ,q}$; these sets are of course random, due to their dependency on $\Pi_{ \ell, q }$. The next claim, whose proof appears in Appendix~\ref{app:proof_lem_bound_on_far_pairs}, shows that the number of far pairs is only $O( \frac{ 1 }{ \eps } )$. 

\begin{claim} \label{clm:bound_on_far_pairs}
$|{\cal F}_{ \ell ,q}^{ \Pi_{ \ell, q } }| \leq \frac{ 11 }{ 10\eps }$.    
\end{claim}

\paragraph{Near pairs are  sub-$\bs{1}$ couples.} An immediate consequence of this result is that there are considerably more near pairs than far ones, since
\[ |{\cal N}_{ \ell ,q}^{ \Pi_{ \ell, q } }| ~~=~~ \left\lfloor \frac{ | {\cal H}_{ \ell, q }  | }{ 2 } \right\rfloor - |{\cal F}_{ \ell ,q}^{ \Pi_{ \ell, q } }| ~~\geq~~ \left\lfloor \frac{ \lfloor |{\cal H}_{ \ell }| / Q \rfloor }{ 2 } \right\rfloor - \frac{ 11 }{ 10\eps } ~~\geq~~ \left\lfloor \frac{ 1 }{ \eps^2 } \right\rfloor - \frac{ 11 }{ 10\eps } ~~=~~ O \left( \frac{ 1 }{ \eps^2 } \right) \ , \]
where the second inequality  follows from inequality~\eqref{eqn:ratio_HQ_eps}. The important observation is that for every near pair,  $( \Pi_{ \ell, q }(2i-1), T_{ \Pi_{ \ell, q }(2i-1)}^{ \Theta_{\ell, q} })$ and $( \Pi_{ \ell, q }(2i), T_{ \Pi_{ \ell, q }(2i)}^{ \Theta_{\ell, q} })$ jointly form a sub-$1$ couple. Indeed, by Claim~\ref{clm:properties_PO2}(3), we already know that the ratio between $T_{ \Pi_{ \ell, q }(2i-1)}^{ \Theta_{\ell, q} }$ and $T_{ \Pi_{ \ell, q }(2i)}^{ \Theta_{\ell, q} }$ is an integer power of $2$. In addition, by definition of near pairs, $\gamma_{ \Pi_{ \ell, q }(2i-1) } \cdot T_{ \Pi_{ \ell, q }(2i-1)}^{ \Theta_{\ell, q} } \in [1,1+\eps) \cdot \gamma_{ \Pi_{ \ell, q }(2i) } \cdot T_{ \Pi_{ \ell, q }(2i)}^{ \Theta_{\ell, q} }$. Consequently, by employing Lemma~\ref{lem:construction_pair}, we obtain the next result.

\begin{corollary} \label{cor:construct_sub1_near}
For every $\{ \Pi_{ \ell, q }(2i-1), \Pi_{ \ell, q }(2i) \} \in {\cal N}_{ \ell ,q}^{ \Pi_{ \ell, q } }$, we can construct in polynomial time dynamic replenishment policies ${\cal P}_{ \Pi_{ \ell, q }(2i-1) }^{\ell}$ and ${\cal P}_{ \Pi_{ \ell, q }(2i)}^{\ell}$ such that:
\begin{enumerate}
    \item {\em Joint occupied space:}
    \[ V_{\max}( {\cal P}_{ \Pi_{ \ell, q }(2i-1) }^{\ell}, {\cal P}_{ \Pi_{ \ell, q }(2i)}^{\ell} ) ~~\leq~~ (1 + \eps) \cdot \frac{ 7 }{ 8 } \cdot \left( \gamma_{ \Pi_{ \ell, q }(2i-1) } \cdot T_{ \Pi_{ \ell, q }(2i-1)}^{ \Theta_{\ell, q} } + \gamma_{ \Pi_{ \ell, q }(2i) } \cdot T_{ \Pi_{ \ell, q }(2i)}^{ \Theta_{\ell, q} } \right) \ . \]

    \item {\em Cost blow-up:}
    \[ \max \left\{ \frac{ C_{ \Pi_{ \ell, q }(2i-1) }( {\cal P}^{\ell}_{ \Pi_{ \ell, q }(2i-1) } ) }{ C_{ \Pi_{ \ell, q }(2i-1) }( T_{ \Pi_{ \ell, q }(2i-1)}^{ \Theta_{\ell, q} } ) }, \frac{ C_{ \Pi_{ \ell, q }(2i) }( {\cal P}^{\ell}_{ \Pi_{ \ell, q }(2i) } ) }{ C_{ \Pi_{ \ell, q }(2i) }( T_{ \Pi_{ \ell, q }(2i)}^{ \Theta_{\ell, q} } ) } \right\} ~~\leq~~ \frac{ 32 }{ 31} \ . \]  
\end{enumerate}
\end{corollary}

\subsection{Overall policy construction} \label{subsec:PO2_sync_policy}

\paragraph{Distinction between commodity types.} Based on the preceding discussion, it follows that the commodities in $\tilde{\cal V}_{\ell} = {\cal H}_{\ell} \cup {\cal L}_{\ell}$ can be classified into four types:
\begin{enumerate}
    \item Those belonging to near pairs across ${\cal H}_{ \ell , 1}, \ldots, {\cal H}_{\ell,Q}$.

    \item Those belonging to far pairs across ${\cal H}_{ \ell , 1}, \ldots, {\cal H}_{\ell,Q}$.

    \item Within each of the subsets ${\cal H}_{ \ell , 1}, \ldots, {\cal H}_{\ell,Q}$ with odd cardinality, there is a single commodity that does not belong to any pair.

    \item The collection of light commodities ${\cal L}_{\ell}$.
\end{enumerate}
We denote these types by ${\cal T}_1^{ \Theta_\ell }$, ${\cal T}_2^{\Theta_\ell }$, ${\cal T}_3^{ \Theta_\ell }$, and ${\cal T}_4$, with the superscript $\Theta_\ell$ emphasizing the randomness inherent to the first three types. Here, $\Theta_\ell = ( \Theta_{ \ell, q } )_{q \in [Q]}$ is the random vector resulting from our power-of-$2$ rounding procedure (see Section~\ref{subsec:pow-rounding}), noting that by construction, its components are mutually independent.

\paragraph{The overall policy.} The final policy we return depends on whether an easily-checkable condition is satisfied or not. To formalize this condition, we remind the reader that for every subset ${\cal H}_{ \ell ,q}$, starting with the SOSI policies $\{ \hat{T}_i \}_{i \in {\cal H}_{ \ell ,q}}$, our power-of-$2$ rounding procedure  created the collection of random policies $\{ T_i^{ \Theta_{\ell, q} } \} _{i \in {\cal H}_{ \ell ,q}}$, independently of other subsets.  With respect to these policies, let ${\cal A}_{\ell}$ be the event where
\begin{equation} \label{eqn:definition_A_ell}
\sum_{i \in {\cal H}_{\ell} } \gamma_i   T_i^{ \Theta_{\ell} } ~~\leq~~ (1 + \eps) \cdot \frac{ 1 }{ \sqrt{2} \ln 2 } \cdot \sum_{i \in {\cal H}_{\ell} } \gamma_i   \hat{T}_i \ .  
\end{equation}
Here, for ease of notation, we make use of $T_i^{ \Theta_{\ell} } = T_i^{ \Theta_{\ell, q} }$, where $q \in [Q]$ is the index for which $i \in {\cal H}_{\ell,q}$. Given this definition, our final policy $\tilde{\cal P}^{ \ell }$ depends on whether ${\cal A}_{ \ell }$ occurs or not:
\begin{itemize}
    \item {\em When ${\cal A}_{\ell}$ occurs}: Our policies for ${\cal T}_1^{ \Theta_\ell }$-commodities are exactly those  obtained via Corollary~\ref{cor:construct_sub1_near}. It is important to point out that, for each such commodity $i \in {\cal T}_1^{ \Theta_\ell }$, due to conditioning on ${\cal A}_{\ell}$, the object we return is the conditional random policy ${\cal P}_i^{\ell+} = [ {\cal P}_i^{\ell} | {\cal A}_{\ell} ]$. For every commodity $i \in {\cal T}_2^{ \Theta_\ell } \cup {\cal T}_3^{ \Theta_\ell }$, its random policy is kept unchanged, meaning that we return ${\cal P}_i^{\ell+} = [T_i^{ \Theta_{\ell} } | {\cal A}_{\ell}]$, again due to conditioning on ${\cal A}_{\ell}$. Finally, for every commodity $i \in {\cal T}_4^{ \Theta_\ell }$, we return its original SOSI policy, i.e., ${\cal P}_i^{\ell+} = \hat{T}_i$, which is of course deterministic.

    \item {\em When ${\cal A}_{\ell}$ does not occurs}: In this case, the policy we employ is a scaled version of $\hat{T}$. Specifically, for every commodity $i \in \tilde{\cal V}_{\ell}$, we return the policy ${\cal P}_i^{\ell-} = \alpha \hat{T}_i$, where $\alpha = \frac{ 7/8 }{ \sqrt{2} \ln 2 }$. 
\end{itemize}

\subsection{Analysis} \label{subsec:analysis_Aell_space_cost} 
In the remainder of this section, we conclude the proof of Theorem~\ref{thm:main_result_Vell_heavy} by arguing that our random replenishment policy $\tilde{\cal P}^{\ell} = \{ \tilde{\cal P}^{\ell}_i \}_{i \in \tilde{\cal V}_{ \ell }}$ guarantees a $((1 + 7\eps) \cdot \frac{ 7/4 }{ \sqrt{2} \ln 2},(1 + \frac{4\eps}{9}) \cdot \frac{ 32/31 }{ \sqrt{2} \ln 2})$-ratio. In other words, we will prove that the maximal space requirement of this policy is almost surely 
\[ V_{\max}( \tilde{\cal P}^{\ell} ) ~~\leq~~ (1 + 6\eps) \cdot \frac{ 7/4 }{ \sqrt{2} \ln 2 } \cdot | \tilde{\cal V}_{\ell} | \cdot \frac{1}{(1+\eps)^{\ell-1}} \cdot{\cal V} \ , \]
and that its expected long-run average cost is 
\[ \exsub{ \Theta_{\ell} }{ C( \tilde{\cal P}^{\ell} ) } ~~\leq~~ \left( 1 + \frac{2\eps}{5}  \right) \cdot \frac{ 32/31 }{ \sqrt{2} \ln 2} \cdot \sum_{i \in \tilde{\cal V}_{ \ell } } C_i( \hat{T}_i ) \ . \]

\paragraph{The likelihood of $\bs{{\cal A}_{\ell}}$.} To go about deriving these bounds, we first  show that the event ${\cal A}_{\ell}$ occurs with probability $1-O(\eps)$,  as stated in Lemma~\ref{lem:success_Aell} below. To understand why this claim makes general sense, let us observe that by Claim~\ref{clm:properties_PO2}(1), we know that $\exsubpar{ \Theta_{\ell} } { T_i^{ \Theta_{\ell} } } = \frac{ 1 }{ \sqrt{2} \ln 2 } \cdot \hat{T}_i$ for every commodity $i \in \tilde{\cal V}_{\ell}$. Therefore, by defining ${\cal A}_{\ell}$ via inequality~\eqref{eqn:definition_A_ell}, this event essentially states that the random variable $\sum_{i \in {\cal H}_{\ell} } \gamma_i T_i^{ \Theta_{\ell} }$ does not deviate much from its expectation, $\frac{ 1 }{ \sqrt{2} \ln 2 } \cdot \sum_{i \in {\cal H}_{\ell} } \gamma_i  \hat{T}_i$. While the rather involved proof of this claim is deferred to Appendix~\ref{app:proof_lem_success_Aell}, it will assist readers in realizing that various choices up until now were meant to guarantee that  appropriate concentration inequalities will be applicable. Among others, these choices include making use of power-of-$2$ rounding, independently drawing $( \Theta_{ \ell, q } )_{q \in [Q]}$, and ensuring that each subset ${\cal H}_{\ell,q}$ is sufficiently large.

\begin{lemma} \label{lem:success_Aell}
$\prsubpar{ \Theta_\ell }{ {\cal A}_{\ell} } \geq 1 - \frac{ \eps }{ 10 }$.
\end{lemma}

\paragraph{Space requirement.} We are now ready to upper-bound the maximal space requirement of our random policy $\tilde{\cal P}^{\ell}$. Once again, the proof of this result is quite technical, and we therefore present its finer details in Appendix~\ref{app:proof_lem_PO2_bound_Vmax}. At a high level, when ${\cal A}_{\ell}$ does not occur, it will turn out that scaling $\hat{T}$ by a factor of $\alpha = \frac{ 7/8 }{ \sqrt{2} \ln 2 }$ indeed does the trick. The complementary case, where ${\cal A}_{\ell}$ occurs, is rather challenging to analyze, and our approach consists of separately bounding each of the commodity types ${\cal T}_1^{ \Theta_\ell }$, ${\cal T}_2^{\Theta_\ell }$, ${\cal T}_3^{ \Theta_\ell }$, and ${\cal T}_4$. Here, the observation that there are considerably more near pairs than far ones (see Section~\ref{subsec:near_far_pairs}) will play an important role. 

\begin{lemma} \label{lem:PO2_bound_Vmax}
$V_{\max}( \tilde{\cal P}^{\ell} ) \leq (1 + 6\eps) \cdot \frac{ 7/4 }{ \sqrt{2} \ln 2 } \cdot | \tilde{\cal V}_{\ell} | \cdot \frac{1}{(1+\eps)^{\ell-1}} \cdot{\cal V}$ almost surely. 
\end{lemma}

\paragraph{Expected cost.} The final piece of our puzzle resides in upper-bounding the expected long-run average cost of $\tilde{\cal P}^{\ell}$, as formally stated in Lemma~\ref{lem:PO2_bound_cost} below. The complete proof of this result is provided in Appendix~\ref{app:proof_lem_PO2_bound_cost}. Toward obtaining the desired bound, we will first observe that when ${\cal A}_{\ell}$ does not occur, the specific policy we pick has negligible effects on the expected cost, primarily since $\prsubpar{ \Theta_\ell }{ \bar{\cal A}_{\ell} } = O(\eps)$. As before, the difficult case to handle is the one where ${\cal A}_{\ell}$ occurs. Here, our approach will provide separate arguments for the four commodity types, showing that even when conditioning on ${\cal A}_{\ell}$, the guarantees of Corollary~\ref{cor:construct_sub1_near} remain essentially unchanged.

\begin{lemma} \label{lem:PO2_bound_cost}
$\exsubpar{ \Theta_{\ell} }{ C( \tilde{\cal P}^{\ell} ) } \leq ( 1 + \frac{ 2\eps }{ 5 }  ) \cdot \frac{ 32/31 }{ \sqrt{2} \ln 2} \cdot \sum_{i \in \tilde{\cal V}_{ \ell } } C_i( \hat{T}_i )$.
\end{lemma}

\section{Concluding Remarks}

In conjunction to resolving long-standing open questions about the economic warehouse lot scheduling problem, for which rigorously analyzable results have been few and far between, we believe that our work introduces an even greater number of topics to be investigated in future research.
The next few paragraphs are intended to single out some of these prospective directions.

\paragraph{The strategic version.} As pointed out by \citet[p.~51]{GallegoSS92}, and as one can discover by browsing through recent literature, the vast majority of academic research around economic warehouse lot scheduling has been dedicated to its  ``tactical'' version, which is precisely the one studied in this paper. Yet, a decent amount of attention has concurrently been given to the so-called ``strategic'' version (see, e.g., \cite{HodgsonL82, ParkY85, Hall88, RosenblattR90}), where  rather than having a warehouse space constraint, this term appears as part of the objective function; namely, we wish to minimize $C({\cal P}) + V_{\max}( {\cal P} )$. It is not difficult to verify that, by guessing the $V_{\max}( {\cal P}^* )$ term of an optimal policy ${\cal P}^*$ and writing an appropriate constraint, we can show that a tactical $\alpha$-approximation can be converted to a strategic $(\sqrt{\alpha} + \eps)$-approximation. Consequently, our  work provides a polynomial-time approximation scheme for constantly-many commodities in this context, as well as a polynomial-time $(\sqrt{2 - \frac{17}{5000}} + \eps)$-approximation for general problem instances. The latter result improves on the well-known $\sqrt{2}$-approximation due to \cite{Anily91} and \cite{GallegoQS96}. It will be particularly interesting to examine whether our algorithmic approach can be enhanced to attain improved guarantees, given the extra leeway of having $V_{\max}( {\cal P} )$ within the objective function rather than within a space constraint.

\paragraph{Efficient implementation?} As formally stated in Theorem~\ref{thm:2_minus_delta}, our main algorithmic finding for general instances consists in devising a $O( | {\cal I} |^{\tilde{O}( 1/\eps^5)} \cdot 2^{ \tilde{O}( 1 / \eps^{35} ) } )$-time construction of a random capacity-feasible policy whose expected long-run average cost is within factor $2-\frac{17}{6250} + \eps$ of optimal. Along these lines, much of the running time exponent can be attributed to avoiding a highly-optimized implementation of our approximation scheme for constantly-many commodities. Even though we have not attempted to eliminate such dependencies, we believe that there might be an alternative way to connect the dots between our main ideas, ending up with a  more efficient implementation. Of course, the potential inaccessibly cost of efforts along these lines is still unknown.

\paragraph{Derandomization?} On a different front, coming up with a deterministic construction of sub-$2$-approximate policies is a very interesting question to be tackled as part of future work. To this end, in relation to  Section~\ref{subsec:pow-rounding}, let us point out a well-known observation regarding the power-of-$2$ rounding procedure of \cite{TeoB01}: Up to scaling, the resulting random policies $\{ T_i^{ \Theta_{\ell, q} } \} _{i \in {\cal H}_{ \ell ,q}}$ jointly have only $O(n)$ possible realizations. That said, we are currently not seeing how a given set of realizations can be efficiently scaled while keeping the Po2-Synchronization Theorem intact. It would be interesting to investigate whether this direction could lead to derandomizing our approach, or perhaps, whether completely different ideas will do the trick. With the scarcity of rigorous results around economic warehouse lot scheduling, we expect this direction to be particularly challenging.

\addcontentsline{toc}{section}{Bibliography}
\bibliographystyle{plainnat}
\bibliography{BIB-Lot-Sizing}

\appendix
\section{Additional Proofs from Section~\ref{sec:exponential_approx}} \label{app:proofs_sec_exponential_approx}

\subsection{Proof of Theorem~\ref{thm:good_cyclic}} \label{app:proof_lem_good_cyclic}

\paragraph{Constructing $\bs{\tilde{\cal P}}$.} For the purpose of deriving this result, let ${\cal P}^*$ be an optimal policy. With respect to ${\cal P}^*$, we define the infinite sequence $\tau_1 < \tau_2 < \cdots$ of time points as follows:
\begin{itemize}
    \item First, $\tau_1$ is the minimal $t > 0$ for which the segment $[0,t]$ has: (a)~At least $\frac{ 1 }{ \eps } + 1$ orders of every commodity $i \in [n]$; and (b)~Exactly $\frac{ 1 }{ \eps } + 1$ orders of some commodity $i \in [n]$.

    \item In turn, $\tau_2$ is the minimal $t > \tau_1$ for which the segment $[\tau_1,t]$ meets conditions~(a) and~(b) above.

    \item The remaining sequence $\tau_3, \tau_4, \ldots$ is defined in a similar way.
\end{itemize}
By recalling how dynamic replenishment policies are defined (see Section~\ref{subsec:model_definition}), it is easy to verify the sequence $\tau_1, \tau_2, \ldots$ is well-defined and tends to infinity. 

Given this sequence, the optimal long-run average cost can be written as $C( {\cal P^*} ) = \lim_{\hat{\kappa} \to \infty} \frac{ C( {\cal P}^*, [0, \tau_{\hat{\kappa}} )) }{ \tau_{\hat{\kappa}} }$, noting that 
$\tau_{\hat{\kappa}} = \sum_{ \kappa \leq \hat{\kappa} } ( \tau_{\kappa} - \tau_{\kappa-1} )$ and $C( {\cal P}^*, [0, \tau_{\hat{\kappa}} )) = \sum_{ \kappa \leq \hat{\kappa} } C( {\cal P}^*, [\tau_{\kappa-1}, \tau_{\kappa} ))$. Consequently, for every $\eps >0$, there exists an index $\kappa = \kappa(\eps)$ for which 
\begin{equation} \label{eqn:proof_lem_good_cyclic_eq1}
\frac{ C( {\cal P^*}, [ \tau_{\kappa}, \tau_{\kappa+1}) ) }{ \tau_{\kappa+1} - \tau_{\kappa} } ~~\leq~~ (1 + \eps) \cdot C( {\cal P^*} ) \ .
\end{equation}
As such, our cyclic policy $\tilde{\cal P}$ is created by duplicating the actions of ${\cal P}^*$ over the segment $[ \tau_{\kappa}, \tau_{\kappa+1})$ across the entire planning horizon $[0,\infty)$. To align any pair of successive cycles, we decrement the last order of each commodity $i \in [n]$, thereby making sure that each  cycle ends with zero inventory. In addition, at the beginning of each cycle, we place an extra order of this commodity, consisting of $I( {\cal P}^*_i, \tau_{\kappa})$ units. From this point on, $\tau_{\mycyc} = \tau_{\kappa+1} - \tau_{\kappa}$ will denote the cycle length of this policy.

\paragraph{Feasibility and cost.} It is easy to verify that our resulting policy $\tilde{\cal P}$ is capacity-feasible, since 
\[ V_{\max}( \tilde{\cal P} ) ~~\leq~~ \max_{\tau \in [ \tau_{\kappa}, \tau_{\kappa+1})}  V( {\cal P}^*, \tau)  ~~\leq~~ V_{\max}( {\cal P}^* ) ~~\leq~~ {\cal V} \ , \]
where the last inequality holds since the optimal policy ${\cal P}^*$ is in particular capacity-feasible. To upper bound the long-run average cost of $\tilde{\cal P}$, let us break this measure into holding and ordering costs as follows:
\begin{itemize}
    \item {\em Holding}: Along a single cycle, we incur a holding cost of ${\cal H}( \tilde{\cal P}, [0,\tau_{\mycyc})) \leq {\cal H}( {\cal P}^*, [ \tau_{\kappa}, \tau_{\kappa+1}))$, due to having $I( \tilde{\cal P}_i, \tau) \leq I( {\cal P}^*_i, \tau_{\kappa} + \tau)$ for all $i \in [n]$ and $\tau \in [0, \tau_{\mycyc})$.

    \item {\em Ordering}: Along a single cycle, we have added at most one extra order per commodity in comparison to ${\cal P}^*$, implying that 
    \[ N( \tilde{\cal P}_i, [0,\tau_{\mycyc})) ~~\leq~~ N( {\cal P}^*_i, [ \tau_{\kappa}, \tau_{\kappa+1})) + 1 ~~\leq~~ (1 + \eps) \cdot  N( {\cal P}^*_i, [ \tau_{\kappa}, \tau_{\kappa+1})) \ , \]
    where the last inequality holds since $N( {\cal P}^*_i, [ \tau_{\kappa}, \tau_{\kappa+1})) \geq \frac{ 1 }{ \eps }$, by definition of $\tau_{\kappa+1}$. Therefore, ${\cal K}( \tilde{\cal P}, [0, \tau_{\mycyc})) \leq (1 + \eps) \cdot {\cal K}( {\cal P}^*,  [ \tau_{\kappa}, \tau_{\kappa+1}))$.
\end{itemize}
By aggregating these two observations, we have
\begin{eqnarray}
C( \tilde{\cal P} ) & = & \frac{  C( \tilde{\cal P}, [0,\tau_{\mycyc})) }{ \tau_\mycyc } \nonumber \\
& \leq & (1 + \eps) \cdot \frac{ C( {\cal P^*}, [ \tau_{\kappa}, \tau_{\kappa+1}) ) }{ \tau_{\kappa+1} - \tau_{\kappa} } \nonumber \\
& \leq & (1 + 3\eps) \cdot C( {\cal P^*} ) \ , \label{eqn:step2_feas_obj}
\end{eqnarray}
where the last inequality holds since $\frac{ C( {\cal P^*}, [ \tau_{\kappa}, \tau_{\kappa+1}) ) }{ \tau_{\kappa+1} - \tau_{\kappa} } \leq (1 + \eps) \cdot C( {\cal P^*} )$ according to~\eqref{eqn:proof_lem_good_cyclic_eq1}.

\paragraph{Proof of item~2.} We first establish an auxiliary claim about how the optimal long-run average cost $C( {\cal P^*})$ is related to ${\cal M}$. The proof of this result appears in Appendix~\ref{app:proof_clm_Cpstar_vs_M}.

\begin{claim} \label{clm:Cpstar_vs_M}
$C( {\cal P^*}) \leq n {\cal M}$.
\end{claim}

We proceed by explaining why the above claim leads to the desired bounds on the cycle length $\tau_{\mycyc}$, specifically showing that $\tau_{\mycyc} \in [\frac{ K_{\max} }{ 2\eps n {\cal M} }, \frac{ 2 n {\cal M} }{ \eps^2 H_{\min} }]$: 
\begin{itemize}
    \item {\em Upper bound}: By definition of $\tau_{\kappa+1}$, we know that there exists at least one commodity $i \in [n]$ with $N( \tilde{\cal P}_i, [0,\tau_{\mycyc})) = \frac{ 1 }{ \eps }$ orders. At least one of these orders consists of at least $\eps \tau_{\mycyc}$ units, meaning that the holding cost of these units is at least $H_i \eps^2 \tau_{\mycyc}^2$. As a result,
    \begin{equation} \label{eqn:proof_good_cyclic_item2} 
    H_i \eps^2 \tau_{\mycyc}^2 ~~\leq~~ C( \tilde{\cal P}, [0,\tau_{\mycyc}] ) ~~\leq~~ (1 + 3\eps) \cdot \tau_{\mycyc} \cdot  C( {\cal P^*} ) ~~\leq~~ 2 \tau_{\mycyc} n {\cal M} \ , 
    \end{equation}
    where the second and third inequalities hold since $\frac{  C( \tilde{\cal P}, [0,\tau_{\mycyc})) }{ \tau_\mycyc } \leq (1 + 3\eps) \cdot C( {\cal P^*} )$, as shown  in~\eqref{eqn:step2_feas_obj}, and since $C( {\cal P^*} ) \leq n {\cal M}$ by Claim~\ref{clm:Cpstar_vs_M}. By rearranging, we indeed get $\tau_{\mycyc} \leq \frac{ 2 n {\cal M} }{ \eps^2 H_{i} } \leq \frac{ 2 n {\cal M} }{ \eps^2 H_{\min} }$.

    \item {\em Lower bound}: Again by definition of $\tau_{\kappa+1}$, we know that every commodity $i \in [n]$ has $N( \tilde{\cal P}_i, [0,\tau_{\mycyc })) \geq \frac{ 1 }{ \eps }$ orders in each cycle. Therefore, $\tilde{\cal P}$ pays an ordering cost of at least $\frac{ K_{\max} }{ \eps }$ in each cycle,  implying that
    \[ \frac{ K_{\max} }{ \eps \tau_{\mycyc } } ~~\leq~~ C( \tilde{\cal P} ) ~~\leq~~ (1 + 3\eps) \cdot C( {\cal P^*} ) ~~\leq~~ 2 n {\cal M} \ . \]
    Here, the second inequality is precisely~\eqref{eqn:step2_feas_obj}, and the third inequality follows from Claim~\ref{clm:Cpstar_vs_M}. Again by rearranging, $\tau_{\mycyc } \geq \frac{ K_{\max} }{ 2\eps n {\cal M} }$.
\end{itemize}

\paragraph{Proof of item~3.} As explained above, we know that for every commodity $i \in [n]$, the policy $\tilde{\cal P}$ places $N( \tilde{\cal P}_i, [0,\tau_\mycyc)) \geq \frac{ 1 }{ \eps }$ orders across the segment $[0,\tau_\mycyc)$, with equality for at least one commodity. In the opposite direction, we obtain an upper bound on $N( \tilde{\cal P}_i, [0,\tau_\mycyc))$ by observing that, since the ordering cost of each commodity $i$ by itself is $K_i \cdot N( \tilde{\cal P}_i, [0,\tau_\mycyc))$, it follows that
\[ K_i \cdot N( \tilde{\cal P}_i, [0,\tau_\mycyc)) ~~\leq~~ C( \tilde{\cal P}, [0,\tau_\mycyc) ) ~~\leq~~ 2 \tau_\mycyc n {\cal M} ~~\leq~~ \frac{ 4 n^2 {\cal M}^2 }{ \eps^2 H_{\min} } \ . \]
Here, the second inequality has already been derived in~\eqref{eqn:proof_good_cyclic_item2}, starting from the second term, and the third inequality holds since $\tau_{\mycyc} \leq \frac{ 2 n {\cal M} }{ \eps^2 H_{\min} }$, as shown in item~2. Rearranging this inequality, we get $N( \tilde{\cal P}_i, [0,\tau_\mycyc)) \leq \frac{ 4 n^2 {\cal M}^2 }{ \eps^2 K_i H_{\min} } \leq \frac{ 4 n^2 {\cal M}^2 }{ \eps^2 K_{\min} H_{\min} }$.

\subsection{Proof of Claim~\ref{clm:Cpstar_vs_M}} \label{app:proof_clm_Cpstar_vs_M}
Focusing on a single commodity $i \in [n]$, let us consider the next formulation, where one wishes to compute a minimum-cost capacity-feasible policy for this commodity by itself:
\begin{equation} \label{eqn:single_com_problem}
\begin{array}{ll}
{\displaystyle \min_{{\cal P}_i}} & C( {\cal P}_i ) \\
\text{s.t.} & V_{\max}( {\cal P}_i ) \leq {\cal V} 
\end{array}
\end{equation}
The important observation is that, by duplicating the arguments employed to prove  Lemma~\ref{lem:sosi_optimal}, it is not difficult to verify  that formulation~\eqref{eqn:single_com_problem} admits a SOSI optimal replenishment policy. 
Consequently, this formulation can be equivalently rephrased as:
\begin{equation} 
\begin{array}{lll}
{\displaystyle \min_{T_i}} & {\displaystyle \frac{ K_i }{ T_i } + H_i T_i} \\
\text{s.t.} & {\displaystyle  \gamma_i T_i \leq {\cal V}} 
\end{array}
\end{equation}
Letting $T_i^*$ be an optimal  ordering interval in this context, by consulting Claim~\ref{clm:EOQ_properties}, it is easy to see that 
$T_i^* = \min \{ \sqrt{K_i/H_i}, {\cal V}/ \gamma_i \}$, noting that this term matches our definition of $T^{\cal V}_i$.

Now, to propose a single capacity-feasible replenishment policy for all commodities, we simply glue together the optimal capacity-feasible SOSI policies for the individual commodities, $T_1^*, \ldots, T_n^*$, and  scale down the combined policy by a factor of $n$ to make it capacity-feasible. This alteration blows up the total cost by a factor of at most $n$ (see, Claim~\ref{clm:EOQ_properties}(3)), and since $C( {\cal P^*})$ is the optimal long-run average cost, we have
\[ C( {\cal P^*} ) ~~\leq~~ n \cdot \sum_{i \in [n]} \left( \frac{ K_i }{ T_i^* } + H_i T_i^* \right) ~=~~ n \cdot \sum_{i \in [n]} \left( \frac{ K_i }{ T^{\cal V}_i } + H_i T^{\cal V}_i \right) ~~=~~ n {\cal M} \ . \]

\subsection{Proof of Theorem~\ref{thm:exist_aligned}} \label{app:proof_lem_exist_aligned}

To establish the existence of a $B$-aligned policy $\hat{\cal P}$ which is near-feasible and near-optimal, we alter $\tilde{\cal P}$ in two  steps. As explained below, the first step is meant to attain property~\ref{item:B_aligned_1} of $B$-aligned policies, whereas the second step will take care of property~\ref{item:B_aligned_2}.

\paragraph{Ensuring property~\ref{item:B_aligned_1}: Zero inventory at $\bs{B_q^-}$-points.} For every frequency class $q \in [Q]$ and commodity $i \in \tilde{\cal F}_q$, we augment $\tilde{\cal P}$ with an additional $i$-order at each and every $B_q^-$-point. It is easy to verify that we can adjust the ordering quantities of $i$-orders such that the resulting policy $\hat{\cal P}^1$ satisfies the next two properties:
\begin{itemize}
    \item We arrive to $B_q^-$-points with zero inventory, i.e., $I( \hat{\cal P}^1_i, b^-) = 0$ for every $b \in B_q^-$, thereby instilling property~\ref{item:B_aligned_1}.

    \item The inventory level of this commodity across the entire cycle is upper-bounded by the analogous level with respect to $\tilde{\cal P}$, namely, $I( \hat{\cal P}^1_i, t) \leq I( \tilde{\cal P}_i, t)$ for every $t \in [0, \tau_\mycyc)$.    
\end{itemize}
As a consequence, $\hat{\cal P}^1$ is capacity-feasible, and its holding cost does not exceed that of $\tilde{\cal P}$. In terms of ordering costs, each commodity $i \in \tilde{\cal F}_q$ previously had $N( \tilde{\cal P}_i, [0,\tau_\mycyc)) \geq (\frac{n}{\eps})^{3(q-1)}$ orders, by definition of $\tilde{\cal F}_q$, whereas the number of newly-added $i$-orders is at most
\[ |B_q^-| ~~=~~ \left( \frac{n}{\eps} \right)^{3q-4}   ~~\leq~~ \frac{ \eps }{ n } \cdot N( \tilde{\cal P}_i, [0,\tau_\mycyc)) \ . \]
Therefore, the combined ordering cost of $\hat{\cal P}^1$ across the entire cycle is
\begin{eqnarray}
{\cal K}( \hat{\cal P}^1, [0,\tau_\mycyc) ) & = & \sum_{q \in [Q]} \sum_{i \in \tilde{\cal F}_q} K_i \cdot N( \hat{\cal P}^1_i, [0,\tau_\mycyc) ) \nonumber \\
& \leq & \sum_{q \in [Q]} \sum_{i \in \tilde{\cal F}_q} K_i \cdot \left( N( \tilde{\cal P}_i, [0,\tau_\mycyc) ) + |B_q^-|\right) \nonumber\\
& \leq & \left( 1 + \frac{ \eps }{ n } \right) \cdot \sum_{q \in [Q]} \sum_{i \in \tilde{\cal F}_q} K_i \cdot N( \tilde{\cal P}_i, [0,\tau_\mycyc)) \nonumber \\
& \leq & (1 + \eps) \cdot {\cal K}( \tilde{\cal P}, [0,\tau_\mycyc) ) \ . \label{eqn:proof_thm_aligned_A}
\end{eqnarray}

\paragraph{Ensuring property~\ref{item:B_aligned_2}: Orders only at $\bs{B_q^+}$-points.} For every frequency class $q \in [Q]$ and commodity $i \in \tilde{\cal F}_q$, the remaining issue with $\hat{\cal P}^1$ is that this policy may be placing $i$-orders in-between points in $B_q^+$. To eliminate this issue, we modify each $i$-order as follow:
\begin{itemize}
    \item Let $[t,t+\Delta)$ be the interval covered by the currently considered $i$-order. Specifically, since $\hat{\cal P}^1_i$ is a zero-inventory ordering policy, it arrives to time $t$ with zero inventory, ordering $\Delta$ units at that time.

    \item {\em Right-lengthening}: We extend the covered interval further to the right, by setting its right endpoint as $\lceil t + \Delta \rceil^{ B_q^+ }$. Here, $\lceil \cdot \rceil^{ B_q^+ }$ is an operator that rounds its argument up to the nearest $B_q^+$-point. This step requires increasing our ordering quantity from $\Delta$ to $\lceil t + \Delta \rceil^{ B_q^+ } - t \leq \Delta + \frac{ \tau_\mycyc }{ (n/\eps)^{3q + 1} }$, where the last inequality holds since the gap between successive $B_q^+$-points is $\frac{ \tau_\mycyc }{ (n/\eps)^{3q + 1} }$.

    \item {\em Left-shortening}: We now shorten the newly-covered interval, $[t,\lceil t + \Delta \rceil^{ B_q^+ })$, by setting its left endpoint as $\lceil t \rceil^{ B_q^+ }$, which requires decreasing our ordering quantity. In the degenerate event where $[t,t+\Delta)$ shrinks into a single point, i.e., $\lceil t \rceil^{ B_q^+ } = \lceil t + \Delta \rceil^{ B_q^+ }$, we discard this order.
\end{itemize}  
It is easy to verify that the resulting policy $\hat{\cal P}^2$ covers the entire cycle $[0,\tau_\mycyc)$, places orders only at $B_q^+$-points, and preserves property~\ref{item:B_aligned_1}. We proceed by analyzing this policy in terms of its occupied space and cost.

\begin{claim} \label{clm:hatP2_near_feasible}
$\hat{\cal P}^2$ is $(1+\eps) $-feasible.
\end{claim} 
\begin{proof}
As explained above, for every frequency class $q \in [Q]$ and commodity $i \in \tilde{\cal F}_q$, right-lengthening $i$-orders may increase the inventory level of this commodity at any point by at most $\frac{ \tau_\mycyc }{ (n/\eps)^{3q + 1} }$, whereas left-shortening may only decrease this level. Therefore, for every $t \in [0,\tau_\mycyc)$,
\begin{equation} \label{eqn:proof_clm_hatP2_near_feasible_eq1}
I( \hat{\cal P}^2_i, t) ~~\leq~~ I( \hat{\cal P}^1_i, t) + \frac{ \tau_\mycyc }{ (n/\eps)^{3q + 1} } \ .   
\end{equation}
Based on this observation, we conclude that the maximal space occupied by $\hat{\cal P}^2$ is at most $(1 + \eps) \cdot {\cal V}$, since
\begin{eqnarray}
V_{\max}( \hat{\cal P}^2 ) & = & \max_{t \in [0,\tau_\mycyc)}  V( \hat{\cal P}^2, t) \nonumber \\
& = & \max_{t \in [0,\tau_\mycyc)}  \sum_{q \in [Q]} \sum_{i \in \tilde{\cal F}_q} \gamma_i \cdot I( \hat{\cal P}^2_i, t) \nonumber \\
& \leq & \max_{t \in [0,\tau_\mycyc)}  \sum_{q \in [Q]} \sum_{i \in \tilde{\cal F}_q} \gamma_i \cdot \left( I( \hat{\cal P}^1_i, t) + \frac{ \tau_\mycyc }{ (n/\eps)^{3q + 1} } \right) \nonumber \\
& \leq & V_{\max}( \hat{\cal P}^1 ) + \eps {\cal V} \label{eqn:proof_thm_aligned_1} \\
& \leq & V_{\max}( \tilde{\cal P} ) + \eps {\cal V} \label{eqn:proof_thm_aligned_2} \\
& \leq & (1 + \eps) \cdot {\cal V} \ . \label{eqn:proof_thm_aligned_3}
\end{eqnarray}     
Here, inequality~\eqref{eqn:proof_thm_aligned_1} is obtained by arguing that $\gamma_i \cdot \frac{ \tau_{\mycyc} }{ (n/\eps)^{3q} } \leq {\cal V}$ for every commodity $i \in \tilde{\cal F}_q$. Indeed, we have $N( \tilde{\cal P}_i, [0,\tau_{\mycyc})) \leq (\frac{n}{\eps})^{3q}$ by definition of $\tilde{\cal F}_q$, implying that our original policy $\tilde{\cal P}$ has at least one $i$-order consisting of at least $\frac{ \tau_{\mycyc} }{ (n/\eps)^{3q} }$ units. Since $\tilde{\cal P}$ is capacity-feasible, we must have $\gamma_i \cdot \frac{ \tau_{\mycyc} }{ (n/\eps)^{3q} } \leq {\cal V}$. Inequality~\eqref{eqn:proof_thm_aligned_2} holds since $I( \hat{\cal P}^1_i, t) \leq I( \tilde{\cal P}_i, t)$ for every $t \in [0, \tau_\mycyc)$, as explained when establishing property~\ref{item:B_aligned_1}. Finally, inequality~\eqref{eqn:proof_thm_aligned_3} follows by recalling that $\tilde{\cal P}$ is capacity-feasible.
\end{proof}

\begin{claim}
$C( \hat{\cal P}^2, [0,\tau_{\mycyc}) ) \leq (1 + \eps) \cdot C( \tilde{\cal P}, [0,\tau_{\mycyc}))$.
\end{claim}
\begin{proof}
First, we can relate the total ordering cost of $\hat{\cal P}^2$ across $[0,\tau_{\mycyc})$ to the analogous cost with respect to $\tilde{\cal P}$ by observing that
\[ {\cal K}( \hat{\cal P}^2, [0,\tau_{\mycyc}) ) ~~\leq~~ {\cal K}( \hat{\cal P}^1, [0,\tau_{\mycyc}) ) ~~\leq~~ ( 1 + \eps) \cdot {\cal K}( \tilde{\cal P}, [0,\tau_{\mycyc})) \ . \]
Here, the first inequality holds since our method for converting  $\hat{\cal P}^1$ into $\hat{\cal P}^2$ does not involve placing additional orders; in fact, some orders may actually be discarded during left-shortening. The second inequality is precisely~\eqref{eqn:proof_thm_aligned_A}.

Moving to consider holding costs, let us recall that inequality~\eqref{eqn:proof_clm_hatP2_near_feasible_eq1} relates between the inventory levels of $\hat{\cal P}^2$ and $\hat{\cal P}^1$, stating that $I( \hat{\cal P}^2_i, t) \leq I( \hat{\cal P}^1_i, t) + \frac{ \tau_\mycyc }{ (n/\eps)^{3q + 1} }$ for every frequency class $q \in [Q]$, commodity $i \in \tilde{\cal F}_q$, and time point $t \in [0,\tau_\mycyc)$. As a result,
\begin{eqnarray}
{\cal H}( \hat{\cal P}^2, [0,\tau_{\mycyc}) ) & \leq & \sum_{q \in [Q]} \sum_{i \in \tilde{\cal F}_q} \left( {\cal H}( \hat{\cal P}^1_i, [0,\tau_{\mycyc}) )  + 2H_i \cdot \frac{ \tau_\mycyc^2 }{ (n/\eps)^{3q + 1} } \right) \nonumber \\
& = & {\cal H}( \hat{\cal P}^1, [0,\tau_{\mycyc}) ) +2 \cdot \sum_{q \in [Q]} \sum_{i \in \tilde{\cal F}_q} H_i \cdot \frac{ \tau_\mycyc^2 }{ (n/\eps)^{3q + 1} } \nonumber  \\
& \leq & {\cal H}( \tilde{\cal P}, [0,\tau_{\mycyc}) ) + 2 \cdot \sum_{q \in [Q]} \sum_{i \in \tilde{\cal F}_q} H_i \cdot \frac{ \tau_\mycyc^2 }{ (n/\eps)^{3q + 1} } \label{eqn:proof_thm_aligned_4} \\
& \leq & {\cal H}( \tilde{\cal P}, [0,\tau_{\mycyc}) ) + \frac{ 2\eps }{ n } \cdot \sum_{q \in [Q]} \sum_{i \in \tilde{\cal F}_q} {\cal H}( \tilde{\cal P}_i, [0,\tau_{\mycyc}) ) \label{eqn:proof_thm_aligned_5} \\
& \leq & (1 + \eps) \cdot {\cal H}( \tilde{\cal P}, [0,\tau_{\mycyc}) ) \ . \nonumber
\end{eqnarray}
Here, inequality~\eqref{eqn:proof_thm_aligned_4} holds since, as explained when establishing property~\ref{item:B_aligned_1}, the holding cost of $\hat{\cal P}^1$ is upper-bounded by that of $\tilde{\cal P}$. Inequality~\eqref{eqn:proof_thm_aligned_5} follows from the next auxiliary claim, whose proof is provided below.

\begin{claim} \label{clm:aux_holding_inequality}
$H_i \cdot \frac{ \tau_\mycyc^2 }{ (n/\eps)^{3q} } \leq {\cal H}( \tilde{\cal P}_i, [0,\tau_{\mycyc}) )$. 
\end{claim}
\end{proof}

\paragraph{Proof of Claim~\ref{clm:aux_holding_inequality}.} We begin by recalling that,  for every $i \in \tilde{\cal F}_q$, the policy $\tilde{\cal P}$ places $N( \tilde{\cal P}_i, [0,\tau_{\mycyc})) \leq (\frac{n}{\eps})^{3q}$ orders for this commodity. Therefore, its holding cost ${\cal H}( \tilde{\cal P}_i, [0,\tau_{\mycyc}) )$ can be lower-bounded by that of the holding-cost-wise cheapest policy for commodity $i$ across the cycle $[0,\tau_{\mycyc})$, subject to placing exactly $N( \tilde{\cal P}_i, [0,\tau_{\mycyc}))$ orders. In other words, our lower bound is the optimum value of:
\[ \begin{array}{ll}
{\displaystyle \min_{{\cal P}_i}} & {\cal H}( {\cal P}_i, [0,\tau_{\mycyc}) ) \\
\text{s.t.} & N( {\cal P}_i, [0,\tau_{\mycyc})) = N( \tilde{\cal P}_i, [0,\tau_{\mycyc}))
\end{array} \]
Based on the proof of Lemma~\ref{lem:sosi_optimal}, this formulation admits a SOSI optimal replenishment policy. In the latter, exactly $N( \tilde{\cal P}_i, [0,\tau_{\mycyc}))$ orders are placed, each consisting of $\frac{ \tau_\mycyc }{ N( \tilde{\cal P}_i, [0,\tau_{\mycyc})) }$ units, leading to a total holding cost of $H_i \cdot \frac{ \tau_\mycyc^2 }{ N( \tilde{\cal P}_i, [0,\tau_{\mycyc})) }$. Therefore,
\[ {\cal H}( \tilde{\cal P}_i, [0,\tau_{\mycyc}) ) ~~\geq~~ H_i \cdot \frac{ \tau_\mycyc^2 }{ N( \tilde{\cal P}_i, [0,\tau_{\mycyc})) } ~~\geq~~ H_i \cdot \frac{ \tau_\mycyc^2 }{ (n/\eps)^{3q} } \ . \]

\section{Additional Proofs from Section~\ref{sec:sub2-approx}}

\subsection{Proof of Lemma~\ref{lem:avg_space_prefix}} \label{app:proof_lem_avg_space_prefix}

Our proof is based on a charging argument, comparing the total average space occupied by the commodities in ${\cal S}_{\mysuff}$ to the analogous quantity with respect to ${\cal V}_{\hat{\ell}}^{\eps}$, where $\hat{\ell}$ is the smallest index of a non-empty class in ${\cal S}_{\mypref}$. For this purpose, in regard to the latter term, we observe  that 
\begin{equation} \label{eqn:LB_sum_gammaI_l1}
\sum_{i \in {\cal V}_{\hat{\ell}}^{\eps}} \gamma_i \cdot \bar{I}( {\cal P}_i^{\eps} ) ~~\geq~~ \frac{ 1 }{ (1 + \eps)^{\hat{\ell} - 1} } \cdot {\cal V} \ ,
\end{equation}
since $\gamma_i \cdot \bar{I}( {\cal P}^{\eps}_i ) \in ( \frac{1}{(1+\eps)^{\hat{\ell}}} \cdot {\cal V}, \frac{1}{(1+\eps)^{\hat{\ell}-1}} \cdot{\cal V}]$ for every commodity $i \in {\cal V}^{\eps}_{ \hat{\ell} }$ and since ${\cal V}^{\eps}_{ \hat{\ell} } \neq \emptyset$.

On the other hand, to upper-bound the total average space occupied by ${\cal S}_{\mysuff}$-commodities, note that 
\begin{eqnarray}
\sum_{\ell \in {\cal S}_{\mysuff}} \sum_{i \in {\cal V}_{\ell}^{\eps}} \gamma_i \cdot \bar{I}( {\cal P}_i^{\eps} ) 
& \leq & \sum_{\ell \in {\cal S}_{\mysuff}} | {\cal V}_{\ell}^{\eps} | \cdot \frac{ 1 }{ (1 + \eps)^{ \ell - 1}} \cdot {\cal V} \label{eqn:proof_lem_avg_space_prefix_0} \\
& \leq & \frac{ 100 }{ \eps^5 } \cdot \sum_{\ell \in {\cal S}_{\mysuff}} \frac{ 1 }{ (1 + \eps)^{ \ell - 1}} \cdot {\cal V} \nonumber \\
& \leq & \frac{ 100 }{ \eps^5 } \cdot \sum_{\ell \geq \ell_\mymid+1}^{\infty} \frac{ 1 }{ (1 + \eps)^{ \ell - 1}} \cdot {\cal V} \label{eqn:proof_lem_avg_space_prefix_1} \\
& \leq & \frac{ 100 }{ \eps^5 } \cdot \frac{ 1 + \eps }{ \eps } \cdot \frac{ 1 }{ (1 + \eps)^{ \ell_{\mymid} }} \cdot {\cal V} \nonumber \\
& \leq & \frac{ 125 }{ \eps^6 } \cdot \frac{ 1 }{ (1 + \eps)^{ \Delta } }  \cdot \frac{ 1 }{ (1 + \eps)^{ \hat{\ell} - 1 }} \cdot {\cal V} \label{eqn:proof_lem_avg_space_prefix_2}  \\
& \leq & \eps \cdot \sum_{i \in {\cal V}_{\hat{\ell}}^{\eps}} \gamma_i \cdot \bar{I}( {\cal P}_i^{\eps} ) \label{eqn:proof_lem_avg_space_prefix_3} \\
& \leq & \eps {\cal V} \ . \nonumber
\end{eqnarray}
Here, inequality~\eqref{eqn:proof_lem_avg_space_prefix_1} holds since ${\cal S}_{\mysuff} = \{ \ell_{\mymid} + 1, \ldots, \ell_M \}$, noting that this sequence of indices is increasing. Inequality~\eqref{eqn:proof_lem_avg_space_prefix_2} follows by recalling that $\ell_{\mymid}$ is the unique index up to which $\Delta$ of the classes ${\cal V}^{\eps}_{\ell_1}, \ldots, {\cal V}^{\eps}_{\ell_{\mymid}}$ are not empty, implying that $\ell_{\mymid} \geq \hat{\ell} + \Delta - 1$. Finally, inequality~\eqref{eqn:proof_lem_avg_space_prefix_2} is obtained by plugging in $\Delta = \lceil \log_{1 + \eps} ( \frac{ 125 }{ \eps^7 } ) \rceil$ and utilizing inequality~\eqref{eqn:LB_sum_gammaI_l1}.

\subsection{Proof of Lemma~\ref{lem:cost_U_dense}} \label{app:proof_lem_cost_U_dense}

We first observe that, since every commodity-vertex $i \in U$ has a unique adjacent edge $(i,\ell) \in {\cal E}^*$, in which case $C_i( \hat{T}_i ) = C_i( \hat{T}_{i \ell} ) = w_{i \ell}$, it follows that the total cost of the policies $\{ \hat{T}_i \}_{i \in U}$ can be written as $\sum_{i \in U}  C_i( \hat{T}_i ) = \sum_{(i,\ell) \in {\cal E}^*} w_{i \ell} = \opt({\cal I})$, where the second equality holds due to the optimality of ${\cal E}^*$. Therefore, to establish the desired claim, it suffices to argue that $\opt({\cal I}) \leq \sum_{i \in U} C_i( {\cal P}^{\eps}_i )$. 

For this purpose, consider the set of edges ${\cal E} = \{ (i,\ell) : i \in {\cal V}_{\ell}^{\eps} \}$, matching each commodity-vertex $i$ to the unique class-vertex $\ell$ for which $i \in {\cal V}_{\ell}^{\eps}$. This edge set is clearly a feasible solution to ${\cal I}$, implying that $\opt({\cal I}) \leq \sum_{(i,\ell) \in {\cal E}} w_{i \ell}$. We conclude the proof by showing that $w_{i \ell} \leq C_i( {\cal P}^{\eps}_i )$ for every edge $(i,\ell) \in {\cal E}$. Indeed, focusing on one such edge, since $i \in {\cal V}_{\ell}^{\eps}$, our definition of the volume class ${\cal V}_{\ell}^{\eps}$ (see Section~\ref{subsec:alg_outline_2minus}) guarantees that the  replenishment policy ${\cal P}^{\eps}_i$ satisfies 
\[ \gamma_i \cdot \bar{I}( {\cal P}^{\eps}_i ) ~~\leq~~ \begin{cases}
\frac{1}{(1+\eps)^{\ell-1}} \cdot{\cal V}, \qquad & \text{if } \ell \in [L] \\
\frac{ \eps }{ n } \cdot {\cal V}, & \text{if } \ell = \infty
\end{cases} \]
The right-hand-side of this inequality is identical to that  of the average space constraint~\eqref{eqn:occupied_space_constraint}. In contrast, the left-hand-side here involves ${\cal P}^{\eps}_i$, which may not be a SOSI policy, whereas the left-hand-side of constraint~\eqref{eqn:occupied_space_constraint} is only concerned with SOSI policies. However, by duplicating the arguments within the proof of Lemma~\ref{lem:sosi_optimal}, it follows that there exists a SOSI policy $T_{i \ell}$ with $\bar{I}( T_{i \ell} ) \leq \bar{I}( {\cal P}^{\eps}_i )$ and $C_i( T_{i \ell} ) \leq C_i( {\cal P}^{\eps}_i )$. Consequently, $T_{i \ell}$ forms a feasible solution to the problem of minimizing $C_i( \cdot )$ subject to constraint~\eqref{eqn:occupied_space_constraint}, for which $\hat{T}_{i \ell}$ is an optimal solution. As a result, we indeed get $w_{ i \ell } = C_i( \hat{T}_{i \ell} ) \leq C_i( T_{i \ell} ) \leq C_i( {\cal P}^{\eps}_i )$.

\subsection{Proof of Lemma~\ref{lem:bound_space_suffix}} \label{app:proof_lem_bound_space_suffix}

To establish the desired claim, we observe that
\begin{eqnarray}
V_{\max}( {\cal P}_{\mysuff} ) & = & \sum_{\ell \in {\cal S}_{\mysuff}} \sum_{i \in \tilde{\cal V}_{\ell}} \gamma_i \hat{T}_i \nonumber \\
& \leq & 2 \cdot \sum_{\ell \in {\cal S}_{\mysuff}} \sum_{i \in \tilde{\cal V}_{\ell}} \left( \frac{1}{(1+\eps)^{\ell-1}} \cdot{\cal V} + \frac{ \eps }{ n } \cdot {\cal V} \right) \label{eqn:proof_space_suffix_1} \\
& \leq & 2 \cdot \sum_{\ell \in {\cal S}_{\mysuff}} | {\cal V}_{\ell}^{\eps} | \cdot \frac{1}{(1+\eps)^{\ell-1}} \cdot{\cal V} + 2\eps {\cal V} \label{eqn:proof_space_suffix_2} \\
& \leq & 4\eps {\cal V} \ . \label{eqn:proof_space_suffix_3}
\end{eqnarray}
Here, inequality~\eqref{eqn:proof_space_suffix_1} is an immediate consequence of property~\ref{prop:hatT_space}. Similarly, inequality~\eqref{eqn:proof_space_suffix_1} follows from property~\ref{prop:tildeV_size}, stating in particular that $|\tilde{\cal V}_{ \ell }| = |{\cal V}_{\ell}^{\eps} |$ for every $\ell \in {\cal S}_{\mysuff}$. Finally, inequality~\eqref{eqn:proof_space_suffix_3} is obtained by recalling that $\sum_{\ell \in {\cal S}_{\mysuff}} | {\cal V}_{\ell}^{\eps} | \cdot \frac{1}{(1+\eps)^{\ell-1}} \cdot{\cal V}$ is precisely the term we are seeing in~\eqref{eqn:proof_lem_avg_space_prefix_0}, which has already been upper-bounded by $\eps {\cal V}$ within the proof of Lemma~\ref{lem:avg_space_prefix}.

\subsection{Proof of Lemma~\ref{lem:main_result_Vell_light}} \label{app:proof_lem_main_result_Vell_light}

In terms of cost, due to setting $\tilde{\cal P}^{\ell}_i = \hat{T}_i$ for every commodity $i \in \tilde{\cal V}_{ \ell }$, we immediately get $C( \tilde{\cal P}^{\ell} ) = \sum_{i \in \tilde{\cal V}_{ \ell } } C_i ( \tilde{\cal P}^{\ell}_i ) = \sum_{i \in \tilde{\cal V}_{ \ell } } C_i ( \hat{T}_i )$. To derive an upper bound on the maximum space requirement of $\tilde{\cal P}^{\ell}$, by recalling that $\{ \hat{T}_i \}_{ i \in \tilde{\cal V}_{ \ell } }$ are SOSI policies, and therefore
\begin{eqnarray}
V_{\max}( \tilde{\cal P}^{\ell} ) & = & \sum_{i \in \tilde{\cal V}_{ \ell } } \gamma_i \hat{T}_i \nonumber \\
& = & 2 \cdot \left( \sum_{i \in {\cal L}_{ \ell } } \gamma_i \cdot \bar{I}( \hat{T}_i ) + \sum_{i \in {\cal H}_{ \ell } } \gamma_i \cdot \bar{I}( \hat{T}_i ) \right) \nonumber \\
& \leq & 2 \cdot \left( \frac{ 3 }{ 4 } \cdot |{\cal L}_{ \ell }| + |{\cal H}_{ \ell }| \right) \cdot \frac{1}{(1+\eps)^{\ell-1}} \cdot{\cal V} \label{eqn:proof_lem_main_result_Vell_light_eq2} \\
& \leq & \frac{ 7 }{ 4 } \cdot | \tilde{\cal V}_{\ell} | \cdot \frac{1}{(1+\eps)^{\ell-1}} \cdot{\cal V} \ . \label{eqn:proof_lem_main_result_Vell_light_eq3}
\end{eqnarray}
Here, to obtain inequality~\eqref{eqn:proof_lem_main_result_Vell_light_eq2}, we note that $\gamma_i \cdot \bar{I}( \hat{T}_i ) \leq \frac{ 3 }{ 4 } \cdot \frac{1}{(1+\eps)^{\ell-1}} \cdot{\cal V}$ for every light commodity, whereas  $\gamma_i \cdot \bar{I}( \hat{T}_i ) \in (\frac{ 3 }{ 4 } \cdot \frac{1}{(1+\eps)^{\ell-1}} \cdot{\cal V}, \frac{1}{(1+\eps)^{\ell-1}} \cdot{\cal V}]$ for every heavy commodity, by definition of ${\cal L}_{ \ell }$ and ${\cal H}_{ \ell }$. Inequality~\eqref{eqn:proof_lem_main_result_Vell_light_eq3} follows by recalling that $|{\cal L}_{ \ell }| + |{\cal H}_{ \ell }| = | \tilde{\cal V}_{ \ell } |$ and that $| {\cal L}_{ \ell } | \geq \frac{ | \tilde{\cal V}_{ \ell } | }{ 2 }$, by the hypothesis of Lemma~\ref{lem:main_result_Vell_light}.

\section{Additional Proofs from Section~\ref{sec:po2-sync}}

\subsection{Proof of Lemma~\ref{lem:construction_pair}} \label{app:proof_lem_construction_pair}

For ease of presentation, rather than directly considering the regime where $\frac{ \gamma_A T_A }{ \gamma_B T_B } \in 1 \pm \eps$, we will assume that  $\gamma_A T_A = \gamma_B T_B$. One can easily run through our proof with $\tilde{\gamma}_A = \frac{ \gamma_B T_B }{ T_A }$, which obviously satisfies $\tilde{\gamma}_A T_A = \gamma_B T_B$. Subsequently, since $\tilde{\gamma}_A \in (1 \pm \eps) \cdot \gamma_A$, by reverting back to the original coefficient $\gamma_A$, we increase the joint occupied space by a factor of at most $1 + \eps$. With this assumption, we will construct dynamic replenishment policies ${\cal P}_A$ and ${\cal P}_B$ that satisfy
\[ V_{\max}( {\cal P}_A, {\cal P}_B ) ~~\leq~~ \frac{ 7 }{ 4 } \cdot ( \gamma_A \cdot \bar{I}( T_A ) + \gamma_B \cdot \bar{I}( T_B ) ) \qquad \text{and} \qquad \max \left\{ \frac{ C_A( {\cal P}_A ) }{ C_A( T_A )}, \frac{ C_B( {\cal P}_B ) }{ C_B( T_B)} \right\} ~~\leq~~ \frac{ 32 }{ 31} \ . \]
For convenience, we assume without loss of generality that $T_B \leq T_A = 1$. Given that $T_A / T_B$ is an integer power of $2$, our proof considers six different cases: $T_B = 1$, $T_B = \frac{ 1 }{ 2 }$, $T_B = \frac{ 1 }{ 4 }$, $T_B = \frac{ 1 }{ 8 }$, $T_B = \frac{ 1 }{ 16 }$, and $T_B = \frac{ 1 }{ 2^k }$ for some $k \geq 5$. 

\paragraph{Case 1: \bstitle{T_B = 1}.} We begin by considering the simplest possible scenario, since it is a convenient opportunity to explain our notation and terminology:
\begin{itemize}
    \item {\em The policy ${\cal P}_A$}: $A$-orders will be placed at the integer time points $0, 1, 2, \ldots$.

    \item {\em The policy ${\cal P}_B$}: $B$-orders will also have a gap of $1$ between successive orders. However, they will be offset by $0.5$, meaning that beyond a single order at time $0$, we place $B$-orders at $0.5, 1.5, 2.5, \ldots$.
    
    \item {\em Maximal space requirement:} Since ${\cal P}_A$ and ${\cal P}_B$ have a joint cycle of length $1$, to detect where their maximal space requirement is located, it suffices to examine a unit-length interval, say $[0.5,1.5)$. Moreover, since the inventory level of each commodity is decreasing between successive orders, the joint maximal space requirement of ${\cal P}_A$ and ${\cal P}_B$ is attained at either $0.5$ or $1$. In these two points, we have:
    \begin{itemize}
        \item $V( {\cal P}_A, {\cal P}_B, 0.5 ) = \frac{ 1 }{ 2 } \cdot \gamma_A  T_A + \gamma_B  T_B  = \frac{ 3 }{ 2 } \cdot ( \gamma_A \cdot \bar{I}( T_A ) + \gamma_B \cdot \bar{I}( T_B ) )$.

        \item $V( {\cal P}_A, {\cal P}_B, 1 ) =  \gamma_A  T_A + \frac{ 1 }{ 2 } \cdot \gamma_B  T_B  = \frac{ 3 }{ 2 } \cdot ( \gamma_A \cdot \bar{I}( T_A ) + \gamma_B \cdot \bar{I}( T_B ) )$. 
    \end{itemize}
    In both cases, we make use of $\gamma_A \cdot \bar{I}( T_A ) = \gamma_B \cdot \bar{I}( T_B )$, or equivalently $\gamma_A  T_A = \gamma_B   T_B$, which is an important relation to remember throughout this analysis.    
    \item {\em Cost}: In terms of cost, since ${\cal P}_A$ and ${\cal P}_B$ are identical to $T_A$ and $T_B$ up to offsets, we have $C_A( {\cal P}_A ) = C_A( T_A )$ and $C_B( {\cal P}_B ) = C_B( T_B )$.

    \item {\em Summary}: $V_{\max}( {\cal P}_A, {\cal P}_B ) = \frac{ 3 }{ 2 } \cdot ( \gamma_A \cdot \bar{I}( T_A ) + \gamma_B \cdot \bar{I}( T_B ) )$ and $\max \{ \frac{ C_A( {\cal P}_A ) }{ C_A( T_A ) }, \frac{ C_B( {\cal P}_B ) }{ C_B( T_B ) } \} = 1$. 
\end{itemize}

\paragraph{Case 2: \bstitle{T_B = \frac{ 1 }{ 2 }}.}
\begin{itemize}
    \item {\em The policy ${\cal P}_A$}: $A$-orders will be placed at $0, 1, 2, \ldots$.

    \item {\em The policy ${\cal P}_B$}: $B$-orders will be placed at $\frac{1}{3}, \frac{5}{6}, 1\frac{1}{3}, 1\frac{5}{6}, \ldots$. From this point on, we will not be mentioning the obvious singular order at time $0$.

    \item {\em Maximal space requirement:} Once again,  ${\cal P}_A$ and ${\cal P}_B$ have a joint cycle of length $1$, and we examine the unit-length interval $[\frac{1}{3},1\frac{1}{3})$, where there are three ordering points to be tested, $\frac{1}{3}$, $\frac{5}{6}$, and $1$:
    \begin{itemize}
        \item $V( {\cal P}_A, {\cal P}_B, \frac{1}{3} ) = \frac{ 2 }{ 3 } \cdot \gamma_A  T_A + \gamma_B  T_B  = \frac{ 5 }{ 3 } \cdot ( \gamma_A \cdot \bar{I}( T_A ) + \gamma_B \cdot \bar{I}( T_B ) )$.

        \item $V( {\cal P}_A, {\cal P}_B, \frac{5}{6} ) = \frac{ 1 }{ 6 } \cdot \gamma_A  T_A + \gamma_B  T_B  = \frac{ 7 }{ 6 } \cdot ( \gamma_A \cdot \bar{I}( T_A ) + \gamma_B \cdot \bar{I}( T_B ) )$.        

        \item $V( {\cal P}_A, {\cal P}_B, 1 ) =  \gamma_A  T_A + \frac{ 2 }{ 3 } \cdot \gamma_B  T_B  = \frac{ 5 }{ 3 } \cdot ( \gamma_A \cdot \bar{I}( T_A ) + \gamma_B \cdot \bar{I}( T_B ) )$. 
    \end{itemize}  

    \item {\em Cost}: Again,  ${\cal P}_A$ and ${\cal P}_B$ are identical to $T_A$ and $T_B$ up to offsets, and therefore $C_A( {\cal P}_A ) = C_A( T_A )$ and $C_B( {\cal P}_B ) = C_B( T_B )$.

    \item {\em Summary}: $V_{\max}( {\cal P}_A, {\cal P}_B ) = \frac{ 5 }{ 3 } \cdot ( \gamma_A \cdot \bar{I}( T_A ) + \gamma_B \cdot \bar{I}( T_B ) )$ and $\max \{ \frac{ C_A( {\cal P}_A ) }{ C_A( T_A ) }, \frac{ C_B( {\cal P}_B ) }{ C_B( T_B ) } \} = 1$. 
\end{itemize}

\paragraph{Case 3: \bstitle{T_B = \frac{ 1 }{ 4 }}.}
\begin{itemize}
    \item {\em The policy ${\cal P}_A$}: For the first time, ${\cal P}_A$ will not be identical to $T_A$, but rather a $\frac{31}{32}$-scaling of this policy. As such, $A$-orders will be placed at $0, \frac{31}{32}, \frac{62}{32}, \frac{93}{32} \ldots$.

    \item {\em The policy ${\cal P}_B$}: Here, for the first time, we will be employing a non-SOSI policy. Specifically, the cycle length of ${\cal P}_B$ will be $\frac{ 31 }{ 32}$, i.e., similar to that of ${\cal P}_A$, and its offset will be $\frac{ 5 }{ 32 }$. Each such cycle will be filled by four $B$-orders, the first of length $\frac{7}{8} \cdot T_B = \frac{ 7 }{ 32 }$, and the next three of length $T_B = \frac{ 1 }{ 4 }$, noting that these quantities indeed sum up to $\frac{ 31 }{ 32}$.

    \item {\em Maximal space requirement:} ${\cal P}_A$ and ${\cal P}_B$ have a joint cycle of length $\frac{ 31 }{ 32}$, and for convenience, we examine the interval $[\frac{ 5 }{ 32 }, \frac{ 36 }{ 32 })$, where there are five ordering points to be tested, $\frac{ 5 }{ 32 }$, $\frac{ 12 }{ 32 }$, $\frac{ 20 }{ 32 }$, $\frac{ 28 }{ 32 }$, and $\frac{ 31 }{ 32 }$:
    \begin{itemize}
        \item $V( {\cal P}_A, {\cal P}_B, \frac{ 5 }{ 32 } ) = \frac{ 26 }{ 31 } \cdot \frac{ 31 }{ 32 } \cdot \gamma_A  T_A + \frac{ 7 }{ 8 } \cdot \gamma_B  T_B  = \frac{ 27 }{ 16 } \cdot ( \gamma_A \cdot \bar{I}( T_A ) + \gamma_B \cdot \bar{I}( T_B ) )$.

        \item $V( {\cal P}_A, {\cal P}_B, \frac{ 12 }{ 32 } ) = \frac{ 19 }{ 31 } \cdot \frac{ 31 }{ 32 } \cdot \gamma_A  T_A + \gamma_B  T_B  = \frac{ 51 }{ 32 } \cdot ( \gamma_A \cdot \bar{I}( T_A ) + \gamma_B \cdot \bar{I}( T_B ) )$.

        \item $V( {\cal P}_A, {\cal P}_B, \frac{ 20 }{ 32 } ) = \frac{ 11 }{ 31 } \cdot \frac{ 31 }{ 32 } \cdot \gamma_A  T_A + \gamma_B  T_B  = \frac{ 43 }{ 32 } \cdot ( \gamma_A \cdot \bar{I}( T_A ) + \gamma_B \cdot \bar{I}( T_B ) )$.

        \item $V( {\cal P}_A, {\cal P}_B, 28 ) = \frac{ 3 }{ 31 } \cdot \frac{ 31 }{ 32 } \cdot \gamma_A  T_A + \gamma_B  T_B  = \frac{ 34 }{ 32 } \cdot ( \gamma_A \cdot \bar{I}( T_A ) + \gamma_B \cdot \bar{I}( T_B ) )$.

        \item $V( {\cal P}_A, {\cal P}_B, \frac{ 31 }{ 32 } ) = \frac{ 31 }{ 32 } \cdot \gamma_A  T_A + \frac{ 5 }{ 8 } \cdot \gamma_B  T_B  = \frac{ 51 }{ 32 } \cdot ( \gamma_A \cdot \bar{I}( T_A ) + \gamma_B \cdot \bar{I}( T_B ) )$.
    \end{itemize}  

    \item {\em Cost}: Since ${\cal P}_A$ is a $\frac{ 31 }{ 32 }$-scaling of $T_A$, we have $C_A( {\cal P}_A ) \leq \frac{ 32 }{ 31 } \cdot C_A( T_A )$. In addition, given the construction of ${\cal P}_B$, this policy is a $\frac{7}{8}$-scaling of $T_B$ on a $\frac{7}{31}$-fraction of its cycle, and it identifies with $T_B$ on the remaining fraction, meaning that
    \[ C_B( {\cal P}_B ) ~~\leq~~ \left( \frac{7}{31} \cdot \frac{ 8 }{ 7 } + \frac{24}{31}  \right) \cdot C_B( T_B ) ~~=~~ \frac{ 32 }{ 31 } \cdot C_B( T_B )  \ .  \]
    
    \item {\em Summary}: $V_{\max}( {\cal P}_A, {\cal P}_B ) = \frac{ 27 }{ 16 } \cdot ( \gamma_A \cdot \bar{I}( T_A ) + \gamma_B \cdot \bar{I}( T_B ) )$ and $\max \{ \frac{ C_A( {\cal P}_A ) }{ C_A( T_A ) }, \frac{ C_B( {\cal P}_B ) }{ C_B( T_B ) } \} \leq \frac{ 32 }{ 31 }$. 
\end{itemize}

\paragraph{Case 4: \bstitle{T_B = \frac{ 1 }{ 8 }}.}
\begin{itemize}
    \item {\em The policy ${\cal P}_A$}: Similarly to case~3, the policy ${\cal P}_A$ will be a $\frac{31}{32}$-scaling of $T_A$, with $A$-orders placed at $0, \frac{31}{32}, \frac{62}{32}, \frac{93}{32} \ldots$.

    \item {\em The policy ${\cal P}_B$}: Similarly to case~3, the cycle length of ${\cal P}_B$ will be $\frac{ 31 }{ 32}$ as well, with an offset of $\frac{ 3 }{ 32 }$. Each such cycle will be filled in left-to-right direction by eight $B$-orders of length: $\frac{ 27 }{ 32 } \cdot T_B$, $\frac{ 19 }{ 20 } \cdot T_B$, $T_B$, $T_B$, $T_B$, $T_B$, $\frac{ 153 }{ 160 } \cdot T_B$, and $T_B$. Since $T_B = \frac{ 1 }{ 8 }$, these quantities indeed sum up to $\frac{ 31 }{ 32}$.

    \item {\em Maximal space requirement:} ${\cal P}_A$ and ${\cal P}_B$ have a joint cycle of length $\frac{ 31 }{ 32}$, and for convenience, we examine the interval $[\frac{ 3 }{ 32 }, \frac{ 34 }{ 32 })$, where there are nine ordering points to be tested, $\frac{ 3 }{ 32 }$, $\frac{ 51 }{ 256 }$, $\frac{ 407 }{ 1280 }$, $\frac{ 567 }{ 1280 }$, $\frac{ 727 }{ 1280 }$, $\frac{ 887 }{ 1280 }$, $\frac{ 1047 }{ 1280 }$, $\frac{ 15 }{ 16 }$, and $\frac{ 31 }{ 32 }$:
    \begin{itemize}
        \item $V( {\cal P}_A, {\cal P}_B, \frac{3 }{ 32 } ) = \frac{ 28 }{ 31 } \cdot \frac{ 31 }{ 32 } \cdot \gamma_A  T_A + \frac{ 27 }{ 32 } \cdot \gamma_B  T_B  = \frac{ 55 }{ 32 } \cdot ( \gamma_A \cdot \bar{I}( T_A ) + \gamma_B \cdot \bar{I}( T_B ) )$.

        \item $V( {\cal P}_A, {\cal P}_B, \frac{ 51 }{ 256 } ) = (1 - \frac{ 51/256 }{31/32}) \cdot \frac{ 31 }{32}\cdot \gamma_A  T_A + \frac{ 19 }{ 20 } \cdot \gamma_B  T_B  = \frac{ 2201 }{ 1280 } \cdot ( \gamma_A \cdot \bar{I}( T_A ) + \gamma_B \cdot \bar{I}( T_B ) )$.

        \item $V( {\cal P}_A, {\cal P}_B, \frac{ 407 }{ 1280 } ) = (1 - \frac{ 407/1280 }{31/32}) \cdot \frac{ 31 }{32}\cdot \gamma_A  T_A + \gamma_B  T_B  = \frac{ 2113 }{ 1280 } \cdot ( \gamma_A \cdot \bar{I}( T_A ) + \gamma_B \cdot \bar{I}( T_B ) )$.

        \item $V( {\cal P}_A, {\cal P}_B, \frac{ 567 }{ 1280 } ) = (1 - \frac{ 567/1280 }{31/32}) \cdot \frac{ 31 }{32}\cdot \gamma_A  T_A + \gamma_B  T_B  = \frac{ 1953 }{ 1280 } \cdot ( \gamma_A \cdot \bar{I}( T_A ) + \gamma_B \cdot \bar{I}( T_B ) )$.

        \item $V( {\cal P}_A, {\cal P}_B, \frac{ 727 }{ 1280 } ) = (1 - \frac{ 727/1280 }{31/32}) \cdot \frac{ 31 }{32}\cdot \gamma_A  T_A + \gamma_B  T_B  = \frac{ 1793 }{ 1280 } \cdot ( \gamma_A \cdot \bar{I}( T_A ) + \gamma_B \cdot \bar{I}( T_B ) )$.

        \item $V( {\cal P}_A, {\cal P}_B, \frac{ 887 }{ 1280 } ) = (1 - \frac{ 887/1280 }{31/32}) \cdot \frac{ 31 }{32}\cdot \gamma_A  T_A + \gamma_B  T_B  = \frac{ 1633 }{ 1280 } \cdot ( \gamma_A \cdot \bar{I}( T_A ) + \gamma_B \cdot \bar{I}( T_B ) )$.
        
        \item $V( {\cal P}_A, {\cal P}_B, \frac{ 1047 }{ 1280 } ) = (1 - \frac{ 1047/1280 }{31/32}) \cdot \frac{ 31 }{32}\cdot \gamma_A  T_A + \frac{ 153 }{ 160 } \cdot \gamma_B  T_B  = \frac{ 1417 }{ 1280 } \cdot ( \gamma_A \cdot \bar{I}( T_A ) + \gamma_B \cdot \bar{I}( T_B ) )$.

        \item $V( {\cal P}_A, {\cal P}_B, \frac{ 15 }{ 16 } ) = (1 - \frac{ 15/16 }{31/32}) \cdot \frac{ 31 }{32}\cdot \gamma_A  T_A +  \gamma_B  T_B  = \frac{ 33 }{ 32 } \cdot ( \gamma_A \cdot \bar{I}( T_A ) + \gamma_B \cdot \bar{I}( T_B ) )$.

        \item $V( {\cal P}_A, {\cal P}_B, \frac{ 31 }{ 32 } ) = \frac{ 31 }{ 32 } \cdot \gamma_A  T_A + \frac{ 3 }{ 4 } \cdot \gamma_B  T_B  = \frac{ 55 }{ 32 } \cdot ( \gamma_A \cdot \bar{I}( T_A ) + \gamma_B \cdot \bar{I}( T_B ) )$.
    \end{itemize}

    \item {\em Cost}: Since ${\cal P}_A$ is a $\frac{ 31 }{ 32 }$-scaling of $T_A$, we have $C_A( {\cal P}_A ) \leq \frac{ 32 }{ 31 } \cdot C_A( T_A )$.  In addition, given the construction of ${\cal P}_B$, this policy is a $\frac{27}{32}$-scaling of $T_B$ on a $\frac{ 27 / 256 }{ 31/32 }$-fraction of its cycle. Along the same lines, ${\cal P}_B$ is a $\frac{19}{20}$-scaling of $T_B$ on a $\frac{ 19 / 160}{ 31/32 }$-fraction, so on and so forth. Therefore,
    \begin{eqnarray*}
    C_B( {\cal P}_B ) & \leq & \left( \frac{ 27 / 256 }{ 31/32 } \cdot \frac{ 32 }{ 27 } + \frac{ 19 / 160}{ 31/32 } \cdot \frac{ 20 }{ 19 } + 5 \cdot \frac{ 1 / 8}{ 31/32 } + \frac{ 153 / 1280 }{ 31/32 } \cdot \frac{ 160 }{ 153 } \right) \cdot C_B( T_B ) \\
    & = &   \frac{ 32 }{ 31 } \cdot C_B( T_B ) \ . 
    \end{eqnarray*}
       
    \item {\em Summary}: $V_{\max}( {\cal P}_A, {\cal P}_B ) = \frac{ 2201 }{ 1280 } \cdot ( \gamma_A \cdot \bar{I}( T_A ) + \gamma_B \cdot \bar{I}( T_B ) )$ and $\max \{ \frac{ C_A( {\cal P}_A ) }{ C_A( T_A ) }, \frac{ C_B( {\cal P}_B ) }{ C_B( T_B ) } \} \leq \frac{ 32 }{ 31 }$.
\end{itemize}

\paragraph{Case 5: \bstitle{T_B = \frac{ 1 }{ 16 }}.}
\begin{itemize}
    \item {\em The policy ${\cal P}_A$}: Once again, the policy ${\cal P}_A$ will be a $\frac{31}{32}$-scaling of $T_A$, with $A$-orders placed at $0, \frac{31}{32}, \frac{62}{32}, \frac{93}{32} \ldots$.

    \item {\em The policy ${\cal P}_B$}: The cycle length of ${\cal P}_B$ will be $\frac{ 31 }{ 32}$ as well. Each such cycle, say $[0, \frac{ 31 }{ 32 })$, is partitioned into three segments: $[0, \frac{ 9 }{ 32 })$, $[\frac{ 9 }{ 32 }, \frac{ 23 }{ 32 })$, and $[\frac{ 23 }{ 32 },\frac{ 31 }{32})$, filled by $B$-orders as follows:
    \begin{itemize}
        \item In the left segment $[0, \frac{ 9 }{ 32 })$, we place six $B$-orders, each of length $\frac{3}{4} \cdot T_B$. 

        \item In the middle segment $[\frac{ 9 }{ 32 }, \frac{ 23 }{ 32 })$, we place seven $B$-orders, each of length $T_B$.

        \item In the right segment $[\frac{ 23 }{ 32 },\frac{ 31 }{32})$, we place three $B$-orders, each of length $\frac{4}{3} \cdot T_B$. 
    \end{itemize}

    \item {\em Maximal space requirement:} ${\cal P}_A$ and ${\cal P}_B$ have a joint cycle of length $\frac{ 31 }{ 32}$, and for convenience, we examine $[0, \frac{ 31 }{ 32 })$. While this interval has quite a few ordering points, we observe that since the inventory level of commodity $A$ is decreasing across $[0, \frac{ 31 }{ 32 })$, and since we repeat the same policy for commodity $B$ in each of the  segments $[0, \frac{ 9 }{ 32 })$, $[\frac{ 9 }{ 32 }, \frac{ 23 }{ 32 })$, and $[\frac{ 23 }{ 32 },\frac{ 31 }{32})$, the maximal space requirement of ${\cal P}_A$ and ${\cal P}_B$ will be attained at one of the points $0$, $\frac{ 9 }{ 32 }$, and $\frac{ 23 }{ 32 }$.
    \begin{itemize}
        \item $V( {\cal P}_A, {\cal P}_B, 0 ) = \gamma_A  T_A + \frac{ 3 }{ 4 } \cdot \gamma_B  T_B  = \frac{ 7 }{ 4 } \cdot ( \gamma_A \cdot \bar{I}( T_A ) + \gamma_B \cdot \bar{I}( T_B ) )$.

        \item $V( {\cal P}_A, {\cal P}_B, \frac{9}{32} ) = \frac{22}{31} \cdot \frac{31}{32} \cdot \gamma_A  T_A + \gamma_B  T_B  = \frac{ 27 }{ 16 } \cdot ( \gamma_A \cdot \bar{I}( T_A ) + \gamma_B \cdot \bar{I}( T_B ) )$.

        \item $V( {\cal P}_A, {\cal P}_B, \frac{23}{32} ) = \frac{8}{31} \cdot \frac{31}{32} \cdot \gamma_A  T_A + \frac{4}{3} \cdot \gamma_B  T_B  = \frac{ 19 }{ 12 } \cdot ( \gamma_A \cdot \bar{I}( T_A ) + \gamma_B \cdot \bar{I}( T_B ) )$.
    \end{itemize}

    \item {\em Cost}: Since ${\cal P}_A$ is a $\frac{ 31 }{ 32 }$-scaling of $T_A$, we have $C_A( {\cal P}_A ) \leq \frac{ 32 }{ 31 } \cdot C_A( T_A )$.  In addition, given the construction of ${\cal P}_B$, this policy is a $\frac{3}{4}$-scaling of $T_B$ on a $\frac{ 9 }{ 31 }$-fraction of its cycle, $\frac{4}{3}$-scaling of $T_B$ on a $\frac{ 8 }{ 31 }$-fraction, and identifies with $T_B$ on the remaining $\frac{ 14 }{ 31 }$-fraction. Therefore,
    \begin{eqnarray*}
    C_B( {\cal P}_B ) & = & \frac{ 9 }{ 31 } \cdot C_B \left( \frac{ 3 }{ 4 } \cdot T_B \right) + \frac{ 8 }{ 31 } \cdot C_B \left( \frac{ 4 }{ 3 } \cdot T_B \right) + \frac{ 14 }{ 31 } \cdot C_B (  T_B )  \\
    & = & \frac{ 8 }{ 31 } \cdot \underbrace{ \left( C_B \left( \frac{ 3 }{ 4 } \cdot T_B \right) + C_B \left( \frac{ 4 }{ 3 } \cdot T_B \right) \right) }_{ = (\frac{4}{3} + \frac{3}{4}) \cdot C_B(T_B) } + \frac{ 1 }{ 31 } \cdot C_B \left( \frac{ 3 }{ 4 } \cdot T_B \right) + \frac{ 14 }{ 31 } \cdot C_B (  T_B ) \\
    & \leq & \left( \frac{ 50 }{ 93 } + \frac{ 1 }{ 31 } \cdot \frac{4}{3} + \frac{ 14 }{ 31 } \right) \cdot C_B (  T_B ) \\
    & = & \frac{ 32 }{ 31 } \cdot C_B (  T_B ) \ .
    \end{eqnarray*}

    \item {\em Summary}: $V_{\max}( {\cal P}_A, {\cal P}_B ) = \frac{ 7 }{ 4 } \cdot ( \gamma_A \cdot \bar{I}( T_A ) + \gamma_B \cdot \bar{I}( T_B ) )$ and $\max \{ \frac{ C_A( {\cal P}_A ) }{ C_A( T_A ) }, \frac{ C_B( {\cal P}_B ) }{ C_B( T_B ) } \} \leq \frac{ 32 }{ 31 }$.
\end{itemize}

\paragraph{Case 6: \bstitle{T_B = \frac{ 1 }{ 2^k }} for \bstitle{k \geq 5}}
\begin{itemize}
    \item {\em The policy ${\cal P}_A$}: $A$-orders will be placed at the integer time points $0, 1, 2, \ldots$.

    \item {\em The policy ${\cal P}_B$}: The cycle length of ${\cal P}_B$ will be $1$ as well. Each such cycle, say $[0, 1)$, is partitioned into three segments: $[0, \frac{ 3 }{ 8 })$, $[\frac{ 3 }{ 8 }, \frac{ 5 }{ 8 })$, and $[\frac{ 5 }{ 8 },1)$, filled by $B$-orders as follows:
    \begin{itemize}
        \item In the left segment $[0, \frac{ 3 }{ 8 })$, we place $\frac{ 1 }{ 2T_B }$ orders, each of length $\frac{3}{4} \cdot T_B$. 

        \item In the middle segment $[\frac{ 3 }{ 8 }, \frac{ 5 }{ 8 })$, we place $\frac{ 1 }{ 4T_B }$ orders, each of length $T_B$. 

        \item In the right segment $[\frac{ 5 }{ 8 },1)$, we place $\frac{ 9 }{ 32T_B }$ orders, each of length $\frac{4}{3} \cdot T_B$. 
    \end{itemize}
    It is important to note that, since $T_B = \frac{ 1 }{ 2^k }$ for $k \geq 5$, we ensure that each of the terms  $\frac{ 1 }{ 2T_B }$,  $\frac{ 1 }{ 4T_B }$, and $\frac{ 9 }{ 32T_B }$ is indeed an integer.

    \item {\em Maximal space requirement:} As in case~5,  since the inventory level of commodity $A$ is decreasing across $[0, 1)$, and since we repeat the same $B$-policy in each of the segments $[0, \frac{ 3 }{ 8 })$, $[\frac{ 3 }{ 8 }, \frac{ 5 }{ 8 })$, and $[\frac{ 5 }{ 8 },1)$, the maximal space requirement of ${\cal P}_A$ and ${\cal P}_B$ will be attained at one of the points $0$, $\frac{ 3 }{ 8 }$, and $\frac{ 5 }{ 8 }$.
    \begin{itemize}
        \item $V( {\cal P}_A, {\cal P}_B, 0 ) = \gamma_A  T_A + \frac{ 3 }{ 4 } \cdot \gamma_B  T_B  = \frac{ 7 }{ 4 } \cdot ( \gamma_A \cdot \bar{I}( T_A ) + \gamma_B \cdot \bar{I}( T_B ) )$.

        \item $V( {\cal P}_A, {\cal P}_B, \frac{3}{8} ) = \frac{5}{8} \cdot \gamma_A  T_A + \gamma_B  T_B  = \frac{ 13 }{ 8 } \cdot ( \gamma_A \cdot \bar{I}( T_A ) + \gamma_B \cdot \bar{I}( T_B ) )$.

        \item $V( {\cal P}_A, {\cal P}_B, \frac{5}{8} ) = \frac{3}{8} \cdot \gamma_A  T_A + \frac{4}{3} \cdot \gamma_B  T_B  = \frac{ 41 }{ 24 } \cdot ( \gamma_A \cdot \bar{I}( T_A ) + \gamma_B \cdot \bar{I}( T_B ) )$.
    \end{itemize}

    \item {\em Cost}: Since ${\cal P}_A$ identifies with $T_A$, we have $C_A( {\cal P}_A ) = C_A( T_A )$.  In addition, given the construction of ${\cal P}_B$, this policy is a $\frac{3}{4}$-scaling of $T_B$ on a $\frac{ 3 }{ 8 }$-fraction of its cycle, $\frac{4}{3}$-scaling of $T_B$ on a $\frac{ 3 }{ 8 }$-fraction, and identifies with $T_B$ on the remaining $\frac{ 1 }{ 4 }$-fraction. Therefore, 
    \[ C_B( {\cal P}_B ) ~~=~~ \frac{ 3 }{ 8 } \cdot \underbrace{ \left( C_B \left( \frac{ 4 }{ 3 } \cdot T_B \right) + C_B \left( \frac{ 3 }{ 4 } \cdot T_B \right) \right) }_{ = (\frac{4}{3} + \frac{3}{4}) \cdot C_B(T_B) } + \frac{ 1 }{ 4 } \cdot C(T_B) ~~=~~  \frac{ 33 }{ 32 } \cdot C_B( T_B ) \ . \]

    \item {\em Summary}: $V_{\max}( {\cal P}_A, {\cal P}_B ) = \frac{ 7 }{ 4 } \cdot ( \gamma_A \cdot \bar{I}( T_A ) + \gamma_B \cdot \bar{I}( T_B ) )$ and $\max \{ \frac{ C_A( {\cal P}_A ) }{ C_A( T_A ) }, \frac{ C_B( {\cal P}_B ) }{ C_B( T_B ) } \} \leq \frac{ 33 }{ 32 }$.
\end{itemize}

\subsection{Proof of Claim~\ref{clm:bound_on_far_pairs}} \label{app:proof_lem_bound_on_far_pairs}

By Claim~\ref{clm:properties_PO2}(2), we know that $T_i^{ \Theta_{\ell, q} } \in [ \frac{ \hat{T}_i }{ \sqrt{2} } , \sqrt{2} \hat{T}_i ]$, for every commodity $i \in {\cal H}_{ \ell ,q}$. In addition, since any such commodity is heavy, $\gamma_i \hat{T}_i = 2\gamma_i \cdot \bar{I}( \hat{T}_i ) \in 2 \cdot [\frac{3}{4},1] \cdot \frac{1}{(1+\eps)^{\ell-1}} \cdot{\cal V}$. It follows that all elements of the sequence~\eqref{eqn:sequence_gamma_T} are bounded within the interval $[\frac{3 }{2\sqrt{2}}, 2\sqrt{2}] \cdot \frac{1}{(1+\eps)^{\ell-1}} \cdot{\cal V}$, whose endpoints differ by a multiplicative factor of $8/3$. Given this bound, we conclude that $|{\cal F}_{ \ell ,q}^{ \Pi_{ \ell, q } }| \leq  \log_{ 1+\eps } (\frac{ 8 }{ 3 }) \leq \frac{ 11 }{ 10\eps }$.

\subsection{Proof of Lemma~\ref{lem:success_Aell}} \label{app:proof_lem_success_Aell}

Let us consider the collection of random variables $X_1, \ldots, X_Q$, where $X_q = \sum_{i \in {\cal H}_{\ell,q} } \gamma_i  T_i^{ \Theta_{\ell, q} }$. Recalling that ${\cal A}_{\ell}$ stands for the event 
\[ \sum_{i \in {\cal H}_{\ell} } \gamma_i   T_i^{ \Theta_{\ell} } ~~\leq~~ (1 + \eps) \cdot \frac{ 1 }{ \sqrt{2} \ln 2 } \cdot \sum_{i \in {\cal H}_{\ell} } \gamma_i   \hat{T}_i \ , \] 
we observe that the left-hand-side of this inequality is exactly $\sum_{q \in [Q]} X_q$. In addition, $X_1, \ldots, X_Q$ are mutually independent, since our power-of-$2$ rounding procedure is employed for each subset ${\cal H}_{\ell,q}$ independently of any other subset. For these random variables, we establish the next auxiliary claim, whose proof is given in Appendix~\ref{app:proof_clm_auxiliary_success_Aell}. For readability, we make use of the shorthand notation $\eta_q = \sum_{i \in {\cal H}_{\ell,q} } \gamma_i \hat{T}_i$, with $\eta_{\max}$ and $\eta_{\min}$ being the maximum and minimum of these quantities over $q \in [Q]$.

\begin{claim} \label{clm:auxiliary_success_Aell}
\begin{enumerate}
    \item $\exsubpar{ \Theta_\ell }{ X_q } = \frac{ \eta_q }{ \sqrt{2} \ln 2 }$.

    \item $X_q \leq \sqrt{2} \eta_q$. 

    \item $\frac{ \eta_{\min}}{ \eta_{\max} } \geq \frac{ 1 }{ 2 }$.
\end{enumerate}
\end{claim}

Given this claim, we argue that $\prsubpar{ \Theta_\ell }{ \bar{\cal A}_{ \ell }} \leq \eps$, by noting that 
\begin{eqnarray}
\prsub{ \Theta_\ell }{ \bar{\cal A}_{ \ell }} & = & \prsub{ \Theta_\ell }{  \sum_{i \in {\cal H}_{\ell} } \gamma_i   T_i^{ \Theta_{\ell} } >  (1 + 2\eps) \cdot \frac{ 1 }{ \sqrt{2} \ln 2 } \cdot \sum_{i \in {\cal H}_{\ell} } \gamma_i   \hat{T}_i } \nonumber \\
& = & \prsub{ \Theta_\ell }{ \sum_{q \in [Q]} \frac{ X_q }{ \sqrt{2}\eta_{\max} }  >  (1 + 2\eps) \cdot \sum_{q \in [Q]} \frac{ \exsubpar{ \Theta_\ell }{ X_q } }{ \sqrt{2}\eta_{\max} } } \label{eqn:proof_lem_success_Aell_1}  \\
& \leq & \exp \left\{ -\frac{ \eps^2 }{3} \cdot \sum_{q \in [Q]} \frac{ \exsubpar{ \Theta_\ell }{ X_q } }{ \sqrt{2} \eta_{\max} } \right\} \label{eqn:proof_lem_success_Aell_2}   \\
& \leq & \exp \left\{ -\frac{ \eps^2 Q }{6\ln 2} \cdot \frac{  \eta_{\min} }{ \eta_{\max} } \right\} \label{eqn:proof_lem_success_Aell_3}   \\
& \leq & \exp \left\{ - 2 \ln \left( \frac{1}{ \eps } \right) \right\} \label{eqn:proof_lem_success_Aell_4} \\
& = & \frac{ \eps }{ 10 }  \ . \nonumber
\end{eqnarray}
Here, equality~\eqref{eqn:proof_lem_success_Aell_1} is obtained by  combining the definitions of $X_q$ and $\eta_q$ along with Claim~\ref{clm:auxiliary_success_Aell}(1). To arrive at inequality~\eqref{eqn:proof_lem_success_Aell_2}, we make use of a dimension-free Chernoff-Hoeffding inequality stated in \citet[Thm.~1.1 and Ex.~1.1]{DubhashiP09}. Specifically, letting $Y_1, \ldots, Y_n$ be independent $[0,1]$-bounded random variables, for every $\eps \in (0,1)$, we have
\[ \pr{ \sum_{i \in [n]} Y_i > (1 + \eps) \cdot \ex{ \sum_{i \in [n]} Y_i } } ~~\leq~~ \exp \left( - \frac{ \eps^2 }{ 3 } \cdot \ex{ \sum_{i \in [n]} Y_i } \right) \ . \]
In this context, it is worth mentioning that $\{ \frac{ X_q }{\sqrt{2}\eta_{\max} } \}_{q \in [Q]}$ are mutually independent. In addition, these random variables are $[0,1]$-bounded, by Claim~\ref{clm:auxiliary_success_Aell}(2). Inequality~\eqref{eqn:proof_lem_success_Aell_3} holds since $\exsubpar{ \Theta_\ell }{ X_q } = \frac{ \eta_q }{ \sqrt{2} \ln 2 } \geq \frac{ \eta_{\min} }{ \sqrt{2} \ln 2 }$, by Claim~\ref{clm:auxiliary_success_Aell}(1). Finally, inequality~\eqref{eqn:proof_lem_success_Aell_4} follows by noting that $Q = \frac{ 20 \ln (1/\eps) }{ \eps^2 }$ and that $\frac{ \eta_{\min}}{ \eta_{\max} } \geq \frac{ 1 }{ 2 }$, by Claim~\ref{clm:auxiliary_success_Aell}(3).

\subsection{Proof of Claim~\ref{clm:auxiliary_success_Aell}} \label{app:proof_clm_auxiliary_success_Aell}

By definition of $X_q$, we have
\[ \exsub{ \Theta_\ell }{ X_q } ~~=~~ \exsub{ \Theta_{ \ell, q } }{ \sum_{i \in {\cal H}_{\ell,q} } \gamma_i   T_i^{ \Theta_{\ell, q} } } ~~=~~ \frac{ 1 }{ \sqrt{2} \ln 2 } \cdot \sum_{i \in {\cal H}_{\ell,q} } \gamma_i  \hat{T}_i ~~=~~ \frac{ \eta_q }{ \sqrt{2} \ln 2 } \ , \] where the second equality holds since $\exsubpar{ \Theta_{\ell, q} } { T_i^{ \Theta_{\ell, q} } } = \frac{ 1 }{ \sqrt{2} \ln 2 } \cdot \hat{T}_i$ , by Claim~\ref{clm:properties_PO2}(1). Similarly,
\[ X_q ~~=~~ \sum_{i \in {\cal H}_{\ell,q} } \gamma_i   T_i^{ \Theta_{\ell, q} } ~~\leq~~ \sqrt{2} \cdot \sum_{i \in {\cal H}_{\ell,q} } \gamma_i   \hat{T}_i ~~=~~ \sqrt{2} \eta_q \ , \]
where the inequality above holds since $T_i^{ \Theta_{\ell, q} } \leq \sqrt{2} \hat{T}_i$,  by  Claim~\ref{clm:properties_PO2}(2).  

Now, in regard to the relation between $\eta_{\max}$ and $\eta_{\min}$, we have on the one hand,
\[ \eta_{\max} ~~=~~ \max_{q \in [Q]} \left\{ \sum_{i \in {\cal H}_{\ell,q} } \gamma_i   \hat{T}_i \right\} ~~\leq~~ \left( \left\lfloor \frac{ |{\cal H}_{ \ell }| }{ Q } \right\rfloor+1 \right) \cdot \frac{2}{(1+\eps)^{\ell-1}} \cdot{\cal V} ~~\leq~~  \frac{ |{\cal H}_{ \ell }| }{ Q } \cdot \frac{2(1+\eps)}{(1+\eps)^{\ell-1}} \cdot{\cal V}  \ . \]
Here, the first inequality is obtained by noting that each subset ${\cal H}_{\ell,q}$ is of size either $\lfloor \frac{ |{\cal H}_{ \ell }| }{ Q } \rfloor$ or $\lceil \frac{ |{\cal H}_{ \ell }| }{ Q } \rceil$, as explained in Section~\ref{subsec:pow-rounding}. In addition, $\gamma_i \cdot \bar{I}( \hat{T}_i ) \leq \frac{1}{(1+\eps)^{\ell-1}} \cdot{\cal V}$ for every commodity $i \in {\cal H}_{\ell,q} \subseteq \tilde{\cal V}_{\ell}$, according to property~\ref{prop:hatT_space}. On the other hand, 
\[ \eta_{\min} ~~=~~ \min_{q \in [Q]} \left\{ \sum_{i \in {\cal H}_{\ell,q} } \gamma_i  \hat{T}_i \right\} ~~\geq~~ \left\lfloor \frac{ |{\cal H}_{ \ell }| }{ Q } \right\rfloor \cdot \frac{3}{2(1+\eps)^{\ell-1}} \cdot{\cal V} ~~\geq~~  \frac{ |{\cal H}_{ \ell }| }{ Q } \cdot \frac{3(1-\eps)}{2(1+\eps)^{\ell-1}} \cdot{\cal V} \ , \]
whether the first inequality holds since $\gamma_i \cdot \bar{I}( \hat{T}_i ) \geq \frac{ 3 }{ 4 } \cdot \frac{1}{(1+\eps)^{\ell-1}} \cdot{\cal V}$, by definition of 
${\cal H}_{\ell}$. Combining these bounds on $\eta_{\max}$ and $\eta_{\min}$, it follows that $\frac{ \eta_{\min}}{ \eta_{\max} } \geq \frac{ 3(1-\eps) }{ 4(1 + \eps) } \geq \frac{ 1 }{ 2 }$.

\subsection{Proof of Lemma~\ref{lem:PO2_bound_Vmax}} \label{app:proof_lem_PO2_bound_Vmax}

\paragraph{Bounding $\bs{V_{\max}( {\cal P}^{\ell-} )}$.} Starting with the easy claim, we derive an upper bound on the maximal space requirement of ${\cal P}^{\ell-}$ by recalling that ${\cal P}_i^{\ell-} = \alpha \hat{T}_i =  \frac{ 7/8 }{ \sqrt{2} \ln 2 } \cdot \hat{T}_i$ for every commodity $i \in \tilde{\cal V}_{\ell}$. Therefore,
\begin{eqnarray*}
V_{\max}( {\cal P}^{\ell-} ) & = & \alpha \cdot \sum_{i \in \tilde{\cal V}_{\ell}} \gamma_i  \hat{T}_i \\
& \leq & 2\alpha \cdot | \tilde{\cal V}_{\ell} | \cdot \frac{1}{(1+\eps)^{\ell-1}} \cdot{\cal V}  \\
& = & \frac{ 7/4 }{ \sqrt{2} \ln 2 } \cdot | \tilde{\cal V}_{\ell} | \cdot \frac{1}{(1+\eps)^{\ell-1}} \cdot{\cal V} \ ,
\end{eqnarray*}
where the inequality above holds since $\gamma_i \cdot \bar{I}( \hat{T}_i ) \leq \frac{1}{(1+\eps)^{\ell-1}} \cdot{\cal V}$, by property~\ref{prop:hatT_space}.

\paragraph{Bounding $\bs{V_{\max}( {\cal P}^{\ell+} )}$.} To upper-bound the random maximal space requirement associated with $\tilde{\cal P}^{\ell+}$, note that
\begin{eqnarray}
V_{\max}( \tilde{\cal P}^{\ell+} ) & \leq & \underbrace{ \left[ \left. \sum_{q \in [Q]} \sum_{ \{ \Pi_{ \ell, q }(2i-1), \Pi_{ \ell, q }(2i) \} \in {\cal N}_{ \ell ,q}^{ \Pi_{ \ell, q } } } V_{\max}( {\cal P}_{ \Pi_{ \ell, q }(2i-1) }^{\ell}, {\cal P}_{ \Pi_{ \ell, q }(2i)}^{\ell} ) \right| {\cal A}_{\ell} \right] }_{ \text{type 1} }  \nonumber \\
&& \mbox{} + \underbrace{ \left[ \left. \sum_{ i \in {\cal T}_2^{ \Theta_\ell } } \gamma_i  T_i^{ \Theta_{\ell} } \right| {\cal A}_{\ell} \right] }_{ \text{type 2} } + \underbrace{ \left[ \left. \sum_{ i \in {\cal T}_3^{ \Theta_\ell } } \gamma_i  T_i^{ \Theta_{\ell} } \right| {\cal A}_{\ell} \right] }_{ \text{type 3} } + \underbrace{ \sum_{ i \in {\cal T}_4 } \gamma_i  \hat{T}_i }_{ \text{type 4} } \ . \label{eqn:proof_lem_PO2_bound_Vmax_eq6}
\end{eqnarray}
We proceed by separately bounding each of these terms.
\begin{enumerate}
    \item {\em Bounding type-1}: By Corollary~\ref{cor:construct_sub1_near}(1), we have
    \begin{eqnarray}
    && \underbrace{ \left[ \left. \sum_{q \in [Q]} \sum_{ \{ \Pi_{ \ell, q }(2i-1), \Pi_{ \ell, q }(2i) \} \in {\cal N}_{ \ell ,q}^{ \Pi_{ \ell, q } } } V_{\max}( {\cal P}_{ \Pi_{ \ell, q }(2i-1) }^{\ell}, {\cal P}_{ \Pi_{ \ell, q }(2i)}^{\ell} ) \right| {\cal A}_{\ell} \right] }_{ \text{type 1} } \nonumber \\
    && \qquad \leq~~ (1 + \eps) \cdot \frac{ 7 }{ 8 } \cdot \left[ \sum_{q \in [Q]} \sum_{ \{ \Pi_{ \ell, q }(2i-1), \Pi_{ \ell, q }(2i) \} \in {\cal N}_{ \ell ,q}^{ \Pi_{ \ell, q } } } \left(  \gamma_{ \Pi_{ \ell, q }(2i-1) } \cdot T_{ \Pi_{ \ell, q }(2i-1)}^{ \Theta_{\ell, q} } \right. \right. \nonumber \\ 
    && \qquad \qquad \qquad \qquad \qquad \qquad \qquad \qquad \qquad \left. \left. \vphantom{ \sum_{ \{ \Pi_{ \ell, q }(2i-1), \Pi_{ \ell, q }(2i) \} \in {\cal N}_{ \ell ,q}^{ \Pi_{ \ell, q } } } } \left. \mbox{} + \gamma_{ \Pi_{ \ell, q }(2i) } \cdot T_{ \Pi_{ \ell, q }(2i)}^{ \Theta_{\ell, q} } \right) \right| {\cal A}_{\ell} \right] \nonumber \\
    && \qquad \leq~~ (1 + \eps) \cdot \frac{ 7 }{ 8 } \cdot \left[ \left. \sum_{i \in {\cal H}_{\ell}} \gamma_i T_i^{ \Theta_{\ell} } \right| {\cal A}_{\ell} \right] \label{eqn:proof_lem_PO2_bound_Vmax_eq1} \\
    && \qquad \leq~~ (1 + 3\eps) \cdot \frac{ 7/8 }{ \sqrt{2} \ln 2 } \cdot \sum_{i \in {\cal H}_{\ell}} \gamma_i  \hat{T}_i \label{eqn:proof_lem_PO2_bound_Vmax_eq2} \\
    && \qquad \leq~~ (1 + 3\eps) \cdot \frac{ 7/4 }{ \sqrt{2} \ln 2 } \cdot | {\cal H}_{\ell} | \cdot \frac{1}{(1+\eps)^{\ell-1}} \cdot{\cal V} \ . \label{eqn:proof_lem_PO2_bound_Vmax_eq3}     
    \end{eqnarray}
    Here, inequality~\eqref{eqn:proof_lem_PO2_bound_Vmax_eq1} utilizes the basic observation that the set of commodities residing in the random collection of near pairs ${\cal N}_{ \ell ,q}^{ \Pi_{ \ell, q } }$ is a subset of ${\cal H}_{\ell,q}$. Inequality~\eqref{eqn:proof_lem_PO2_bound_Vmax_eq2} follow by recalling that ${\cal A}_{\ell}$ stands for the event where $\sum_{i \in {\cal H}_{\ell} } \gamma_i T_i^{ \Theta_{\ell} } \leq (1 + \eps) \cdot \frac{ 1 }{ \sqrt{2} \ln 2 } \cdot \sum_{i \in {\cal H}_{\ell} } \gamma_i   \hat{T}_i$. Finally, inequality~\eqref{eqn:proof_lem_PO2_bound_Vmax_eq3} holds since $\gamma_i \cdot \bar{I}( \hat{T}_i ) \leq \frac{1}{(1+\eps)^{\ell-1}} \cdot{\cal V}$, by property~\ref{prop:hatT_space}.
    
    \item {\em Bounding type-2}: To bound this term, we recall that ${\cal T}_2^{ \Theta_\ell }$ stands for the collection of commodities that belong to far pairs across ${\cal H}_{ \ell , 1}, \ldots, {\cal H}_{\ell,Q}$. Therefore, by Claim~\ref{clm:bound_on_far_pairs}, we almost surely have $| {\cal T}_2^{ \Theta_\ell } | = 2 \cdot \sum_{q \in [Q]} | {\cal F}_{ \ell ,q}^{ \Pi_{ \ell, q } } | \leq \frac{ 11Q }{ 5\eps }$. Consequently,
    \begin{eqnarray}
    \underbrace{ \left[ \left. \sum_{ i \in {\cal T}_2^{ \Theta_\ell } } \gamma_i  T_i^{ \Theta_{\ell} } \right| {\cal A}_{\ell} \right] }_{ \text{type 2} } & \leq & \sqrt{2} \cdot \left[ \left. \sum_{ i \in {\cal T}_2^{ \Theta_\ell } } \gamma_i  \hat{T}_i \right| {\cal A}_{\ell} \right] \label{eqn:proof_lem_PO2_bound_Vmax_eq3.5} \\    
    & \leq & \frac{ 22 \sqrt{2} Q }{ 5\eps } \cdot \frac{1}{(1+\eps)^{\ell-1}} \cdot{\cal V} \label{eqn:proof_lem_PO2_bound_Vmax_eq4} \\
    & \leq & 4\eps \cdot |{\cal H}_{\ell}| \cdot \frac{1}{(1+\eps)^{\ell-1}} \cdot{\cal V} \label{eqn:proof_lem_PO2_bound_Vmax_eq5} \\
    & \leq & 4\eps \cdot |\tilde{\cal V}_{\ell}| \cdot \frac{1}{(1+\eps)^{\ell-1}} \cdot{\cal V} \ . \nonumber
    \end{eqnarray}
    Here, inequality~\eqref{eqn:proof_lem_PO2_bound_Vmax_eq3.5} holds since $T_i^{ \Theta_{\ell} } \leq \sqrt{2} \hat{T}_i$ almost surely, by Claim~\ref{clm:properties_PO2}(2). Inequality~\eqref{eqn:proof_lem_PO2_bound_Vmax_eq4} is obtained by recalling that $\gamma_i \cdot \bar{I}( \hat{T}_i ) \leq \frac{1}{(1+\eps)^{\ell-1}} \cdot{\cal V}$, by property~\ref{prop:hatT_space}. Finally, inequality~\eqref{eqn:proof_lem_PO2_bound_Vmax_eq5} follows due to having $\lfloor \frac{ |{\cal H}_{ \ell }| }{ Q } \rfloor \geq \frac{ 2 }{ \eps^2 }$, as shown in~\eqref{eqn:ratio_HQ_eps}.

    \item {\em Bounding type-3}: To bound this term, we recall that ${\cal T}_3^{ \Theta_\ell }$ contains at most one commodity from each of the subsets ${\cal H}_{ \ell , 1}, \ldots, {\cal H}_{\ell,Q}$, implying that $|{\cal T}_3^{ \Theta_\ell }| \leq Q$. As a result, based on argument similar to those of the previous item,
    \[ \underbrace{ \left[ \left. \sum_{ i \in {\cal T}_3^{ \Theta_\ell } } \gamma_i  T_i^{ \Theta_{\ell} } \right| {\cal A}_{\ell} \right] }_{ \text{type 3} } ~~\leq~~ 2 \sqrt{2} Q \cdot \frac{1}{(1+\eps)^{\ell-1}} \cdot{\cal V} ~~\leq~~ \sqrt{2}\eps^2 \cdot |\tilde{\cal V}_{\ell}| \cdot \frac{1}{(1+\eps)^{\ell-1}} \cdot{\cal V} \ . \]

    \item {\em Bounding type-4}: By definition, we know that  $\gamma_i \cdot \bar{I}( \hat{T}_i ) \leq \frac{ 3 }{ 4 } \cdot \frac{1}{(1+\eps)^{\ell-1}} \cdot{\cal V}$ for every commodity $i \in {\cal L}_{\ell}$. Hence,
    \[ \underbrace{ \sum_{ i \in {\cal T}_4 } \gamma_i  \hat{T}_i  }_{ \text{type 4} } ~~\leq~~ \frac{ 3 }{ 2 } \cdot | {\cal L}_{\ell} | \cdot  \frac{1}{(1+\eps)^{\ell-1}} \cdot{\cal V} \ .  \]
\end{enumerate}
In summary, by plugging the above-mentioned bounds into inequality~\eqref{eqn:proof_lem_PO2_bound_Vmax_eq6}, we conclude that 
\begin{eqnarray*}
V_{\max}( {\cal P}^{\ell+} ) & \leq &  (1 + 3\eps) \cdot \frac{ 7/4 }{ \sqrt{2} \ln 2 } \cdot | {\cal H}_{\ell} | \cdot \frac{1}{(1+\eps)^{\ell-1}} \cdot{\cal V}  \\
&& \mbox{} + 4\eps \cdot |\tilde{\cal V}_{\ell}| \cdot \frac{1}{(1+\eps)^{\ell-1}} \cdot{\cal V} \\
&& \mbox{} + \sqrt{2}\eps^2 \cdot |\tilde{\cal V}_{\ell}| \cdot \frac{1}{(1+\eps)^{\ell-1}} \cdot{\cal V} \\
&& \mbox{} + \frac{ 3 }{ 2 } \cdot | {\cal L}_{\ell} | \cdot  \frac{1}{(1+\eps)^{\ell-1}} \cdot{\cal V} \\
& \leq &  (1 + 6\eps) \cdot  \frac{ 7/4 }{ \sqrt{2} \ln 2 } \cdot | \tilde{\cal V}_{\ell} | \cdot \frac{1}{(1+\eps)^{\ell-1}} \cdot{\cal V} \ .
\end{eqnarray*}

\subsection{Proof of Lemma~\ref{lem:PO2_bound_cost}} \label{app:proof_lem_PO2_bound_cost}

\paragraph{The cost of $\bs{{\cal P}^{\ell-}}$.} Since ${\cal P}^{\ell-}$ is a determinstic policy, with ${\cal P}_i^{\ell-} = \alpha \hat{T}_i$ for every commodity $i \in \tilde{\cal V}_{\ell}$, we have
\begin{equation} \label{eqn:proof_lem_PO2_bound_cost_eq4}
C( {\cal P}^{\ell-} ) ~~=~~ \sum_{ i \in \tilde{\cal V}_{\ell} } C_i( \alpha \hat{T}_i ) ~~\leq~~ \max \left\{ \alpha, \frac{ 1 }{ \alpha } \right\} \cdot \sum_{ i \in \tilde{\cal V}_{\ell} } C_i( \hat{T}_i ) ~~\leq~~ 2 \cdot \sum_{ i \in \tilde{\cal V}_{\ell} } C_i( \hat{T}_i ) \ ,
\end{equation}
where the first inequality follows from Claim~\ref{clm:EOQ_properties}(3), and the second inequality is obtained by recalling that $\alpha =  \frac{ 7/8 }{ \sqrt{2} \ln 2 } \approx 0.892$. 

\paragraph{The expected cost of $\bs{{\cal P}^{\ell+}}$.} We obtain an upper bound on the expected long-run average cost of the policy ${\cal P}^{\ell+}$ by initially overlooking the  conditioning on ${\cal A}_{\ell}$. With respect to the latter, we have
\begin{eqnarray}
&& \exsub{ \Theta_{\ell} }{ C( {\cal P}^{\ell} ) } \nonumber \\
&& \qquad \leq~~ \underbrace{ \exsub{ \Theta_{\ell} }{ \sum_{q \in [Q]} \sum_{ \{ \Pi_{ \ell, q }(2i-1), \Pi_{ \ell, q }(2i) \} \in {\cal N}_{ \ell ,q}^{ \Pi_{ \ell, q } } } \left( C_{ \Pi_{ \ell, q }(2i-1) }( {\cal P}^{\ell}_{ \Pi_{ \ell, q }(2i-1) } ) + C_{ \Pi_{ \ell, q }(2i) }( {\cal P}^{\ell}_{ \Pi_{ \ell, q }(2i) } ) \right) } }_{ \text{type 1} } \nonumber \\
&& \qquad \qquad  \mbox{} + \underbrace{ \exsub{ \Theta_{\ell} }{ \sum_{ i \in {\cal T}_2^{ \Theta_\ell } \cup {\cal T}_3^{ \Theta_\ell }  } C_i( T_i^{ \Theta_{\ell }} ) } }_{ \text{types 2 and 3} } + \underbrace{ \sum_{ i \in {\cal T}_4 } C_i( \hat{T}_i ) }_{ \text{type 4} }\nonumber \\
&& \qquad \leq~~ \frac{ 32 }{ 31 } \cdot \exsubnop{ \Theta_{\ell} } \left[ \sum_{q \in [Q]} \sum_{ \{ \Pi_{ \ell, q }(2i-1), \Pi_{ \ell, q }(2i) \} \in {\cal N}_{ \ell ,q}^{ \Pi_{ \ell, q } } }  \Big(  C_{ \Pi_{ \ell, q }(2i-1) }( T_{ \Pi_{ \ell, q }(2i-1)}^{ \Theta_{\ell, q} } )  \right. \nonumber \\
&& \qquad \qquad \qquad \qquad \left. \vphantom{\sum_{q \in [Q]} \sum_{ \{ \Pi_{ \ell, q }(2i-1), \Pi_{ \ell, q }(2i) \} \in {\cal N}_{ \ell ,q}^{ \Pi_{ \ell, q } } } } 
\mbox{} + C_{ \Pi_{ \ell, q }(2i) }( T_{ \Pi_{ \ell, q }(2i)}^{ \Theta_{\ell, q} } ) \Big) \right] + \exsub{ \Theta_{\ell} }{ \sum_{ i \in {\cal T}_2^{ \Theta_\ell } \cup {\cal T}_3^{ \Theta_\ell }  } C_i( T_i^{ \Theta_{\ell }} ) }  + \sum_{ i \in {\cal T}_4 } C_i( \hat{T}_i ) \label{eqn:proof_lem_PO2_bound_cost_eq1} \\
&& \qquad \leq~~ \frac{ 32 }{ 31 } \cdot \exsub{ \Theta_{\ell} }{ \sum_{ i \in {\cal H}_{\ell} } C_i( T_i^{ \Theta_{\ell}} ) } + \sum_{ i \in {\cal L}_{\ell} } C_i( \hat{T}_i )\nonumber \\
&& \qquad =~~ \frac{ 32/31 }{ \sqrt{2} \ln 2 } \cdot \sum_{ i \in {\cal H}_{\ell} } C_i( \hat{T}_i ) + \sum_{ i \in {\cal L}_{\ell} } C_i( \hat{T}_i ) \label{eqn:proof_lem_PO2_bound_cost_eq2} \\
&& \qquad \leq~~ \frac{ 32/31 }{ \sqrt{2} \ln 2} \cdot \sum_{i \in \tilde{\cal V}_{ \ell } } C_i( \hat{T}_i ) \ . \label{eqn:proof_lem_PO2_bound_cost_eq5}
\end{eqnarray}
Here, inequalities~\eqref{eqn:proof_lem_PO2_bound_cost_eq1} and~\eqref{eqn:proof_lem_PO2_bound_cost_eq2} respectively follow from Corollary~\ref{cor:construct_sub1_near}(2) and Claim~\ref{clm:properties_PO2}(1).

Now, to derive an upper bound on the expected cost of the conditional policy ${\cal P}^{\ell+} = [ {\cal P}^{\ell} | {\cal A}_{\ell} ]$, we observe that
\[ \exsub{ \Theta_{\ell} }{ C( {\cal P}^{\ell} ) } ~~\geq~~ \prsub{ \Theta_\ell }{ {\cal A}_{\ell} } \cdot \exsub{ \Theta_{\ell} }{ \left. C( {\cal P}^{\ell} ) \right| {\cal A}_{\ell} }  ~~\geq~~ \left( 1 - \frac{ \eps }{ 10 } \right) \cdot \exsub{ \Theta_{\ell} }{ C( {\cal P}^{\ell+} ) } \ ,  \]
where the last inequality follows from Lemma~\ref{lem:success_Aell}. Therefore,
\begin{equation} \label{eqn:ex_PLplus_UB}
\exsub{ \Theta_{\ell} }{ C( {\cal P}^{\ell+} ) } ~~\leq~~ \left( 1 + \frac{ \eps }{ 5 } \right) \cdot \frac{ 32/31 }{ \sqrt{2} \ln 2} \cdot \sum_{i \in \tilde{\cal V}_{ \ell } } C_i( \hat{T}_i ) \ .
\end{equation}

\paragraph{Putting everything together.} We remind the reader that the policy $\tilde{\cal P}^{ \ell }$ we return is precisely ${\cal P}^{\ell+}$ when the event ${\cal A}_{\ell}$ occurs; otherwise, it identifies with the deterministic policy ${\cal P}^{\ell-}$. Therefore, 
\begin{eqnarray}
\exsub{ \Theta_{\ell} }{ C( \tilde{\cal P}^{\ell} ) } & = & \prsub{ \Theta_\ell }{ {\cal A}_{\ell} } \cdot \exsub{ \Theta_{\ell} }{ C( {\cal P}^{\ell+} ) } + \prsub{ \Theta_\ell }{ \bar{\cal A}_{\ell} } \cdot  C( {\cal P}^{\ell-} ) \nonumber \\
& \leq & \exsub{ \Theta_{\ell} }{ C( {\cal P}^{\ell+} ) } + \prsub{ \Theta_\ell }{ \bar{\cal A}_{\ell} } \cdot  C( {\cal P}^{\ell-} ) \nonumber \\
& \leq & \left( 1 + \frac{ \eps }{ 5 } \right) \cdot \frac{ 32/31 }{ \sqrt{2} \ln 2} \cdot \sum_{i \in \tilde{\cal V}_{ \ell } } C_i( \hat{T}_i ) + \frac{ \eps }{ 5 } \cdot \sum_{ i \in \tilde{\cal V}_{\ell} } C_i( \hat{T}_i ) \label{eqn:proof_lem_PO2_bound_cost_eq3} \\
& \leq & \left( 1 + \frac{ 2\eps }{ 5 } \right) \cdot \frac{ 32/31 }{ \sqrt{2} \ln 2} \cdot \sum_{i \in \tilde{\cal V}_{ \ell } } C_i( \hat{T}_i ) \ , \nonumber
\end{eqnarray}
where inequality~\eqref{eqn:proof_lem_PO2_bound_cost_eq3} is obtained by recalling that $\prsubpar{ \Theta_\ell }{ \bar{\cal A}_{\ell} } \leq \frac{ \eps }{10}$, according to Lemma~\ref{lem:success_Aell}, and by plugging-in~\eqref{eqn:proof_lem_PO2_bound_cost_eq4} and~\eqref{eqn:ex_PLplus_UB}.

\end{document}